\documentclass[12pt,a4paper,fleqn]{article}

\usepackage[margin=1.25in]{geometry}
\usepackage{setspace}
\usepackage{endnotes}
\let\footnote=\endnote
\usepackage{newtxtext}
\usepackage{newtxmath}
\usepackage{amsmath}
\usepackage{latexsym}
\usepackage{natbib}
\usepackage{xcolor}
\usepackage[pdftex]{graphicx}
\usepackage{pgf}
\usepackage{tikz}
\usetikzlibrary {arrows,decorations}
\usetikzlibrary {shapes}
\usetikzlibrary{decorations.markings}
\usetikzlibrary{positioning,calc,fit}
\usepackage{url}
\usepackage{booktabs}
\usepackage{afterpage}
\usepackage{pgfplots}
\pgfplotsset{compat=1.18}
\usepgfplotslibrary{groupplots}

\newif\ifstandalone
\standalonefalse
\newcommand{\dmsection}[1]{\subsection{#1}}
\newcommand{\dmsubsection}[1]{\subsubsection*{#1}}
\providecommand{\betah}{\widehat{\beta}}
\providecommand{\Sigmah}{\widehat{\Sigma}}
\providecommand{\E}{\mathbb{E}}
\providecommand{\Var}{\operatorname{Var}}
\providecommand{\Cov}{\operatorname{Cov}}

\providecommand{\ME}{\mathrm{ME}}

\newif\ifblinded
\blindedfalse 

\newif\ifarxiv
\arxivtrue

\newcommand{\PP}{\mathrm{P}}
\newcommand{\EE}{\mathrm{E}}

\DeclareMathAlphabet{\mathcal}{OMS}{cmsy}{m}{n}

\setcitestyle{aysep={}} 

\title
{Marginal effects for the stochastic actor-oriented model}
\author%
{\ifblinded
\else
Daniel Gotthardt\thanks{Corresponding author: Daniel Gotthardt, Department of
Social Sciences, Universit\"{a}t Hamburg,
\texttt{daniel.gotthardt@uni-hamburg.de}.}\\
\textit{University of Hamburg}
\and
Christian Steglich\\
\textit{University of Groningen}\\
\textit{Link\"{o}ping University}
\fi
}

\date{\ifblinded\else\today\fi}

\makeatletter
\newcommand{\pdf@pagelayout}{}
\makeatother
\newcommand{\glasgowAmpCreationHi}{6.4}
\newcommand{\glasgowAmpCreationLo}{3.9}
\newcommand{\glasgowAmpMaintenanceHi}{2.3}
\newcommand{\glasgowAmpMaintenanceLo}{1.0}

\newcommand{\glasgowHomophilyMaintenance}{7.6}

\newcommand{\glasgowInteractionMinistepMaintenance}{-13.5}
\newcommand{\glasgowInteractionPeriodMaintenance}{-13.7}
\newcommand{\glasgowInteractionSDcreation}{2.8}
\newcommand{\glasgowInteractionSDmaintenance}{-12.9}
\newcommand{\glasgowInteractionSDoverall}{1.2}
\newcommand{\glasgowPredCreation}{0.43}
\newcommand{\glasgowPredMaintenance}{88.1}
\newcommand{\glasgowPredPeriodCreation}{1.65}
\newcommand{\glasgowPredPeriodMaintenance}{47.5}
\newcommand{\glasgowRecipFDcreation}{3.9}
\newcommand{\glasgowRecipFDmaintenance}{21.4}
\newcommand{\glasgowRecipFDoverall}{5.7}
\newcommand{\glasgowRecipMinistepMaintenance}{23.3}
\newcommand{\glasgowRecipPeriodMaintenance}{53.1}
\newcommand{\glasgowTransTripCreatRecipOneTTfour}{15.2}
\newcommand{\glasgowTransTripCreatRecipOneTTzero}{4.3}

\newcommand{\glasgowTransTripFDcreation}{1.0}
\newcommand{\glasgowTransTripFDmaintenance}{7.4}
\newcommand{\glasgowTransTripFDoverall}{1.7}
\newcommand{\glasgowTransTripMaintRecipZeroTTtwo}{4.3}
\newcommand{\glasgowTransTripMaintRecipZeroTTzero}{23.8}

\newcommand{\expDgpSameX}{2.0}

\newcommand{\expDgpW}{1.0}

\newcommand{\expN}{50}

\newcommand{\expOneAmeRecipGapSd}{3.55}
\newcommand{\expOneAmeRecipPct}{-30.3}
\newcommand{\expOneAmeRecipPctAbs}{30.3}

\newcommand{\expOneAmeTransTripPct}{-21.7}
\newcommand{\expOneAmeTransTripPctAbs}{21.7}

\newcommand{\expOneEndRecipGapSd}{0.67}
\newcommand{\expOneEndRecipPct}{+1.3}

\newcommand{\expOneEndTransTripGapSd}{1.34}
\newcommand{\expOneEndTransTripPct}{-3.5}

\newcommand{\expOneMaxMcse}{0.55}
\newcommand{\expOneNreps}{107}
\newcommand{\expOnePiKFull}{0.0631}

\newcommand{\expOnePiKRed}{0.0607}

\newcommand{\expOneRateFullTwo}{6.95}

\newcommand{\expOneRatePct}{+42.9}
\newcommand{\expOneRatePctAbs}{42.9}

\newcommand{\expOneRateRedTwo}{9.92}

\newcommand{\expOneThetaRecipGapSd}{4.91}
\newcommand{\expOneThetaRecipPct}{-22.6}

\newcommand{\expOneThetaTransTripPct}{-23.7}

\newcommand{\expTwoAmeTransTripPct}{-9.9}

\newcommand{\expTwoEndTransTripBelow}{2}

\newcommand{\expTwoEndTransTripPct}{+20.3}

\newcommand{\expTwoNreps}{115}

\newcommand{\expTwoRatePct}{+43.3}

\newcommand{\expTwoThetaTransTripPct}{-11.7}

\newcommand{\expWaves}{2}

\begin{document}

\maketitle

\begin{abstract}
Parameter estimates of stochastic actor-oriented models for network
evolution are notoriously hard to interpret. Especially a comparison of
parameter estimates is problematic, both between different models for the
same data set and between the same model for different data sets. Effect size
measures based on the calculation of marginal effects are proposed as an alternative 
to standardize the scale of interpretation but can and should not 
universally erase the effects of differences in scale and heterogeneity. 
Simulation experiments show how an omitted effect displaces coefficients and
marginal effects differently, and an analysis of empirical friendship network
data finds that reciprocity and triadic closure reinforce one another when ties
are created but partially substitute for one another when ties are maintained,
a contrast the coefficients do not reveal.
\end{abstract}

\vfill

\noindent
\ifarxiv
\emph{Preprint. Comments welcome.}
\else
\emph{Working paper, do not quote or distribute without permission of the authors.}
\fi

\newpage

\section{Introduction}
\label{S_Introduction}
Sociocentric social network data map the direct, bilateral dependence between
actors in a given social system. Classical examples are friendship networks
among all students in a school class or professional advice-seeking networks
among all employees of an organization. Social science theories on the
structure and dynamics of such networks can be empirically assessed with the
help of stochastic network models. Two prominent model families are the
\emph{exponential random graph model} \citep[ERGM,][]{ERGMbook} for
cross-sectional data and the \emph{stochastic actor-oriented model}
\citep[SAOM,][]{Snijders01,AMAND} for longitudinal data. Both model families
express the probability of a bilateral relationship (a \emph{network tie})
between two actors based on the characteristics of the actors and their
surrounding network neighborhood. For example, a friendship tie between two
students in a school class may depend on whether the students have the same
gender (a case of \emph{homophily}) or the number of common friends they have
among the other students (a case of \emph{closure}).

The ERGM and the SAOM enjoy increasing popularity among applied researchers
in a variety of social science disciplines. They often are interpreted as
generalizations of binary and multinomial logistic regression, respectively,
which carefully relax the strong independence assumptions of these regression
models, thus facilitating the analysis of (by definition, interdependent)
network data. The increased use of these models goes hand in hand with an
increasing demand for a richer interpretation of the analysis results, going
beyond simple significance testing of the model parameters. In particular,
effect sizes on the probability-scale such as marginal effects
\citep{LongMustillo2021} have not been available so far for the SAOM. 
Only very recently, some such measures were developed for the ERGM
\citep{Duxbury2023, Duxbury2026}.
For longitudinal data, and the SAOM in particular, analogous work has still been
lacking. It gets introduced in this article.

Interpretation problems that nonlinear models generally suffer from are still
insufficiently solved for stochastic network models. They are partially rooted in
features of the link function between the outcome variable \emph{network tie}
and the linear combination of predictors used in a model specification. For
the ERGM, this link function is the logit known from binary logistic
regression. For the SAOM, it is the multinomial logit known from multinomial
logistic regression -- but this is not the full story, as will be seen below.
The problems are caused by different flavors of \emph{unobserved
heterogeneity}. The same model parameter value will have a different impact
on probabilities for a network tie, depending not only on who
sends or receives the tie and which other variables are included in the model
specification (heterogeneity of marginal effects), but also on how much
variation remains unexplained (scaling). The main practical consequence for
applied research is that the direct comparison of parameter estimates is most
often inappropriate. This holds for the comparison of estimates obtained from
fitting different (e.g., nested) models to the same data set. It also holds for
the comparison of results obtained by fitting the same model to different data
sets, the classical case of meta-analysis. Despite being problematic, such
comparisons are regularly part of published research employing stochastic
network models. It is hard to assess what the consequences of this suboptimal
practice have been in the past, but at the same time it is safe to say that
it is better to replace it with improved practice in the future. This article
will make a critical step in this direction.

\subsection{Related work}
Recent advances in effect size interpretation for network models have focused 
primarily on the ERGM. Early attempts at probability-scale effect sizes used
simulation-based goodness-of-fit plots \citep{HGH2008}, marginal effects
visualizations \citep{SohnEtAl2020}, or average posterior probabilities for
Bayesian ERGMs \citep{EstevezEtAl2022}. \citet{Duxbury2023} first formalized
the scaling problem for ERGMs via a latent variable representation, showing that
coefficients absorb residual variation and that this biases effect size
interpretation, cross-model comparisons, and interaction coefficients, including
homophily, even when omitted variables are uncorrelated with included predictors.
As a remedy, the same paper introduced probability-scale marginal effect measures
for ERGMs based on partial derivatives and binary contrasts\footnote{The
marginal effects for ERGM are implemented in the R package
\texttt{ergMargins} \citep{Duxbury2023a}.}. \citet{DuxburyWertsching2023} extended
this to pooled ERGM (meta-regression and block diagonal estimation), showing that
scaling there introduces qualitatively new problems: unlike in the single-network
case, coefficients pooled across networks with heterogeneous latent error variance
can yield the wrong sign and significance. They further show that computing marginal
effects within a random-effects or block diagonal framework substantially reduces
this pooling bias.  \citet{Duxbury2026} raises a different limitation of the same
measures.  Because network statistics are built from one another, a higher-order
configuration cannot be moved while the lower-order structures nested within it
stay at their observed values, so a marginal effect computed that way describes a
network that could not occur.  He proposes structural effect measures that let
those induced changes propagate and that separate the total from the partial
contribution of a mechanism.  We keep deliberately to the partial reading here,
for reasons set out in Section~\ref{S_QOI}, and return to the open question in
Section~\ref{S_Extensions}.

Beyond ERGMs, econometric binary choice models with explicit network structure and 
heterogeneous rational expectations, such as those of \citet{LeeEtAl2014a} and 
\citet{LiuZhou2017}, have developed direct and indirect marginal effect concepts 
for equilibrium outcome probabilities. Although these approaches are conceptually 
related, they focus on system-level equilibrium outcome probabilities and
spillover effects, differing from the decision-opportunity-conditional (ministep)
SAOM marginal effects developed here.

To date, marginal effect measures have not been established for the most
commonly applied model family for analyzing network dynamics, the SAOM.
Part of the reason lies in the more complicated probability model. Unlike
the autoregressive temporal ERGM variants, it is explicitly formulated
as a continuous-time model. It makes a point of distinguishing between 
the observed changes and the unobserved data-generating process.
In addition, it has separate model components for the timing of
opportunities for change and the actual change. This gives rise to
a correspondingly richer differentiation of marginal effect measures,
which can either incorporate the timing aspects, or explicitly exclude
them to focus on conditional change probabilities. In this paper,
we develop marginal effects for the SAOM in the latter sense, and then
carry them to the level of a whole observation period, where the timing
component re-enters through the number of opportunities an actor receives.
\subsection{Contribution}
Building on the recent developments for ERGM marginal effects, this 
paper develops, for stochastic actor-oriented models (SAOMs),
interpretable effect size measures on the probability-scale.
The key idea is to treat marginal effects as \emph{quantities of interest}
(QOIs) in the sense of \citet{KingEtAl2000} and to compute them from a fitted
SAOM, rather than to interpret the SAOM coefficients themselves as effect sizes.

Our contributions are threefold: (i) we provide clear, probability-scale 
marginal-effect definitions for the SAOM choice component, (ii) we explicitly 
derive and formalize these as quantities of interest for SAOMs, and (iii) we 
present an extended discussion of scaling issues, showing how they affect the 
interpretation and comparability of both coefficients and marginal effects.
We also propose three practically relevant ways to aggregate these effects:
over actors at observed-wave networks, over model-implied intermediate decision
situations along simulated chains, and over a whole observation period, where the
contrast is carried to the probability that a tie is present at the end.
Together, these probability-scale effect size measures allow a different type of
comparison than coefficients do: across nested specifications of one data set
they are the more stable summary, most of all at the period level, and across
data sets they report the magnitude an effect actually attains in each
environment rather than abstracting from it.

\subsection{Overview}
To guide the reader, we briefly outline the structure of the paper and how each 
section builds on the previous ones.  
Section~\ref{S_SAOM} formally defines the SAOM as a continuous-time 
Markov chain and establishes the model's basic concepts and notation,
including ministep choice probabilities.
Section~\ref{S_Interpretation} motivates probability-scale interpretation,
introduces the quantities-of-interest (QOI) framework, clarifies why SAOM
coefficients are not the effect sizes researchers are typically after, and defines
marginal effects, first and second differences, at a single ministep decision.
Section~\ref{S_Aggregation} averages those into the quantities we propose as QOIs:
effect sizes over actors and decision contexts, and a period-level contrast that
carries the same comparison to the end of an observation period.
Section~\ref{S_Scaling} examines how scaling, unobserved heterogeneity,
and the evaluation context affect the comparability of SAOM coefficients and
probability-scale effects across model specifications and data sets.
Section~\ref{S_Example} illustrates the full pipeline of deriving probability-scale 
QOIs from theoretical assumptions, estimating the SAOM, and then computing
marginal effects and contrasting them with the SAOM coefficients
in an analysis of the Glasgow friendship data \citep{WestSweeting96}, 
showing that the distinctions drawn in the preceding sections matter 
for substantive interpretation.
Section~\ref{S_Discussion} summarizes the findings, and
Section~\ref{S_Extensions} discusses extensions, limitations, and directions for
future research.

\section{Towards marginal effects for the SAOM}
\label{S_SAOM}

The stochastic actor-oriented model (SAOM) is a continuous-time Markov chain
(CTMC) model for the evolution of a sociocentric (`complete') network for
which data are available in a sequence of network observations at a few
discrete time points \citep{Snijders96, Snijders01}. For 
simplicity and without loss of generality, consider the case of two time
points $t_1 < t_2$ at which a binary directed social network is
observed among $n$ actors that constitute a stable social system. These
network observations are denoted by $x_1, x_2$ and can be encoded as
adjacency matrices, elements of the outcome space
\begin{equation}
    \mathcal{X} \,=\, \big\{ x_{ij} \in \{ 0,1 \} \mid 1 \leq i,j \leq n ;
     i \neq j ; x_{ii} = 0 \text{ for all } i \big\} \ . \label{outcomespace}
\end{equation}
The binary \emph{tie variables} $x_{ij}$ indicate whether a directed
tie from actor $i$ to actor $j$ is present ($x_{ij}=1$) or absent
($x_{ij}=0$); note that diagonal entries $x_{ii}$ are constrained to zero
here, reflecting that they do not encode any interpersonal relationship.
In CTMC notation, the SAOM for the observed data is a stochastic
process $\{X(t) \in \mathcal{X} ; t \in [ t_1 , t_2 ]\}$.  The first
observation is not modelled but conditioned on, so the data enter through the
transition probability
\begin{equation}
    P(X(t_2)=x_2 | X(t_1)=x_1) \ , \label{likelihood}
\end{equation}
which, read as a function of the model parameters at the observed $x_1$ and
$x_2$, is the likelihood.
Assuming right-continuity and the existence of left limits on the time
dimension (\emph{c\`adl\`ag} property), define the time-dependent
\emph{transition intensity} between two states
$x', x'' \in \mathcal{X}$ as
\begin{equation}
    q_t(x', x'') \,=\, \lim_{\varepsilon \, \downarrow \, 0} 
    \frac{P\{X(t) = x'' \mid X(t-\varepsilon) = x'\}}
    {\varepsilon} \ , \qquad x'' \neq x' \ . \label{transition}
\end{equation}
A CTMC is fully characterized by its transition intensities. The SAOM
therefore can be fully sketched by spelling out what these look
like. A first simplification is to drop the time dependence of transition
intensities by assuming \emph{time-homogeneity}: they are constant over time
within an observation period.

Because the state space is finite, the process
can be decomposed into a discrete-time
Markov chain (DTMC) on the same state space and a state-dependent family of
exponentially distributed (`memoryless') waiting times between the
time points at which this DTMC is moving. For the SAOM,
these partial models are designed and formulated in an \emph{actor-oriented}
way. The aim of this is to facilitate operationalization of expectations
about actor behavior, derived from social science theory, in terms of
model parameters. To this end, define the \emph{Hamming distance} as a
metric on $\mathcal{X}$,
\begin{equation}
    d^\text{H}(x', x'') \,=\, \sum_{i \neq j} |x'_{ij}-x''_{ij}| \ , \label{hamming}
\end{equation}
and restrict the moves of the process to the immediate neighborhood in
$\mathcal{X}$. This so-called \emph{ministep assumption} implies that
the transition intensities between any two states 
$x', x'' \in \mathcal{X}$ are zero when these states are
non-adjacent, i.e., when $d^\text{H}(x', x'') \geq 2$.

\paragraph{Toggling and staying}
For adjacent states ($d^\text{H}(x', x'')=1$), it is possible to
identify a unique pair $\langle i,j \rangle$ of actors for which the
binary tie variable $x_{ij}$ differs between $x'$ and $x''$. We say that
this tie variable \emph{toggles}, either from zero to one (tie creation)
or from one to zero (tie dissolution), in a formula, $x''_{ij} = 1- x'_{ij}$.
For these transitions, let $q_{ij}(x')$ denote the corresponding
transition intensity.\footnote{Note that $x''$ need not occur in this
notation because due to the binary nature of the tie variables, it is
implicitly defined by $x'$ and the pair $\langle i,j \rangle$.}
The remaining transitions for which intensities are still undefined 
refer to situations where the process
stays in the same state ($d^\text{H}(x', x'')=0$ and hence $x'=x''$).
Transition intensities for this case of staying in the same state
are already implicitly defined by
the above as $ q_0(x') = - \sum_{i \neq j}q_{ij}(x')$.\footnote{The 
intuition behind this is that the transition
intensities for $x'' \neq x'$ are derivatives of transition probabilities as per
Equation~\ref{transition}, while the staying intensity is the derivative of
$P\{X(t)=x' \mid X(t-\varepsilon)=x'\}-1$. For any given state of origin $x'$, the
transition probabilities over all possible destination states,
including the possibility to stay at $x'$, must sum up to constant one
- and the derivative of a constant is zero.}

The SAOM now injects `actor-orientedness' by expressing transition
intensities to neighboring states as \emph{choice probabilities}.
For the same actor $i$, transition intensities $q_{ij}(x)$ are assumed
to be proportional to probabilities $p_{ij}(x)$ resulting from a
discrete choice between the $n-1$ options to change the tie from $i$
to another actor $j \neq i$ and the $n$th option to not change any
tie, abbreviated for convenience as $p_{ii}(x)$.
This can be written as $q_{ij}(x)=\lambda_i(x)p_{ij}(x)$, where
$\lambda_i(x)$ is the proportionality factor, implicitly defined by
the $p_{ij}(x)$ summing up to one over the $n$ decision outcomes.

\paragraph{Terminology} The probability $p_{ij}(x)$, for
$i \neq j$, is called the \emph{toggle probability}.
The (conditional, predicted) \emph{tie probability} is defined as
\begin{equation}
    r_{ij}(x) \,=\, 
    \begin{cases}
      p_{ij}(x), & \text{if}\ x_{ij}=0 \qquad \text{``create new tie''} \\
      1-p_{ij}(x), & \text{if}\ x_{ij}=1 \qquad \text{``not destroy old tie''}
    \end{cases}
    \label{tieprob}
\end{equation}

\paragraph{Transition rates and expected impact}
Between ministeps, the network state is constant; it changes only at
decision opportunities. The transition rates~$q_{ij}(x)$ defined above
 are the transition intensities of the CTMC $X(t)$
\citep[cf.][Def.~2.1.1]{Norris97}: $q_{ij}(x)$ governs the rate at
which tie~$(i,j)$ is toggled in state~$x$.
The factorization $q_{ij}=\lambda_i\cdot p_{ij}$ separates
\emph{how often} actor~$i$ gets to decide ($\lambda_i$) from \emph{what
they choose} ($p_{ij}$).

For tie variable $x_{ij}$ $(i \neq j)$, the expected rate of change is
$q_{ij}\,d_{ij}$ with $d_{ij}=1-2x_{ij}$: the transition rate times
the direction of change.\footnote{
  Equivalently, $\lambda_i(r_{ij}-x_{ij})$ where $r_{ij}$ is the
  expected post-ministep tie state.}
For actor~$i$'s outdegree, summing over alters gives
$\lambda_i(\sum_{j \neq i:x_{ij}=0}p_{ij}-\sum_{j:x_{ij}=1}p_{ij})$ which is the 
difference between the probability of creating a new tie and the probability of
destroying an old tie, multiplied by the rate of decision opportunities.
More generally, for any function of the network state $g(x)$ the expected
instantaneous rate of change is given by the action of the
infinitesimal generator $\mathcal{G}$ of the CTMC $X(t)$ on $g$:
\begin{equation}
\mathcal{G}g(x) \;=\;
\sum_i \lambda_i(x)\sum_{j} p_{ij}(x)\,\bigl[g(x^{\sim ij})-g(x)\bigr],
\label{eq:generator}
\end{equation}
where $x^{\sim ij}$ denotes network $x$ modified by toggling tie variable
$x_{ij}$ when $i \neq j$, or unmodified in the case $i=j$. The cases $g(x)=x_{ij}$
and $g(x)=\sum_j x_{ij}$ were calculated above for illustration. 

\paragraph{Estimation}
The fitting of the model to an empirical data set amounts to the identification
of parameters' point estimates and their precisions (frequentist
tradition), or the sampling from their posterior distributions (Bayesian
tradition). Examples of frequentist estimators would be those parameter values
under which the data set is most likely (\emph{maximum likelihood} estimation)
or under which the observed model features of the data are expected
(\emph{method-of-moments} estimation).  The latter is what we assume throughout.

\paragraph{Data-anchoring through estimation}
The generator connects model and data. Under method-of-moments (MoM)
estimation, the fitted model must approximately reproduce the observed
change in the target statistics~$S$ between waves.
Since every target statistic is a particular function~$g$ of the
network state, Dynkin's formula gives
\begin{equation}
\mathbb{E}\bigl[g(X(T))\bigr]-g(x_0)
\;=\;
\mathbb{E}\!\left[\int_0^T
  \mathcal{G}g(X(t))
  \,dt\right],
\label{eq:dynkin}
\end{equation}
so the MoM criterion pins the time-integrated generator, and hence
the transition rates~$q_{ij}$, to the observed data.\footnote{
  This follows from the Kolmogorov backward equation
  $\mathbf{P}'(t)=Q\mathbf{P}(t)$
  \citep[Thm.~2.8.3]{Norris97}; see also
  \citet[Thm.~3.5.1]{EthierKurtz86}.}
For maximum-likelihood estimation, the same transition rates enter the
likelihood through both the holding-time and the choice components,
providing an analogous data constraint.

\subsection{Multinomial logistic regression}
\label{S_ministep_choice}

Adding another layer of modeling, the probabilities $p_{ij}(x)$ for toggling tie
variable $x_{ij}$ (if $j \neq i$) or not to change the network (if $j=i$),
which we already interpreted as outcomes of probabilistic decision making
by actor $i$, now get parameterized by a \emph{discrete choice model}
\citep{BenAkivaLerman1985, Train2009, Glasgow2022}.
Specifically, the SAOM assumes a logistic link function that transforms a linear
combination of predictors that jointly evaluate the choice set into choice
probabilities. This \emph{objective function} is defined as
\begin{equation}
f_i(j; x, \beta)  \,=\,  \sum_k \beta_k s_{ki}(j; x) \, ,
\label{eq:fi}
\end{equation}
where the predictors $s_{k}$ are called \emph{effect statistics} and their weights
$\beta$ are called the \emph{model parameters}. The most simple type of effect
statistics depends only on what the network would be directly after the decision.
We then speak of \emph{evaluation effects}, and we can drop explicit
notational dependence on $j$ and write simplified 
$s_{ki}(j;x)=s_{ki}(x^{\sim ij})$. When the objective function
consists only of evaluation effects, it is called \emph{evaluation function}, 
and we can write $f_i(j; x, \beta) = f_i(x^{\sim ij}, \beta)$.

The selection of effect statistics into the objective function is the main
technical task (called \emph{model specification}) of researchers who apply
the SAOM to evaluate a data set in terms of social science theories.
With each selected statistic, the researcher adds one dimension
on which the CTMC can differentiate between the different toggle options.
These can be interpreted as single dimensions of what actors strive for
(when the corresponding parameter is positive) or avoid (when it is negative)
in their network decisions.
Among the most frequently used evaluation effects are
subgraph counts in the actor's personal network. For example, the degree to
which an actor is involved in reciprocated relationships can be expressed by
the \emph{reciprocity effect} $s^\text{rec}_{i}=\sum_j x_{ij}x_{ji}$.
For directed friendship networks, most theories would lead one to hypothesize a
positive reciprocity parameter (``actors strive for mutually acknowledged
friendships'').

The objective function in total can be interpreted as expressing how satisfied an
actor is with a toggle or, in the case of a specification with only evaluation
effects, with the resulting network. 
Using the logistic link function, the toggle
probabilities can now be expressed as depending on model parameters,
\begin{equation}
p_{ij}(x, \beta)
\;=\;
\frac{\exp\!\left(f_i(j; x,\beta)\right)}
{\sum_h \exp\!\left(f_i(h; x,\beta)\right)}
\;=\;
\frac{\exp\!\left(\sum_k \beta_k s_{ki}(j; x)\right)}
{\sum_h \exp\!\left(\sum_k \beta_k s_{ki}(h; x)\right)} \,.
\label{eq:clogit}
\end{equation}

\paragraph{Concrete ministep illustration}
We use a small stylized ministep decision situation 
(an explicit adjacency matrix and a binary ego covariate) to illustrate change 
and choice probabilities. The same baseline network 
state and parameters are used for all plots below. The baseline network state is the adjacency matrix $x$
(rows are senders, columns are receivers), shown here with the binary actor
covariate~$v$ alongside it:
{\small
\begin{equation}
x
= \bordermatrix{
& i & j & h_1 & h_2 & h_3 & h_4 \cr
i   & 0 & 0 & 1 & 1 & 0 & 0 \cr
j   & 0 & 0 & 0 & 0 & 0 & 0 \cr
h_1 & 1 & 1 & 0 & 0 & 0 & 0 \cr
h_2 & 0 & 1 & 0 & 0 & 0 & 0 \cr
h_3 & 0 & 0 & 0 & 1 & 0 & 0 \cr
h_4 & 1 & 0 & 0 & 0 & 0 & 0 \cr
}
\qquad\qquad
v
= \bordermatrix{
& \cr
& 0 \cr
& 1 \cr
& 1 \cr
& 0 \cr
& 0 \cr
& 0 \cr
}
\label{eq:ministep_adj}
\end{equation}
}

\begin{figure}[t]
\centering
\includegraphics[width=\linewidth]{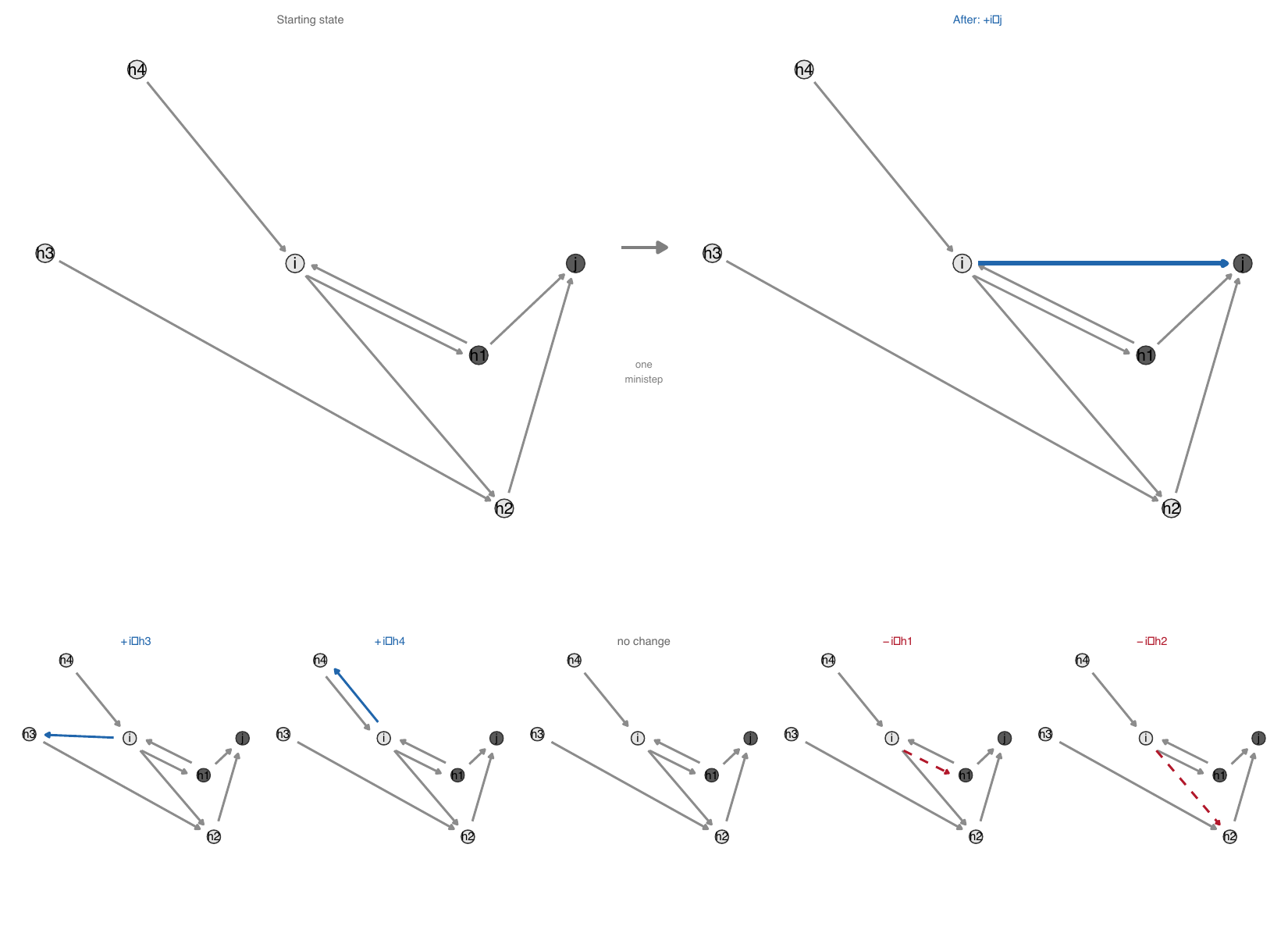}
\caption{Starting network state $x$ (left) alongside the focal choice $i\to j$
(right), both shown at full size (top row); all remaining adjacent states
reachable by actor $i$ in one ministep shown smaller below, ordered as other
creations, no-change, and dissolutions. Bold blue arrow = focal tie created;
thin blue arrow = other creation; red dashed arrow = tie dissolved.
Node fill encodes a binary ego covariate.}
\label{fig:ministep_network}
\end{figure}

Figure~\ref{fig:ministep_network} shows the starting state $x$ alongside all
adjacent states reachable by actor $i$ in a single ministep: ego $i$ can
create any absent outgoing tie, dissolve any existing one, or choose no
change. This is the decision set over which the probabilities in
\eqref{eq:clogit} are defined.

\paragraph{Change statistics}
For parameter interpretation purposes, (\ref{eq:clogit}) can be reformulated in four
steps. First, consider that the addition of a constant term to the objective function
does not change the probabilities.
This way, the decision option not to change the network ($j=i$) can be used as a natural
reference point for the objective function, against which all other options
($j \neq i$) are compared, and we get
\begin{equation}
p_{ij}(x, \beta)
\;=\;
\frac{\exp\!\left(\sum_k \beta_k \tau_{ij,k}(x) \right)}
{\sum_h \exp\!\left(\sum_k \beta_k \tau_{ih,k}(x) \right)} \, ,
\label{eq:toggleprobs}
\end{equation}
where $\tau_{ij,k}(x) = s_{ki}(j; x) - s_{ki}(i; x)$ are called \emph{toggle statistics},
they indicate how much each effect statistic changes when tie variable $x_{ij}$
gets toggled. Second, using this notation and noting that by construction $\tau_{ii,k}(x)=0$,
the log-probability-ratio between the decision to toggle tie $x_{ij}$ and the decision
$ii$ not to toggle anything can be expressed as
\begin{equation}
\ln\left( \frac{p_{ij}(x, \beta)} {p_{ii}(x, \beta)} \right) \,=\,
\sum_k \beta_k \tau_{ij,k}(x) \, .
\label{eq:logoddstau}
\end{equation}
From this equation it can be seen that when comparing any two decision options that
actor $i$ can choose from in a given situation, and when assuming \emph{ceteris
paribus} that these options differ by only one unit in only one toggle statistic
$\tau_k$, then their log-odds-ratio differs by exactly $\beta_k$. In other words,
model parameters are interpretable as \emph{conditional log-odds}. Unfortunately,
this interpretation remains unsatisfactory for several reasons. Most problematic
is that the conditions are actually quite heterogeneous, and the same log-odds ratio
can mean very different probability outcomes. An almost equally awkward aspect is
that toggling can refer to tie creation but also to tie dissolution, i.e., cover quite
opposite outcomes. When toggling the same tie variable twice, the second toggle
reverses the first. Toggles do not obey the intuitive `more is more' principle
\citep{Malmsteen2012}; they are an unusual, awkward-to-communicate QoI.

In an effort to ameliorate, a sense of order can be introduced. A toggle that creates
a tie is considered a step forward, a toggle that dissolves one is a step backward.
Toggle statistics in forward steps will be counted positively, in backward steps they
count negatively. This leads us to \emph{change statistics} as commonly used in
exponential random graph modelling \citep{ERGMbook}. We define them in the third step as 
\begin{equation}
 \delta_{ij,k}(x) \,=\,
 \begin{cases}
  \hphantom{-}\tau_{ij,k}(x)  &\qquad \text{if } x_{ij} = 0 \text{, i.e., ``toggle creates a tie''} \\
  -\tau_{ij,k}(x) &\qquad \text {if } x_{ij} = 1 \text{, i.e.,  ``toggle dissolves a tie''}\ .
 \end{cases}
\label{eq:changestat}
\end{equation}
The change statistic indicates by how much an effect statistic increases when a tie gets
created, and by how much it decreases when it gets dissolved. Evidently, the information
available in change statistics is equivalent to the information available in toggle statistics,
given the current state of the CTMC, i.e., knowing which ties exactly are present and absent. 
This play with signs allows, in the final fourth step, to re-write the log-probability-ratio
of toggling correspondingly as
\begin{equation}
\ln\left( \frac{p_{ij}(x, \beta)} {p_{ii}(x, \beta)} \right) \,=\,
 \begin{cases}
  \hphantom{-}\sum_k \beta_k \delta_{ij,k}(x) \qquad \text {if toggle creates a tie} \\
  -\sum_k \beta_k \delta_{ij,k}(x) \qquad \text {if toggle dissolves a tie} \, ,
 \end{cases}
\label{eq:logoddsdelta}
\end{equation}
which shows that parameter estimates $\beta_k$ can be interpreted (1) as the effect
of a \emph{ceteris paribus} one unit increase in the change statistics $\delta_k$ on
the log-probability-ratio of creating a tie, but also (2) as the effect of a \emph{ceteris
paribus} one unit increase in the change statistic against the log-probability-ratio of dissolving a tie. In one sentence, $\beta$ is the log-probability-ratio
of the tie being present after the decision, no matter if it just was created or
already existed before.

We finish this paragraph with the observation that for \emph{evaluation effects}
the toggle statistic is antisymmetric,
$\tau_{ij,k}(x) = -\tau_{ij,k}(x^{\sim ij})$: creating a tie increases the
objective function by the same amount as dissolving the tie decreases it, so a
single change statistic serves both margins.

\paragraph{Specification as theory}
Based on estimation results, the SAOM allows the assessment of
evidence for social science theory. Any particular specification of an
objective function, i.e., any particular selection of effect statistics,
stands here for a potentially different theoretical understanding of their data
set, and researchers may be interested in a comparison of different such
understandings.

\paragraph{The scale of the objective function}
As shown by \citet{McFadden74}, it is possible to interpret the probabilities (\ref{eq:clogit})
as resulting from \emph{random utility maximization}. Concretely for our case, this means,
\begin{equation}
 p_{ij}(x, \beta) = \Pr \big( \forall h \neq j: F_i(j; x, \beta) > F_i(h; x, \beta) \big) \, ,
 \label{eq:ru}
\end{equation}
where $F_i(j; x, \beta) = f_i(j; x,\beta)+\varepsilon_{ij}$ is the so-called random utility
of actor $i$ to toggle tie variable $x_{ij}$. It consists of the objective function as
deterministic part and an i.i.d. Gumbel$(0,1)$ error term $\varepsilon$ as random part,
typically interpreted as `systematic' and `unexplained' parts.
This interpretation will become important in Section~\ref{S_Scaling}. We note here already
that the standard deviation of the random part $\varepsilon$, which is $\pi / \sqrt{6} \approx 1.283$,
acts as a natural comparison scale for the size differences in the deterministic part $f_i$.

\section{Interpreting the SAOM}
\label{S_Interpretation}

As the considerations about log-odds and log-probabilities already hinted at, 
interpreting SAOM parameters as substantive ``effect sizes'' is difficult.
This becomes more consequential as
researchers increasingly (i) go beyond significance testing, (ii) seek
interpretable summaries on outcome-relevant scales, and (iii) compare findings
across model specifications or across data sets. The reason is not specific to
SAOMs: in many modern empirical settings with complex data and nonlinear models,
parameters are defined on latent scales and their practical implications depend
on the evaluation context. In SAOMs the issue is amplified because the model is
formulated in continuous time at the level of \emph{latent decision
opportunities}: parameters govern both how frequently opportunities arise (rate
components) and how actors choose among ``tie toggle''
alternatives when an opportunity occurs (choice components). Neither component
type is, by itself, a probability-scale summary of between-wave network change.
This paper develops probability-scale quantities of interest for the choice
component, first conditional on the decision opportunity in which they arise
and then carried to the end of an observation period.  Summaries on the count
scale, such as the expected number of ties created between waves, bring in
further assumptions about the time scale and the interplay of rate and choice
components, and we do not pursue them here.
Moreover, as in other discrete choice settings, the choice component is only
identified up to a scale normalization of the random component, which has
long been emphasized in the discrete-choice literature 
\citep[e.g.,][]{BenAkivaLerman1985, Train2009, SwaitLouviere1993}.

We advocate that interpretation targets for SAOMs should be driven by the 
substantive question, not by what happens to be a model coefficient 
\citep[cf.][]{LundbergEtAl2021, KingEtAl2000}.
\citet{LundbergEtAl2021} emphasize that estimands and estimation
are not the same: the theoretical quantity a researcher is after should be 
clarified before choosing an estimation strategy. Like other generative models,
the SAOM encodes theoretical concepts directly into model specifications,
so that the differentiation between estimand and estimation is
traditionally de-emphasized.
The SAOM's multinomial logit choice model occupies a dual role.
On the one hand, the model was grounded in a random-utility
framework \citep{Snijders96, Snijders01}, and there is a tradition of
interpreting its parameters as reflecting actors' preferences (e.g., 
a positive reciprocity coefficient as a preference for mutual ties) or
actors' satisfaction with potential or actual relational states.
On the other hand, many applied researchers adopt the multinomial logit primarily
as a tractable functional form for expressing choice outcomes, and their
substantive interest lies less in the objective function itself than in its
\emph{implications}: how the mechanisms encoded in the model affect predicted
behavior \citep[see e.g., ][]{BravoEtAl2022, DazaKreuger2021,
HovestadtLorenz2025, PasquarettaEtAl2016}.

This motivates making the QOI framing announced in the introduction precise for
the SAOM.  The target quantities are probability-scale
quantities $
h(\lambda, \beta;\mathcal{E})$ derived from a fitted model under 
explicitly stated conditioning and averaging choices, where $\lambda$ and
$\beta$ are the SAOM's rate and objective function parameters, and where
$\mathcal{E}$ describes the so-called \emph{evaluation context}.
For the SAOM, it consists of the set of \emph{decision situations} at which the
ministep choice model is evaluated: a focal actor $i$, a network state $x(t)$,
and the associated choice set (all feasible tie toggles, plus the no-change
option). Throughout, we use ``context'' as shorthand for these ingredients.
In applied work, we ultimately want to report 
not only that mechanisms are ``positive'' or ``negative,'' but \emph{how 
strongly} they affect predicted behavior: How much do reciprocity and closure 
affect the probability that a tie is created or dissolved? Does the effect of 
closure differ between situations that are also characterized by reciprocation, and
those that are not? Et cetera.

\subsection{QOIs for dynamic networks: probabilities and expected change}
\label{S_QOI}

Substantive theories about social network dynamics are naturally phrased in
terms of in-principle-observable outcomes: which ties tend to form or
persist? And how do these tendencies differ across conditions such as actor or dyad 
characteristics? We start with a short overview of potentially useful quantities
that qualify as ``effect size'' measures for network dynamics.

Consider two network observations $x(t_1)$ and $x(t_2)$ and a binary
dyad-level condition~$c_{ij}$ based on the network or covariate state at~$t_1$
(e.g., $c_{ij}=1$ when $j$ reciprocates $i$'s tie, or when $i$ and $j$
share a covariate value).
The \emph{descriptive difference} is the difference in tie proportions at $t_2$
between dyads that satisfy the condition ($c_{ij}=1$) and those that do not ($c_{ij}=0$).
This quantity requires no model, any researcher can compute it from two
waves of panel data. It answers: ``Are ties more prevalent where the
condition holds?'' But it reflects \emph{all} mechanisms at once:
The opportunity to reciprocate a tie that existed at $t_1$ may -- as a
condition -- coincide with a higher likelihood of observing that
new, reciprocating tie at $t_2$. But the reason need not be the operation of
a \emph{reciprocity mechanism}. The tie could just as well come into existence
because it is more transitively embedded, or because correlated covariates 
favour both reciprocation and tie presence.

As explained above, the SAOM model is `generative' in the sense that it
is a process model for network changes. It generates observed network paterns
at a second observation moment, given a start of the change process at an initial
observation moment. The change process consists of a sequence
of unobserved ministeps, each governed by a multinomial choice model
that combines all mechanisms simultaneously. Because each ministep is the
outcome of a decision and hence relies on information about a full decision
situation, we can define adjusted, mechanism-specific
counterfactual contrasts that isolate a focal mechanism's
contribution from the influences of all others.
These per-ministep contrasts live at a latent level (ministeps are not
directly observed).
To build these adjusted contrasts, we need the model's per-ministep
choice probabilities and a way to isolate each mechanism's contribution.

Interpretation requires comparing model implications under alternative conditions.
We use \emph{counterfactual evaluations}: model-based contrasts that hold a 
decision situation fixed and compare predicted probabilities under 
alternative values of a focal network statistic or covariate \citep[e.g.,][]{LongMustillo2021}.
Each such contrast is a ceteris-paribus, mechanism-specific analogue of the
same condition-present-versus-absent comparison that defines the descriptive
difference, but now isolated within a single decision situation rather than
confounded across the full network.

Fixing the decision situation is not yet enough to isolate a mechanism,
however.  The tie $j\to i$ reciprocates $i\to j$, but where some $h$ with
$i\to h\to j$ exists it is also the closing tie of a three-cycle, so switching
it on would move the reciprocity and the cyclicity statistics together and the
resulting shift in probability would mix the two preferences.  That mixture is
a reasonable quantity in its own right (it is what the appearance of the tie
does), but it is not what reciprocation contributes.  We therefore move the
focal statistic and leave such companions at the values they actually have.%
\footnote{Contrasts that let a change act along one route while holding others
at the values they take in its absence are \emph{path-specific} effects
\citep{AvinEtAl2005}.  Which companions count as part of the mechanism rather
than as separate routes is a modelling judgement, stated rather than derived;
Section~\ref{S_localToIntensity} returns to the question at the period level.}
Terms that are \emph{functions} of the focal statistic are a different matter
and move with it: an interaction says how large the focal effect is in this
decision situation, so holding it fixed would report the effect for a situation
other than the one at hand.  What ties companions together here is the
objective function rather than the network: statistics the specification builds
from one another move together, while constraints on which networks can exist
are not imported \citep[the criterion that governs the structural effects
of][]{Duxbury2026}.

No network realises the resulting combination, because the network state in
which the contrast held would move the other statistics too, which is the
bundling the contrast exists to undo.  It is
also what settles an otherwise open question, since being consistent about an
added tie has no natural stopping point. It would alter $i$'s two-paths to
other alters, then $j$'s own alternatives, then the density every actor faces.
The boundary is set by the question rather than by the network: we asked what
reciprocity contributes, so the reciprocity statistic is what moves. The
contrast is therefore defined on the statistics that enter the evaluation
rather than on a network state: it is not the probability that would obtain if
$x_{ji}$ were switched off, which would move the other statistics with it.

We add the qualifier ``evaluation'' here because this construction is \emph{not} a
counterfactual network state and its evolution. We name a change but do not simulate
a second CTMC path under it, isolating instead the mapping from a single
mechanism into a probability contrast at a fixed decision situation,
while the model accounts for the remaining interdependencies.
\citet{GelmanPardoe2007} draw the same distinction for regression models, using
\emph{predictive comparison} to mark a summary of model structure rather than
an estimated causal effect. A proper counterfactual would involve a re-simulation 
of the full CTMC under changed parameters, and actually address a different set of 
questions than simple effect size assessment, because it blends choice effects 
with dynamic feedback, affecting conditions under which later choices are made.
We return to such extensions in the discussion. Working this way keeps the complication of dynamic
feedback responding to the alternative conditions at bay, and focuses on
comparisons of scenarios in the same set of choice situations as closely as possible within the SAOM decision model.
For any predictor $z$, a natural summary is the first difference 
$\Delta \PP$ in the probability of toggling a tie when $z$ 
changes (e.g., from 0 to 1), holding the evaluation context fixed;
interaction effects can be summarized by second
differences \citep[e.g.,][]{AiNorton2003, ClarkGolder2023, LongMustillo2021, Mize2019}.

Because SAOM dynamics unfold through an unobserved sequence of ministeps between
$t_1$ and $t_2$, the evaluation context $\mathcal{E}$ must be made explicit: one 
must state what distribution of decision situations is used for aggregation 
(observed-wave networks, simulated intermediate networks, creation-risk vs.\ 
dissolution-risk sets, opportunity-weighting choices).

Several established summaries---log-odds and odds ratios, semi-standardized
effects, relative importance measures \citep{IndlekoferBrandes2013}, and
entropy-based concentration \citep{Snijders04}---are useful complements but
target different objects than the probability-scale first (and second)
differences defined below \citep[overview:][]{AMAND}.%
\footnote{We also distinguish our local counterfactuals
(changing a focal statistic at a fixed decision situation) from macro-level 
counterfactual analyses that modify a fitted model and trace network-level implications 
\citep[cf.][]{Chabot2024, Duxbury2023b, SnijdersSteglich15}.}

\subsection{Defining marginal effects for the SAOM decision model}
\label{S_Explanation}

From the perspective of multinomial logit, the SAOM choice component is a model
with alternative-varying predictors: in each decision situation, the alternatives 
are the outgoing toggles $(i\to j)$ plus no change, and the predictors are the 
corresponding change statistics. Unlike standard applications with a fixed set of 
substantively stable alternatives (e.g., car/bus/bike), SAOM ``alternatives'' are 
ego-specific and their attributes vary across egos and 
across time, because they are functions of the evolving network state.

Looking at (\ref{eq:clogit}) or (\ref{eq:toggleprobs}) and noticing that the decision
maker $i$ may differ between decisions, but is constant within a decision,
it is natural to first investigate 
marginal-effect summaries for effects that differ between the alternatives $j$ 
in the choice set \citep[``alternative-specific``,][]{Glasgow2022,ClarkGolder2023}.
In what follows, discrete differences for this case are developed,
followed by explicit averaging over a stated collection of 
decision makers $i$.

We define the marginal effect at quite minuscule granularity: 
For effect $k$ in network state $x$, assuming that actor $i$ is the decision maker
and $j$ is the alternative of which the choice is being contemplated,
we perturb the change statistic $\delta_{ij,k}(x)$ by $\Delta$.  What this does
to the choice is fully described by the vector of shifts it induces in the
objective-function values across actor~$i$'s choice set,
$\Delta\boldsymbol{f}_{i}=(\Delta f_{ih})_{h\in A_i(x)}$; we write
$p_{ij}(x;\Delta\boldsymbol{f}_{i})$ for the choice
probability~\eqref{eq:toggleprobs} evaluated at the shifted values.  For an
alternative-varying statistic this reduces to a single moving entry,
$\Delta f_{ij}=(1-2x_{ij})\,\beta_k\Delta$ with all others at zero, and that is
the case developed below; an ego-specific statistic such as a gender sender
effect cannot be altered for one alter alone and therefore requires a genuine
vector change.%
\footnote{This \emph{alter-specific} perturbation is the natural counterfactual
for an alternative-varying statistic, and it is the one used throughout.  An
\emph{ego-specific} (alter-invariant) predictor needs a different construction,
because its change statistic is constant across alters within a decision
situation, $\delta_{ij,k}(x)=\delta_{i,k}(x)$ for all $j\neq i$: adding $\Delta$
for a single alter is then not a coherent ego-level counterfactual.  Changing
the ego attribute implies an \emph{ego-wide} shift, raising all creation
alternatives by $+\beta_k\Delta\delta_k$ and lowering all dissolution
alternatives by the same amount, while the no-change option is unaffected
because $\tau_{ii,k}(x)=0$ by construction.  Both modes are implemented, with
the applicable one inferred from the effect metadata and overridable; for
ego-specific effects the impact can also be summarised at the actor level, as
the shift in the total probability of creating rather than dissolving a tie at
the next opportunity, which we do not report here.}
When two effects respond to the same toggle their contributions add in the same
vector, so a joint perturbation is one manipulation rather than two.

\begin{figure}[t]
\centering
\includegraphics[width=\textwidth]{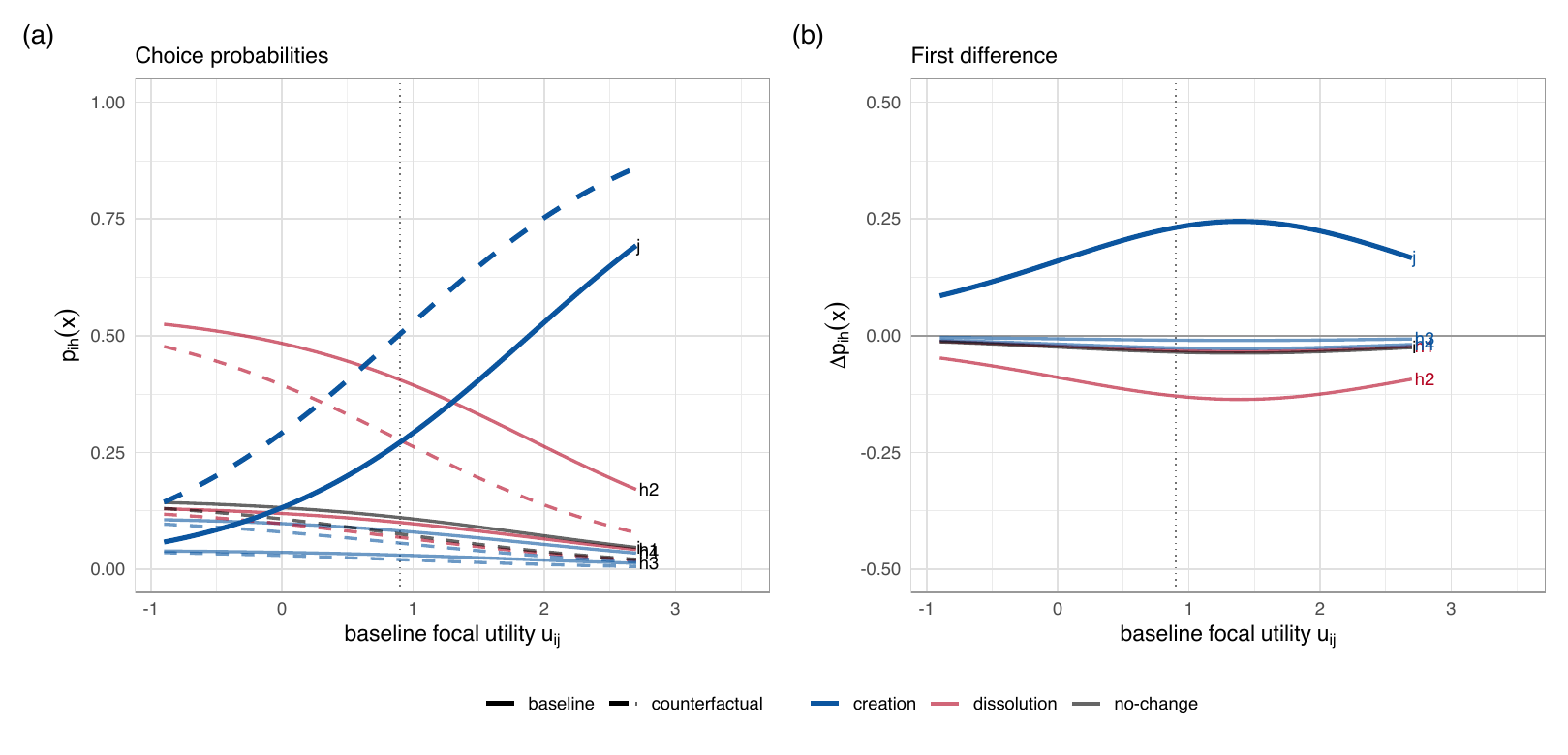}
\caption{Cross-alternative effects in a SAOM ministep choice set: shifting a
focal reciprocity statistic for $i\to j$ changes the focal toggle probability and,
through the shared denominator, also shifts the probabilities of other toggles and
the no-change option. The left panel shows predicted probabilities; the right panel shows the
corresponding probability changes $\Delta p$.}
\label{fig:altspecific_cross}
\end{figure}

Figure~\ref{fig:altspecific_cross} illustrates an alter-specific focal contrast: changing a
single effect statistic $s_{k,i}(j; x)$ for one focal alternative (e.g., reciprocity for
$i\to j$) while holding the rest of the decision situation fixed.
Because all alternatives share the same denominator in \eqref{eq:clogit}, the focal preference shift
changes the focal probability and mechanically redistributes probability mass across all
other alternatives (cross-alternative dependence).

For each effect $k$ we evaluate a discrete first difference with a fixed step size
$\Delta_k$ (typically $\Delta_k=1$ makes sense for effect statistics that are subgraph counts).
Holding all other effect statistics fixed, we compare the baseline value of effect statistic
$s_{k,i}(j;x)$ to the counterfactual value $s_{k,i}(j; x)+\Delta_k$.

The \emph{marginal effect} on the \textbf{creation} probability (evaluated
at $x_{ij}=0$) is the first difference
\begin{equation}
\mathrm{ME}^{+}_{ij,k}(\Delta \mid x)
\;=\;
p_{ij}\!\left(x;\,\Delta\boldsymbol{f}_{i}\right)
-
p_{ij}\!\left(x\right),
\qquad (x_{ij}=0).
\label{eq:MEdef_creation}
\end{equation}
That is, $\mathrm{ME}^{+}_{ij,k}$ is the increase in the
probability that $i$ creates the tie $i\to j$ under the assumption that
the focal change statistic $\delta_{ij,k}$ would be higher by the value
$\Delta$ than it actually is, holding the rest of the decision
situation fixed. This we propose as the model-based, single-mechanism counterpart of the 
descriptive difference introduced in Section~\ref{S_QOI}.
For instance, the dark blue lines in Figure~\ref{fig:altspecific_cross} show the predicted value 
for creating the non-reciprocated tie $i\to j$ and how it would shift if $j$ already reciprocated. The other 
lines demonstrate, that there is still a cross-alternative effect lowering the probability of toggling any 
another tie or not changing ego's ties at all. Importantly, only the focal statistic is perturbed; 
statistics for other alters $h\neq j$ remain unchanged, isolating the decision-level contrast from path-wise 
feedback.

Everything below is stated for this case.  The dissolution effect
$\mathrm{ME}^{-}_{ij,k}$ is the same contrast evaluated where the tie is present
($x_{ij}=1$); it needs no separate definition, because the factor $(1-2x_{ij})$
in $\Delta f_{ij}$ flips the sign of the shift, so a mechanism that raises the
probability of creating a tie lowers that of dissolving one by the same shift in
the objective function.  In practice the two differ in empirical relevance: in
sparse networks, creation contexts vastly outnumber dissolution contexts, and
existing ties tend to be more embedded, which shifts baseline probabilities and
therefore how sensitive marginal effects are to a given shift.

Because only the focal alternative's value moves, the incremental
multinomial-logit update \citep[p.~114]{BenAkivaLerman1985} gives the first
difference a transparent closed form,
\begin{equation}
\mathrm{ME}_{ij,k}(\Delta \mid x) = \frac{p_{ij}(x)(1-p_{ij}(x))\big(\exp\!\left(\Delta f_{ij}\right)-1\big)}
{1+p_{ij}(x)\big(\exp\!\left(\Delta f_{ij}\right)-1\big)},
\label{eq:ME_mlogitupdate}
\end{equation}
where $p_{ij}(x)(1-p_{ij}(x))$ is the logit slope, which determines how much the
excess odds ratio $\exp(\Delta f_{ij})-1$ affects the choice probability.  This
is the quantity the second differences below turn on: the same shift in the
objective function maps into a larger or smaller probability change depending on
where the baseline sits, so it is informally the ``room to move'' a mechanism
has.  We call it the logit slope from here on.

\paragraph{Interactions as second differences.} On the probability scale, we
define ``interaction'' (moderation) as a discrete \emph{second-difference marginal effect}:
does the (first-difference) marginal effect of one focal statistic depend on the
level of another? Fix a decision situation (state $x$ and ego $i$) and a focal
alternative $(i\to j)$. For two focal effects $k$ and $l$ with step sizes
$\Delta\delta_k$ and $\Delta\delta_l$, define the creation-context interaction marginal
effect as a difference in first differences,
\begin{equation}
\begin{split}
\mathrm{ME}^{(2)+}_{ij,k \cdot l}(\Delta\delta_{ij,k},\Delta\delta_{ij,l}\mid x)
&:=
p_{ij}\!\left(x;\,\boldsymbol{\delta}_{i,k}(x)+\Delta\delta_{ij,k},\,\boldsymbol{\delta}_{i,l}(x)+\Delta\delta_{ij,l}\right)\\
&\quad-p_{ij}\!\left(x;\,\boldsymbol{\delta}_{i,k}(x),\,\boldsymbol{\delta}_{i,l}(x)+\Delta\delta_{ij,l}\right)\\
&\quad-p_{ij}\!\left(x;\,\boldsymbol{\delta}_{i,k}(x)+\Delta\delta_{ij,k},\,\boldsymbol{\delta}_{i,l}(x)\right)\\
&\quad+p_{ij}\!\left(x;\,\boldsymbol{\delta}_{i,k}(x),\,\boldsymbol{\delta}_{i,l}(x)\right)
\\
&=\Delta^{(2)} p^{+}_{ij}(\Delta\delta_{ij,k},\Delta\delta_{ij,l}\mid x)
\end{split}
\label{eq:SecondDiff_def}
\end{equation}
where $\Delta^{(2)} p^{+}_{ij}(\Delta\delta_{ij,k},\Delta\delta_{ij,l}\mid x)$ is the corresponding
second difference in the creation probability for alternative $j$.
The analogous dissolution-context object is obtained by evaluating at $x_{ij}=1$,
where the same change statistic enters with the opposite sign.

With the second-difference definition \eqref{eq:SecondDiff_def} in hand, it is useful to
separate three distinct reasons why a probability-scale second difference can be nonzero
\footnote{\citet{BloomeAng2022} show formally under which differences in baseline probabilities 
the sign for average partial effects based on second derivatives differ from odds 
ratios and risk ratios. See in particular Appendix A3 for the multinomial logit case.}.
First, a \emph{conceptual} interaction is a moderation in the
objective function (a nonzero interaction coefficient), which changes the 
objective function differently across the two contrasts being compared. 
Second, a \emph{mechanical} interaction is a nonzero probability-scale
second difference that arises purely from the nonlinearity of the logit mapping and
differences in the baseline logit slope: even when the interaction coefficient is zero, the same
preference first difference generally maps into different $\Delta p$ across
contexts, and is constant only where the baseline slope happens to agree
\citep[``compression''; e.g.,][]{Rainey2016}. Third, a \emph{scaling-induced}
interaction can arise when the effective scale varies across actors or contexts
(heteroscedasticity), so the same contrast on the objective function is
effectively rescaled (via $f/\sigma$) and therefore maps differently into first
differences and their second differences; we return to this in Section~\ref{S_Scaling}.

\section{From decisions to average marginal effects}
\label{S_Aggregation}

A marginal effect as defined above belongs to a single decision situation: one
actor, one network state, one choice set.  Since no study can report one number
per ministep, applied work necessarily reports summaries.  This
section averages those decision-level quantities into effect sizes, states which
distribution of decision situations an average runs over, and then carries the
same contrast to the end of an observation period, where the opportunity rate
re-enters.

\subsection{Decision margins and actor-level averages}

More \textbf{conditional} marginal effects (e.g., dyad- and state-specific,
conditional on actor $i$ and decision situation $x$) are most informative for
interpretation and model diagnostics: they show where, for whom, and under which
local configurations an effect implies which probability changes.
More \textbf{average} marginal effects (e.g., averages over actors and/or
network states) are probability-scale \emph{summaries} (aggregates) of these
heterogeneous local effects. These summaries are typically less sensitive to
idiosyncratic features of any \emph{single} decision situation, but they remain
explicitly dependent on the distribution of decision situations over which they
average. When defined over comparable decision-situation distributions, average
effects are better suited for comparisons across model specifications or across
data sets.

In particular, averaging can mitigate some of the volatility induced by strong
nonlinearity and context dependence, even though both local and averaged effects
can still depend on scaling and on which parts of the state space are
represented. For transparency, we therefore (i) define conditional QOI first and
(ii) then state explicitly what is averaged over (creation-risk vs.\ 
dissolution-risk sets; observed-wave decision situations vs.\ model-implied
intermediate decision situations; and any opportunity-weighting choices).

Equation~\eqref{eq:MEdef_creation} defines marginal effects \emph{conditional on a specific
decision situation} (state $x$ and actor $i$). We first define creation and
dissolution summaries, which make the toggle-specific asymmetries explicit, and
then the joined tie-presence QOI used for substantive reporting. To make the QOI explicit 
by decision margin, for a given state $x$ and actor $i$,
define the creation- and dissolution-risk sets
\[
C_i(x)=\{j\neq i: x_{ij}=0\},
\qquad
D_i(x)=\{j\neq i: x_{ij}=1\}.
\]

For alter-varying statistics, we typically use a common one-step perturbation for all
alternatives in the decision situation, i.e., $\Delta\delta_{ij,k}\equiv\Delta\delta_k$.
The \emph{ministep-average creation marginal effect} is
\begin{equation}
\overline{\mathrm{ME}}^{+}_{i,k}(\Delta \delta_k \mid x)
=
\frac{1}{|C_i(x)|}
\sum_{j\in C_i(x)}
\mathrm{ME}^{+}_{ij,k}(\Delta \delta_{ij,k} \mid x),
\label{eq:ME_actoravg_creation}
\end{equation}
interpreted as the average probability change \emph{among ties that could be created} 
by actor $i$ at state $x$.

The analogous \emph{ministep-average dissolution marginal effect} is
\begin{equation}
\overline{\mathrm{ME}}^{-}_{i,k}(\Delta \delta_k \mid x)
=
\frac{1}{|D_i(x)|}
\sum_{j\in D_i(x)}
\mathrm{ME}^{-}_{ij,k}(\Delta \delta_{ij,k} \mid x),
\label{eq:ME_actoravg_dissolution}
\end{equation}
interpreted as the average probability change \emph{among ties that could be dissolved}.

Conceptually, the aggregation proceeds in layers. Dyad-level (alternative-specific)
contrasts $\mathrm{ME}^{\pm}_{ij,k}(\cdot\mid x)$ are first aggregated within a
decision situation to actor-level objects (risk-set averages). These
actor-level quantities are then averaged over a stated
collection (distribution) of decision situations, yielding reported effect-size
summaries such as $\mathrm{AME}^{\text{obs}}$ and $\mathrm{AME}^{\text{sim}}$.

Because the SAOM sign flip $d_{ij}=(1-2x_{ij})$ implies that the same
mechanism typically increases creation while decreasing dissolution,
directly averaging the two margins produces mechanical attenuation.
Instead, we define a joined QOI on the \emph{tie-presence scale}:
the expected tie state~$r_{ij}(x)$~\eqref{tieprob} immediately after
the ministep.  This is a natural target because the descriptive
difference of Section~\ref{S_QOI} (whether ties are more
\emph{prevalent} at~$t_2$ where a condition holds) is itself a
tie-state quantity that reflects both creation of new ties and
retention of existing ones.  The ME on the expected tie state is thus
the model-based, per-ministep counterpart of the all-inclusive tie
proportion.  A key property: because the pre-ministep state is
a fixed~0 or~1, the marginal effect on the expected tie
state~$r_{ij}$ is identically the marginal effect on the expected
per-ministep change, $\mathbb{E}[\Delta x_{ij}]=r_{ij}-x_{ij}$.

The dyad-level joined marginal effect on tie presence is the first difference of
the tie probability $r_{ij}(x)$ itself,
\begin{equation}
\mathrm{ME}^{\mathrm{tie}}_{ij,k}(\Delta \delta_{ij,k} \mid x)
=
r_{ij}\!\left(x;\,\Delta\boldsymbol{f}_{i}\right)-r_{ij}\!\left(x\right)
=
\begin{cases}
\mathrm{ME}^{+}_{ij,k}(\Delta \delta_{ij,k} \mid x), & x_{ij}=0,\\
-\mathrm{ME}^{-}_{ij,k}(\Delta \delta_{ij,k} \mid x), & x_{ij}=1.
\end{cases}
\label{eq:ME_tie_cases}
\end{equation}
The case form follows from~\eqref{tieprob}: at $x_{ij}=1$ the tie probability is
$1-p_{ij}$, so differencing it returns the negative of the dissolution-probability
difference.  The sign flip on the dissolution margin is therefore not a
convention imposed for comparability but a consequence of working on the
tie-presence scale, and it delivers a common ``higher = more likely to be
present'' reading: for an absent tie, $\mathrm{ME}^{\mathrm{tie}}$ is the effect
on creation probability; for a present tie, it is the effect on retention
probability.

To aggregate across actor $i$'s outgoing dyads at state $x$, we define 
the \emph{ministep-average} joined marginal effect as
\begin{equation}
\overline{\mathrm{ME}}^{\mathrm{tie}}_{i,k}(\Delta \delta_k\mid x)
:=
\frac{1}{n-1}\sum_{j\neq i}\mathrm{ME}^{\mathrm{tie}}_{ij,k}(\Delta \delta_k\mid x).
\label{eq:ME_actoravg_tie}
\end{equation}

\subsection{Which decision situations to average over}

Which decision situations an average runs over is precisely the evaluation
context~$\mathcal{E}$ of Section~\ref{S_QOI}, and it has to be stated rather than
assumed \citep{KingEtAl2000, GelmanPardoe2007}.  We distinguish two natural
choices. Every
quantity in the chain is a difference of expected tie indicators---conditional on
one decision situation for the local ones, and on the state at $t_1$ for the
period-level one---so all of them share the scale of the descriptive difference
of Section~\ref{S_QOI}.

\paragraph{(1) Effect size based on observed data points.}
Let $\mathcal{T}$ index the observed networks at which estimation
periods begin; for an $M$-wave panel these are the first $M-1$
observed waves, so that each $x_t$ is paired with the rate
parameter~$\lambda^{(t)}$ of the period it initiates.\footnote{
    Evaluating at period-start networks ensures that
    $\lambda^{(t)}\cdot\mathrm{AME}^{\text{obs}}_k(x_t)$ is a
    well-defined generator contribution
    (Section~\ref{S_localToIntensity}): both the opportunity rate and the
    choice contrast refer to the same observed state.  Alternative
    schemes such as including the final observed wave or using
    trapezoidal weights across period boundaries would break this
    clean pairing but may be useful for purely descriptive
    snapshot summaries.}
At an observed network $x_t$ we do not observe which actor has an opportunity next.
We therefore treat \emph{all} actors $i=1,\ldots,n$ as potential ministeps at $x_t$,
compute $\overline{\mathrm{ME}}_{i,k}(\Delta \delta_k \mid x_t)$ for each, and average across
$t$ and $i$:
\begin{equation}
\mathrm{AME}^{\text{obs}}_{k}(\Delta \delta_k)
=
\frac{1}{|\mathcal{T}|\,n}
\sum_{t\in\mathcal{T}}
\sum_{i=1}^{n}
\overline{\mathrm{ME}}_{i,k}(\Delta \delta_k \mid x_t).
\label{eq:AMEobs}
\end{equation}

$\mathrm{AME}^{\text{obs}}_{k}$ averages decision-level marginal effects 
evaluated at the observed-wave networks;%
\footnote{Actors enter with equal weight $1/n$.  Under the model the next
opportunity belongs to actor~$i$ with probability
$\lambda_i(x_t)/\sum_h\lambda_h(x_t)$, so equal weighting is model-consistent
when rates are constant across actors, as in the specification used in
Section~\ref{S_Example}.  With actor-dependent rate functions it is instead a
design choice, counting each actor's decision situation once.}
it answers: “How large are these probability contrasts in the decision situations 
that could occur directly after the given network observation?”

Because it is based on observed networks, $\mathrm{AME}^{\text{obs}}_{k}$ is
fast to compute and easy to report. It can also be meaningfully refined by
conditioning or stratifying on observable features of the decision context
(e.g., reciprocity status, degree, closure, or overall preference) to visualize and 
interpret the nonlinear, context-dependent heterogeneity of the local effects. 
For comparison
across studies or specifications, however, such conditional summaries should be
treated cautiously: they are sensitive to how decision contexts are defined
and to which parts of the state space are represented.

\paragraph{(2) Effect size based on model-implied decision sequences.}
A distinctive SAOM feature is that the sequence of ministeps between the observations is unobserved. 
A natural target for effect-size reporting is therefore an average over decision
situations that are \emph{implied by the fitted model}.  Averaging along
simulated ministep sequences rather than at observed states is the step
\citet{IndlekoferBrandes2013} take for their relative-importance measures.  What
differs here is what gets averaged: we perturb a focal change statistic at each
decision rather than comparing objective functions with and without an effect.
Let $\{(x^{(m)}, i^{(m)})\}_{m=1}^M$
denote a simulated sequence of $M$ ministeps. We then define
\begin{equation}
\mathrm{AME}^{\text{sim}}_{k}(\Delta \delta_k)
=
\frac{1}{M}\sum_{m=1}^{M}
\overline{\mathrm{ME}}_{i^{(m)},k}(\Delta \delta_k \mid x^{(m)}).
\label{eq:AMEsim}
\end{equation}
In words, $\mathrm{AME}^{\text{sim}}_{k}$ averages the same decision-level 
effects over model-implied decision situations
visited during ministeps between waves; it answers: “How large are these probability 
contrasts in the decision situations the fitted SAOM says typically occur 
during ministeps between observations?”

For comparability, $\mathrm{AME}^{\text{sim}}_{k}$ is often the more natural
summary: it averages over the model-implied distribution of intermediate decision
situations rather than conditioning only on the observed networks. As with any
summary, meaningful comparisons still require that the underlying decision
contexts (risk sets, weights, and the relevant parts of the state space) are
defined in comparable ways.

Observe that these averages are not conditional on a single network state, but on
the averaging distribution. They can still differ substantially.
For example, if closure effects mainly matter in highly 
clustered intermediate configurations that occur between waves,
then $\mathrm{AME}^{\text{sim}}$ can be noticeably larger than 
$\mathrm{AME}^{\text{obs}}$, even when both are computed from the same fitted model.

\paragraph{Conditioning on decision-context features.}
All summaries above can be stratified by restricting the averaging to
alternatives (or actors) that share a common decision-context feature
(e.g., reciprocity status, degree, or closure level). Such conditional
AMEs reveal heterogeneity across contexts and are useful for
visualization and diagnostics, but they should be reported alongside
their empirical prevalence (e.g., the number of qualifying dyads) and
treated cautiously for cross-study comparisons, because they are
sensitive to how strata are defined.

\subsection{From per-opportunity contrasts to period-level effects}
\label{S_localToIntensity}

The per-opportunity MEs defined above are adjusted,
mechanism-specific counterfactual contrasts, but they live at the
latent ministep level.  If ministeps were observed, one could directly
compare these ``mini-impacts'' against their empirical counterparts;
but ministeps are not observed.  This section connects the adjusted
contrasts to the level at which we \emph{do} have an observable between-wave 
difference in tie proportions.%
\footnote{One intermediate step is worth naming but not pursuing here.  The
transition rate $q_{ij}(x)=\lambda_i(x)\,p_{ij}(x)$ separates how often
actor~$i$ decides from what they choose, so multiplying a per-opportunity
effect by $\lambda_i(x)$ turns it into a rate of change,
$\lambda_i(x)\,\mathrm{ME}_{ij,k}(\Delta\delta_k\mid x)$.  This is the
marginal-effect analogue of the generator~\eqref{eq:generator} and describes
how fast mechanism~$k$ shifts tie probabilities at a single instant.  It
remains a snapshot at one network state rather than a between-wave summary,
which is what we are after.} As \citet{BloomeAng2022} argue, cumulative 
probability differences can also substantially be more important than
instantaneous probability differences, if long term outcomes are the focus
rather than decision mechanisms.

We keep throughout to the narrowest reading of what isolating a mechanism
requires: the focal effect's own statistic moves, terms that are functions of
it move with it, and everything else is held at the value it actually took,
including, now that a whole period is in view, the trajectory of the rest of
the network.

\paragraph{A single tie variable.}
The state space $\mathcal{X}$ has $2^{n(n-1)}$ elements, so~\eqref{eq:generator}
cannot be propagated directly.  But the question we want to answer is about a
\emph{single} tie variable, and there the model collapses: only actor~$i$'s
ministeps can change $x_{ij}$, so if we follow the rest of the network along its
simulated trajectory, the focal dyad is a two-state chain.  It is created with
the toggle probability~\eqref{eq:toggleprobs} evaluated at $x_{ij}=0$ and
dissolved with the same probability evaluated at $x_{ij}=1$,
\begin{equation}
c_{ij}(x) \;=\; p_{ij}(x)\big|_{x_{ij}=0},
\qquad
d_{ij}(x) \;=\; p_{ij}(x)\big|_{x_{ij}=1}.
\label{eq:branchprobs}
\end{equation}
Between ministeps the network is constant, so the chain is determined by its
behaviour at $i$'s opportunities.  Let $\sigma_1<\dots<\sigma_K$ index those
opportunities in $[t_1,t_2]$, write $c_{ij,k}$ and $d_{ij,k}$ for the two
probabilities at $x(\sigma_k)$, and note that the count $K$ is random across
simulated paths.

\paragraph{Propagating the tie probability.}
What we need at $t_2$ is the quantity already defined at a single ministep: the
tie probability~\eqref{tieprob}, the chance that the tie is present.  What has
to change is its conditioning.  At a ministep it reads $c_{ij,k}$ if the focal
tie is absent and $1-d_{ij,k}$ if it is present, conditioning on a state we
know.  Over a period we do not know it: after the first opportunity the state
of the dyad is itself a probability.  Carrying that probability forward from
$\pi_{ij,0}=x_{ij}(t_1)$,
\begin{equation}
\pi_{ij,k}
\;=\;
\underbrace{\bigl(1-\pi_{ij,k-1}\bigr)c_{ij,k}}_{\text{absent, created}}
\;+\;
\underbrace{\pi_{ij,k-1}\bigl(1-d_{ij,k}\bigr)}_{\text{present, retained}},
\label{eq:piRecursion}
\end{equation}
which are exactly the two branches of~\eqref{tieprob}, weighted by the
probability of being in each state rather than selected by the realised one.
The endpoint $\pi_{ij,K}$, the value reached after the last of the $K$
opportunities actor~$i$ receives in the period, is the model-implied probability
that the tie is present at~$t_2$.

The recursion carries a built-in attenuation, visible in the two weights.
Creation acts only on the probability the dyad has \emph{not} already
accumulated, and retention only on the probability it has, so each opportunity
works on what is left to move.  What matters for $\pi_{ij,K}$ is accordingly
not how often a mechanism was active but how much of its influence survives to
the end of the period.%
\footnote{In survival terms $\pi_{ij,K}$ is a cumulative incidence: a bounded
probability of being in a state, not a rate accumulated over time.}
This sub-linearity in the number of opportunities is what makes the quantity
behave as it does when the fitted rate differs across specifications, which
Section~\ref{S_Scaling} takes up.

\paragraph{The period-level marginal effect.}
The effect of mechanism~$k$ on tie existence over the period is the contrast in
the endpoint,
\begin{equation}
\mathrm{ME}^{\pi}_{ij,k}(\Delta\delta_k)
\;=\;
\pi_{ij,K}(\delta_k+\Delta\delta_k)-\pi_{ij,K}(\delta_k),
\label{eq:MEpi}
\end{equation}
with the perturbation applied at \emph{every} one of $i$'s opportunities in
$[t_1,t_2]$.  This is a sustained intervention in the sense of the g-methods
literature \citep{Robins1986,HernanRobins2020}: a policy maintained over the
period, with the outcome propagated under it rather than held at its factual
value.  Because $\pi_{ij,K}$ is a probability, $\mathrm{ME}^{\pi}_{ij,k}$ is
bounded in $[-1,1]$ and is stated in the units of the descriptive difference of
Section~\ref{S_QOI}: percentage points of tie presence at~$t_2$.  Averaging over
dyads gives $\overline{\mathrm{ME}}^{\pi}_{k}$, the mechanism's adjusted
contribution to the expected tie proportion at the second observation.

\paragraph{What it conditions on.}
A mechanism held at a different value for a whole period would, in the model,
change how the network around the dyad evolves.  The
recursion~\eqref{eq:piRecursion} does not allow that: the rest of the network
follows the path it actually took, and enters only by setting $c_{ij,k}$ and
$d_{ij,k}$ at each opportunity.

This is the isolation of Section~\ref{S_QOI} carried over time, and deliberate
for the same reason: a network allowed to respond would bring the other
mechanisms with it, which is what the contrast exists to prevent.  The price is
that no single run of the model produces this combination.%
\footnote{Contrasts with this structure are established objects: the mediation
literature calls them \emph{natural} \citep{Pearl2001} or \emph{pure}
\citep{RobinsGreenland1992} direct effects, the special case of the
path-specific contrast of Section~\ref{S_QOI} in which what is held fixed lies
downstream in time rather than alongside it in the same evaluation.  That no
single realisation of the process yields one is a property of the class, not of
the present construction \citep{RobinsRichardson2010}.}
$\mathrm{ME}^{\pi}$ is not the difference between two simulations but is read
off one simulated path under two values of the focal statistic,%
\footnote{Neither the per-opportunity nor the period-level quantity therefore
requires re-estimating the model or simulating new chains.  Both are evaluated
on the ministep sequence the fitted model already produces: the per-opportunity
contrasts read the toggle probabilities~\eqref{eq:toggleprobs} at each
opportunity, and~\eqref{eq:piRecursion} carries one scalar per dyad forward
through the same pass.  The cost is a constant factor over the simulation
itself rather than an additional order.}
combining an outcome under one condition with everything it would have
influenced held at the value it took under another.

A second and smaller approximation concerns actor~$i$'s own choice set rather
than the rest of the network.  Actor~$i$ takes one action per opportunity, so
letting the focal dyad respond while $i$'s other ties stay as they were would
require $i$ to act twice, and the recursion does not resolve this.  Competition
among the alternatives is retained, since each $c_{ij,k}$ and $d_{ij,k}$
in~\eqref{eq:branchprobs} is normalised over the whole choice set, but the
focal dyad's outcome is treated as independent of the rest rather than as one
branch of the same draw.%
\footnote{The discrepancy is the probability that two of $i$'s alternatives
both change at one opportunity, of order $p_{ij}p_{ih}$ per pair, and is small
wherever the no-change option dominates.  It grows if the focal change
statistic itself depends on $i$'s other ties, which varies by specification.}

\paragraph{Closing the loop: from latent building blocks to an
observable-level decomposition.}
The restrictions set out above share a single purpose.  Fixing the decision
situation, perturbing one mechanism at a time, and holding the rest of the
network to the path it took are all ways of isolating what one mechanism
contributes, at the price of describing a world that no single run of the model
produces.  What isolation buys is that the same contrast can then be read at
more than one level without changing what it means: at a decision opportunity
as a per-ministep probability difference, over a period as
$\overline{\mathrm{ME}}^{\pi}_k$, and, because the latter is stated in the
units of the descriptive difference of Section~\ref{S_QOI}, directly alongside
that comparison.  The descriptive difference remains a different quantity,
since it confounds all mechanisms and compares across observed conditions where
the period-level effect isolates one mechanism for the same dyads.  The two are
nonetheless commensurable, so the distance between them is itself interpretable
rather than an artefact of units.  Table~\ref{tab:notation} collects the
marginal-effect quantities defined so far, at each of these levels.

\begin{table}[t]
\centering
\caption{Notation glossary for marginal-effect quantities.  Superscripts
mark the margin ($+$\,=\,creation, $-$\,=\,dissolution,
$\mathrm{tie}$\,=\,joined tie-presence); the overline marks averaging
over alters; \textsf{AME} marks averaging over decision contexts, and
$\pi$ marks propagation over a period. Subscripts index the dyad ($ij$),
actor~($i$), and mechanism~($k$).  Short forms suppress
the conditioning arguments ($\Delta\delta_k$, $x$, $x_t$, $x^{(m)}$);
see the defining equations for full notation.}
\label{tab:notation}
\footnotesize
\setlength{\tabcolsep}{3pt}
\begin{tabular}{@{}llll@{}}
\toprule
Symbol & Short form & Scope & Meaning \\
\midrule
\multicolumn{4}{@{}l}{\textit{Per-ministep, dyad-level}} \\[2pt]
$\mathrm{ME}^{+}_{ij,k}$ & $p_{ij}(\delta+\Delta)-p_{ij}(\delta)$ & dyad & Creation first difference \eqref{eq:MEdef_creation} \\
$\mathrm{ME}^{-}_{ij,k}$ & (same contrast at $x_{ij}=1$) & dyad & Dissolution first difference \eqref{eq:MEdef_creation}\\
$\mathrm{ME}^{\mathrm{tie}}_{ij,k}$ & $\mathrm{ME}^{+}$ if absent, $-\mathrm{ME}^{-}$ if present & dyad & Joined tie-presence ME \eqref{eq:ME_tie_cases}\\
$\mathrm{ME}^{(2)+}_{ij,k\cdot l}$ & $\mathrm{ME}^{+}_k(\delta_l\!+\!\Delta_l)-\mathrm{ME}^{+}_k(\delta_l)$ & dyad & Interaction (second difference) \\
\midrule
\multicolumn{4}{@{}l}{\textit{Per-ministep, actor-level (avg.\ over alters)}} \\[2pt]
$\overline{\mathrm{ME}}^{+}_{i,k}$ & $\frac{1}{|C_i|}\!\sum_{j\in C_i}\mathrm{ME}^{+}_{ij,k}$ & actor & Risk-set avg.\ creation ME \eqref{eq:ME_actoravg_creation}\\
$\overline{\mathrm{ME}}^{-}_{i,k}$ & $\frac{1}{|D_i|}\!\sum_{j\in D_i}\mathrm{ME}^{-}_{ij,k}$ & actor & Risk-set avg.\ dissolution ME \eqref{eq:ME_actoravg_dissolution}\\
$\overline{\mathrm{ME}}^{\mathrm{tie}}_{i,k}$ & $\frac{1}{n-1}\!\sum_{j}\mathrm{ME}^{\mathrm{tie}}_{ij,k}$ & actor & Avg.\ joined ME over all alters \eqref{eq:ME_actoravg_tie}\\
\midrule
\multicolumn{4}{@{}l}{\textit{Averaged over decision contexts (effect sizes)}} \\[2pt]
$\mathrm{AME}^{\text{obs}}_{k}$ & $\tfrac{1}{|\mathcal{T}|n}\sum_{t,i}\overline{\mathrm{ME}}_{i,k}$ & avg.\ over waves & Observed-wave average \eqref{eq:AMEobs}\\
$\mathrm{AME}^{\text{sim}}_{k}$ & $\tfrac{1}{M}\sum_m\overline{\mathrm{ME}}_{i^{(m)}\!,k}$ & avg.\ over ministeps & Simulated-path average \eqref{eq:AMEsim}\\
\midrule
\multicolumn{4}{@{}l}{\textit{Propagated over a period (between-wave contributions)}} \\[2pt]
$\mathrm{ME}^{\pi}_{ij,k}$ & $\pi_{ij,K}(\delta_k{+}\Delta\delta_k)-\pi_{ij,K}(\delta_k)$ & per dyad & Period-level tie-pres.\ ME \eqref{eq:MEpi}\\
$\overline{\mathrm{ME}}^{\pi}_{k}$ & $\frac{1}{n(n-1)}\!\sum_{i\neq j}\mathrm{ME}^{\pi}_{ij,k}$ & avg.\ over dyads & Period-level decomposition term\\
\bottomrule
\end{tabular}
\end{table}

\subsection{Post-estimation and uncertainty}
Every quantity defined above is a function $\widehat\tau=h(\widehat\beta)$ of the
fitted parameters, so reported point estimates are plug-in quantities evaluated
at $\widehat\beta$ (analytically where closed forms exist).  For observed-wave
summaries such as $\mathrm{AME}^{\text{obs}}$, $h(\cdot)$ evaluates the effects
at the observed $x_t$ and averages.  For model-implied summaries such as
$\mathrm{AME}^{\text{sim}}$, and for the period-level effect, $h(\cdot)$
additionally averages along simulated ministep sequences; there Monte Carlo is
part of computing the estimand itself, and we use enough internal simulations
that the approximation error is small \citep{Rainey2017}.

Uncertainty in $\widehat\beta$ can then be carried into $\widehat\tau$ in two
ways.  They target the same standard error and differ only in how the
propagation is done, so either may be reported; which is preferable is mostly a
question of cost.

The first is \emph{resampling}: draw parameter vectors from the estimated
sampling distribution of $\widehat\beta$ and re-evaluate $h(\cdot)$ at each
draw, taking the spread of the results as the sampling distribution of
$\widehat\tau$.  This is a parametric, model-based bootstrap
\citep{EfronTibshirani1994}, closely related to the Krinsky--Robb procedure for
nonlinear functions of estimated parameters
\citep{KrinskyRobb1986,KrinskyRobb1991} and to standard practice for
post-estimation quantities in nonlinear models
\citep{Herron1999,KingEtAl2000}.  It is also the frequentist counterpart of
propagating a posterior: a Bayesian analysis would push draws from the posterior
of $\widehat\beta$ through the same $h(\cdot)$, and the computation is identical
once the draws are in hand.  It makes no smoothness assumptions beyond
those already implicit in the model and needs no derivatives, but for
path-averaged quantities it re-simulates the chains at every draw, which
dominates the cost.  Appendix~\ref{A_uncertainty} gives the practical recipe.

The second is the \emph{delta method}: because $h(\cdot)$ is smooth, the
standard error follows analytically as
$\sqrt{\nabla h^\top\widehat\Sigma\,\nabla h}$, with no draws and hence no
re-simulation.  For observed-wave and first-difference quantities the gradient
is available in closed form from the multinomial-logit Jacobian.  For quantities
averaged along simulated paths a second term arises, since the parameters shift
not only the choice probabilities but the distribution of states the process
visits; it is recoverable from the complete-data score statistics estimation
already produces \citep{SchweinbergerSnijders07a}, again without further
simulation.  The period-level effect has no closed-form gradient at present and
is differentiated numerically, which is slower but yields the same quantity.
Appendix~\ref{app:deltaMethod} gives the derivations and their correspondence to
the implementation.\footnote{The marginal-effect functions are available on the
\texttt{sienaMargins} branch of \url{https://github.com/stocnet/rsiena}.
Workshop materials demonstrating these functions are available at
\url{https://github.com/Kaladani/sienaMargins-workshop}.}

\section{Scaling, heterogeneity, and comparison of effects}
\label{S_Scaling}

Section~\ref{S_Interpretation} defined probability-scale QOIs at a single
decision, and Section~\ref{S_Aggregation} averaged them over decision situations
and carried the tie probability over a period to obtain effects on tie presence
at~$t_2$. The remaining question is how coefficients and 
the different marginal effects behave when researchers attempt to compare them: 
Do coefficients attenuate as in binary-logit models, and how does this compare to the 
probability-scale QOIs? We must clearly distinguish between two different but related
scenarios: comparing the same model across different datasets (the true
heterogeneity problem), and comparing nested models within the same dataset
(the omitted variable problem).

\subsection{Scenario 1: Real differences in predictability across groups}
\label{S_scaling_groups}

We will first briefly tackle the behavior under pure differences in predictability
across groups. In the binary-logit literature, this has been phrased as a 
group-comparison problem, that can either be within one model where we are 
analyzing interaction effects, or across datasets where we are analyzing 
differences in mechanisms or try to get an estimate of the average effect and 
heterogeneity in a meta-analysis \citep{Allison1999, Mood2010, KarlsonEtAl2012, 
Breen2018, KuhaMills2018}.

To setup this scenario, we first formulate the 
scale-aware SAOM decision model. Condition on actor $i$ having an opportunity 
at network state $x$, with choice set 
$A_i(x)$ comprising the $n-1$ toggles $i \to j$ and the no-change option. 
Keeping an explicit scale parameter $\sigma$, 
the choice probabilities are:
\begin{equation}
    p_{ij}(x;\beta,\sigma_{\text{group}}) = 
        \frac{\exp\left(f_{ij}(x;\beta)/\sigma_{\text{group}}\right)}
        {\sum_{h\in A_i(x)} \exp\left(f_{ih}(x;\beta)/\sigma_{\text{group}}\right)}.
\label{eq:clogit_sigma}
\end{equation}
The variance of the random component $\varepsilon_{ij}$ is then
$\sigma_{\text{group}}^2\,\pi^2/6$ \citep[p.~40]{Train2009}, but the standard
conditional logit and SAOM fix $\sigma_{\text{group}}^2 = 1$. If ministeps were observed and
we could directly fit a conditional logit model, only the ratio $f_{ij}(x;\beta)/\sigma_{\text{group}}$ 
would be identified, and coefficients would scale by a factor of 
$1/\sigma_{\text{group}}$ \citep{BenAkivaLerman1985, Train2009}. That means that the 
estimated group coefficients e.g. in different datasets represent differences 
in preference \emph{and} predictability \citep{SwaitLouviere1993}, 
thereby leading to the same group-comparison problem as in binary logit.

\begin{figure}[h]
\centering
\includegraphics[width=\textwidth]{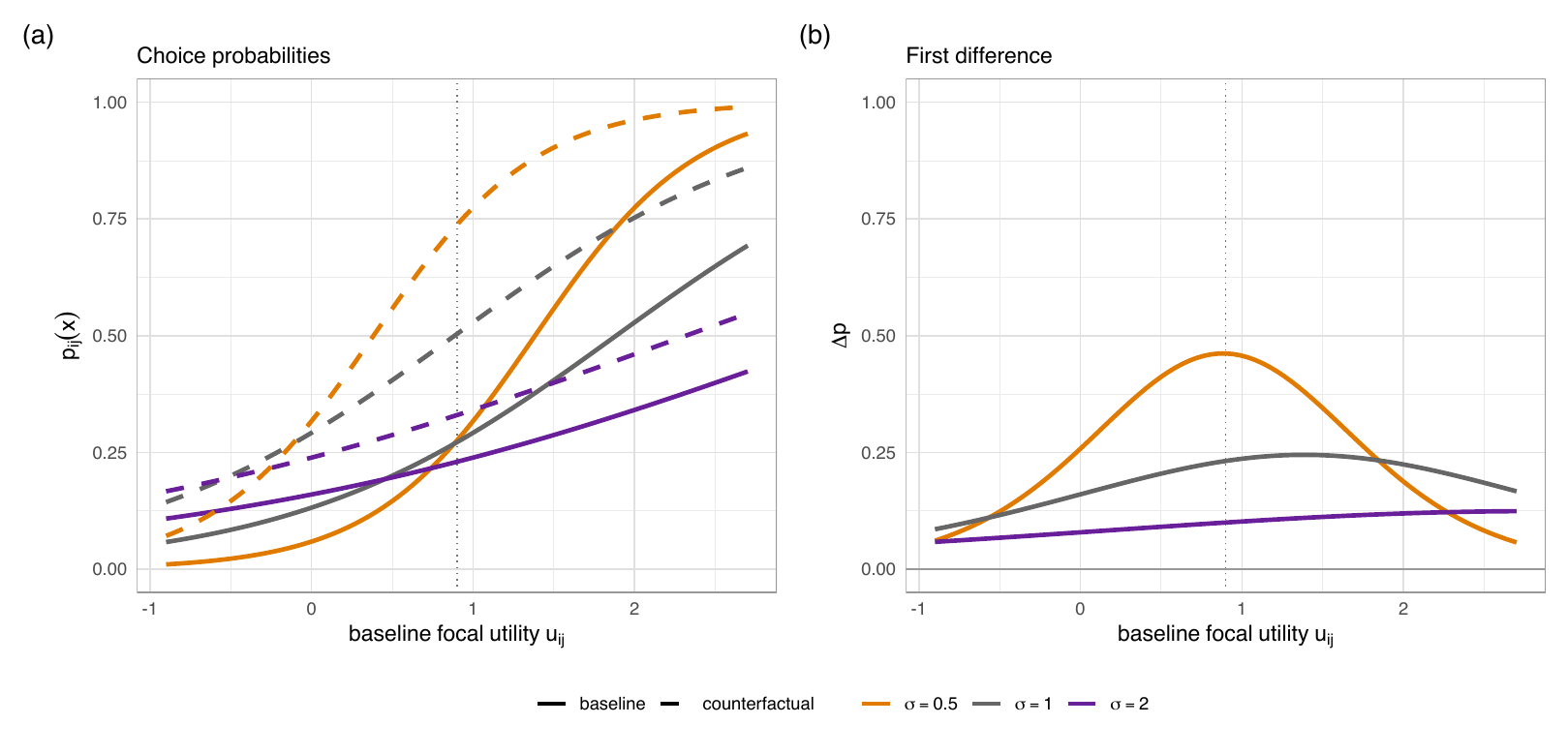}
\caption{Scaling in the multinomial logit: varying $\sigma$ changes how a fixed 
focal preference contrast translates into predicted probabilities. The left panel
shows the focal toggle probability under the baseline and counterfactual given
a range of objective function values; the right panel shows the corresponding focal discrete difference.}
\label{fig:altspecific_scaling}
\end{figure}

\paragraph{Scaling in the ministep illustrations.}
We reuse the stylized decision situation from Section~\ref{S_ministep_choice} 
to illustrate this dynamic. Figure~\ref{fig:altspecific_scaling} shows how the 
same focal counterfactual maps to the probability scale under different values of 
$\sigma$ (smaller $\sigma$ implies more deterministic choices and larger 
probability-scale contrasts). The key point for interpretation is that 
 changing $\sigma$ changes the steepness of the 
 mapping from a fixed preference contrast into a probability contrast: 
 larger $\sigma$ (more residual variability) flattens probability differences.

Crucially, the SAOM features an additional compensatory mechanism not found in
standard conditional logits: the continuous-time rate parameter $\lambda$. 
When predictors are omitted, the remaining choices become less deterministically
tied to the observed target statistics. To compensate for these "less certain" 
choice probabilities, the model often inflates the rate parameter, 
forcing simulated actors to take more ministeps to achieve the observed 
macroscopic network evolution.

However, if the ministeps are not observed, only the observed changes between
networks are identified and the choice model is pinned to the data only via
the generator $\mathcal{G}$, so that by fixing $\sigma_{\text{group}}=1$,
both the choice and rate parameters via transition rate 
$q_{ij}(x)=\lambda_i(x)\,p_{ij}(x;\beta)$ can adjust to differences in
predictability. That can lead to a very different impact of scaling on 
ministep ($\mathrm{AME}^{\text{obs}}$,
$\mathrm{AME}^{\text{sim}}$) and period-level marginal effects
$\overline{\mathrm{ME}}^{\pi}$, which captures joined choice and rate
effects over the whole period(s): the rate enters not as a multiplier
but through the number of opportunities $K$ over which
\eqref{eq:piRecursion} is iterated.

How exactly difference between data-sets will affect SAOM and different QOIs is left
for future work, but since scaling differences reflect genuine predictability
differences, (A)MEs \emph{should not be invariant}. If two groups genuinely differ 
in the noise of their environment, the impact of their preferences on actual decisions outcomes 
that we try to capture with probability-scale QOIs are genuinely different. 

Consider comparing network formation between a highly resourced corporate board 
and a grassroots social movement. The corporate board may possess nearly perfect 
information; their decision-making is highly decisive, meaning a latent preference 
translates to a massive shift in tie-formation and maintenance probability. 
The grassroots movement operates in a highly stochastic environment influenced 
by random encounters. Their underlying preference for reciprocity might be 
identical to the corporate board's, but the \emph{impact} of that preference is 
severely dampened by environmental noise. 

In that sense also the identified coefficients are representing a "predictability-adjusted" 
preference, a measure of how strongly a mechanism cuts through the ambient noise 
of a specific environment \citep[for an argument in favor of this as the true estimand see, e.g.,][]{KuhaMills2018}. 
This also applies to all interaction-like coefficients, 
they incorporate the predictability differences, i.e. a homophily effect can also
be based on a difference in predictability \citep{Duxbury2023}. Probability-scale second 
differences will be affected by the same mechanism, and a difference in sign
between them and the standard coefficient need not indicate scaling: the
compression mechanisms discussed above produce such reversals on their own, and
the two cannot be told apart from the sign alone.

If the goal is uncovering the true underlying preference differences or separating
predictability and relative preference, then the 
standard coefficient is misleading but no post-estimation strategy can resolve that. 
Researchers would need to attempt to isolate scale via heteroscedastic 
(mixed-effects) choice models \citep{FiebigEtAl2010, HessTrain2017, Tutz2021} or need to
try to measure latent preferences directly \citep{KuhaMills2018}.

If the researcher's ultimate goal is to understand the \emph{effective predictive impact} on 
the actors' decision, the AME completes the translation. It projects the 
noise-adjusted preference onto the probability scale, providing a fully standardized, 
behaviorally accurate metric of how actors actually behave given their specific 
environmental constraints.

\subsection{Scenario 2: Estimation differences under omitted variables}
\label{S_scaling_omitted}

When researchers compare two nested SAOM specifications that differ only by a 
(seemingly uncorrelated) omitted predictor, they often expect the focal effect to 
remain stable—as in linear models.  In logit models, this does not hold: 
omitting a predictor increases residual variance, 
and because the scale parameter is implicitly fixed at $\sigma=1$, the remaining 
coefficients contract \citep{Breen2018, KarlsonEtAl2012, Mood2010}.

AMEs are more stable in binary logit because they offset 
smaller coefficients by accounting for the increased sensitivity of baseline 
probabilities \citep{LongMustillo2021, Mize2019, Mood2010}. This improved stability 
is discussed in epidemiology as the collapsibility of risk differences/ratios 
versus the non-collapsibility of odds ratios \citep{Greenland2021, Greenland2021a}, 
and appears to hold in ERGMs as well \citep{Duxbury2023, Duxbury2026}.

For the multinomial/conditional logits underlying SAOM, the literature is thinner. 
However, in multinomial choice models, omitting predictors reallocations probability 
mass across alternatives in a complex way, systematically attenuating remaining 
coefficients except under special conditions that do not apply to standard 
conditional logit \citep{Lee1982, DahlbergEklof2003, CastoEtAl2013}. If the omitted 
predictor is correlated with other predictors or with 
alternative-specific terms, this violates the Independence of Irrelevant 
Alternatives (IIA) assumption \citep{McFadden74}, and the problem is not just 
an inflated error term but a distorted choice geometry \citep{DahlbergEklof2003}.

Both experiments build on the same structural core ($\expWaves$ observations
on $\expN$ actors, the same starting network, and outdegree, reciprocity and
transitive triplets at the same values) and differ in the covariate effect
added to it.  The two covariate effects are set to approximately the same
semi-standardized size, so what differs between the experiments is the kind of
predictor that is omitted rather than how much it moves the evaluation
function.  The structural values are in the range of
those typically estimated for friendship networks in school classes and
similar settings; Appendix~\ref{app:experiment_tables} reports them together
with each experiment's omitted effect, the densities the two processes
generate, and the estimation settings.

Every replication is estimated under both specifications, and each comparison
is formed within a replication before being averaged.  The pairing removes
variation common to the two fits, so the ratios vary less than the levels they
are built from and require no separate adjustment for the dependence between
them; what remains is disagreement about the omission's own effect. Our experiments therefore intentionally
omit a predictor to analyze how SAOM’s Methods-of-Moments estimation—which only
targets observed network statistics—translates this misspecification into
coefficient and AME bias. We consider two cases:

1. An orthogonal directed dyadic covariate $W_{ij}$ 
(iid, undirected, independent of the network and its statistics) at $t_0$.

2. A monadic covariate $Z_i$ with a homophily effect (sameX), that are uncorrelated
with the observed network $x$ at $t_0$.

\paragraph{Experiment 1: Orthogonal dyadic covariate $W$}

\begin{figure}[!ht]
\centering
\includegraphics[width=0.92\textwidth]{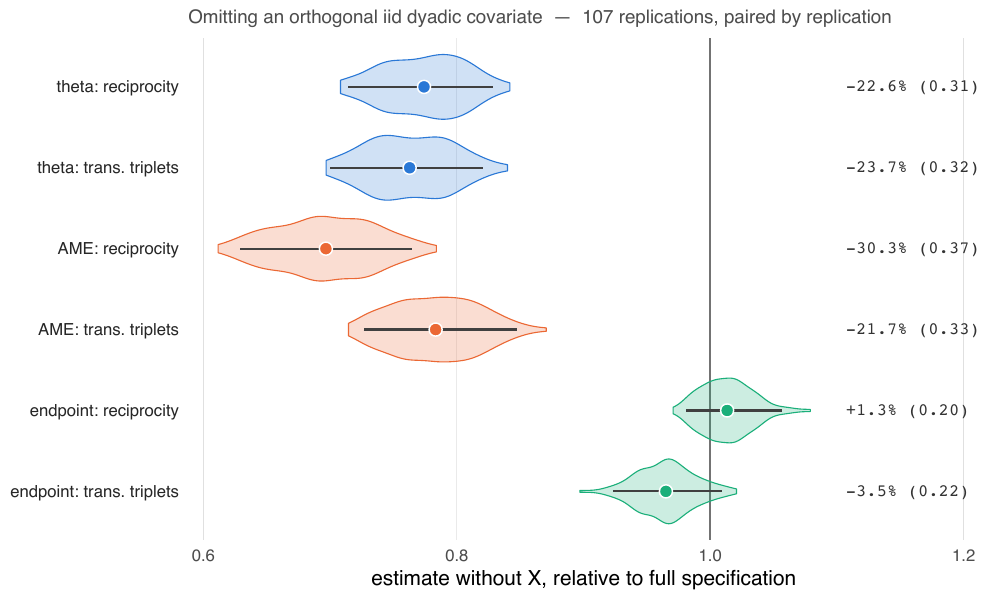}
\caption{Omitting an orthogonal iid dyadic covariate: each estimate under the
reduced specification, relative to the same replication under the full one.
Violins show the distribution over \expOneNreps\ replications, points the mean
and bars the 95\% range, labelled with the mean change and its Monte Carlo
standard error; the rule at $1$ marks an omission that changed nothing.  The
structural coefficients attenuate by about a fifth and the per-ministep
$\mathrm{AME}^{\mathrm{sim}}$ by more ($\expOneAmeRecipPctAbs\%$ for
reciprocity, $\expOneAmeTransTripPctAbs\%$ for transitive triplets), in every
replication.  The period-level endpoint $\overline{\mathrm{ME}}^{\pi}$ is
barely displaced for either effect, which distinguishes it from a property of
one of them, and for reciprocity not even consistent in sign.}
\label{fig:scaling_W_experiment_effect_ratios}
\end{figure}

The omitted predictor here is a continuous dyadic covariate $W_{ij}$, drawn iid
standard normal and independent of the network and of its statistics at $t_0$.
It enters the generating model through a direct dyadic covariate effect
(\texttt{X}) of size $\expDgpW$.  Because it is orthogonal to everything the
estimator targets, there is no structural gap for the remaining effects to
fill, and the misspecification is one of residual variance alone.  We omit it
when estimating and track how coefficients and AMEs respond.
Figure~\ref{fig:scaling_W_experiment_effect_ratios} shows the distribution of 
ratios of estimated coefficients and AMEs to their full-model values across
\expOneNreps\ replicates.

Dropping the orthogonal covariate inflates residual variability, attenuating all 
structural coefficients by roughly the same proportion (approx. $23\%$), 
showing scaling-like behavior. The rate parameter absorbs much of the 
omitted covariate, inflating by approx. $\expOneRatePctAbs\%$; the density coefficient, the
rate and the levels behind every ratio plotted here are reported in
Table~\ref{tab:experiment_ratios}. Neither the
$\mathrm{AME}^{\text{obs}}$ at observation moments nor the average over
simulated ministeps ($\mathrm{AME}^{\text{sim}}$) can compensate. The structural
effects on probability \emph{do} weaken when the covariate is dropped, and the
choice geometry is why: added noise drives probabilities toward the uniform
value $1/K$ rather than toward the steep part of the logistic curve where the
compensation found in binary logit comes from.
Appendix~\ref{app:scaling_details} gives the derivation.

The period-level effect $\overline{\mathrm{ME}}^{\pi}$ engages the inflated
rate directly: because $\lambda$ absorbs much of the omitted covariate, actors
receive more opportunities and the recursion~\eqref{eq:piRecursion} is iterated
more often. The average number of opportunities an actor receives over the
period rises from $\expOneRateFullTwo$ to $\expOneRateRedTwo$
($\expOneRatePct\%$); because a directed dyad $(i,j)$
is exposed at exactly actor~$i$'s opportunities, that is also how many steps
the recursion takes. This is a direct operationalization
of the data-anchoring constraint: because the estimation framework must satisfy
the global moment conditions for total network evolution via the generator
$\mathcal{G}$, the inflation of $\lambda$ operates as a counterweight to the
per-step choice attenuation.

Crucially, those opportunities do not enter as a multiplier.  Each acts only on
the probability still available to move, so more of them also means more
damping, and the endpoint does not simply scale with the opportunity count.
Against coefficient biases of $\expOneThetaRecipPct\%$ and
$\expOneThetaTransTripPct\%$ and per-step biases of $\expOneAmeRecipPct\%$ and
$\expOneAmeTransTripPct\%$, the period-level effect is biased by
$\expOneEndRecipPct\%$ for reciprocity and $\expOneEndTransTripPct\%$ for
triadic closure.  The largest Monte Carlo standard error anywhere in the
experiment is $\expOneMaxMcse$ percentage points
\citep[Table~\ref{tab:experiment_ratios};][]{MorrisEtAl2019}, so none of these
is an artefact of the number of replications.  The residual displacement is
small but not zero, nor should it be: the cancellation balances two separately
moving quantities rather than following from an identity.  It is, however,
negligible in the comparison a researcher would actually make.  Scaled by how
much the between-specification gap varies from dataset to dataset, which is
what a standard error of that difference estimates, the coefficient and
$\mathrm{AME}^{\mathrm{sim}}$ gaps stand at $\expOneThetaRecipGapSd$ and
$\expOneAmeRecipGapSd$, the period-level gap at $\expOneEndRecipGapSd$ for
reciprocity and $\expOneEndTransTripGapSd$ for transitive triplets.  A test of
the difference between the two specifications would therefore detect the first
two reliably and the last barely at all.  The claim is relative stability, not
invariance.

Only the \emph{average} tie probability is directly pinned by estimation,
through the outdegree target, and it barely moves across specifications
($\expOnePiKFull$ versus $\expOnePiKRed$).  The contrast is not pinned in that way, and we
have no derivation for its stability; the pinned level does not supply one
either, since a fixed average constrains where the process operates without
constraining how responsive it is there.  What we would point to instead is
the counterweight above.  A sustained contrast depends on two things: how
much the perturbation moves the per-step probability, and how many
opportunities it is applied at.  In the experiment reported here the orthogonal
omission moves them in opposite directions, attenuating the first while
inflating the second.  The
cancellation is therefore conditional rather than general: it requires the
omission to cost each mechanism a similar share of its influence.  An omission
absorbed by one mechanism instead would raise that mechanism's effect size at
the same time as the rate, and the two would compound rather than cancel.

\paragraph{Experiment 2: Homophily (sameX) with structural entanglement}

\begin{figure}[!ht]
\centering
\includegraphics[width=0.92\textwidth]{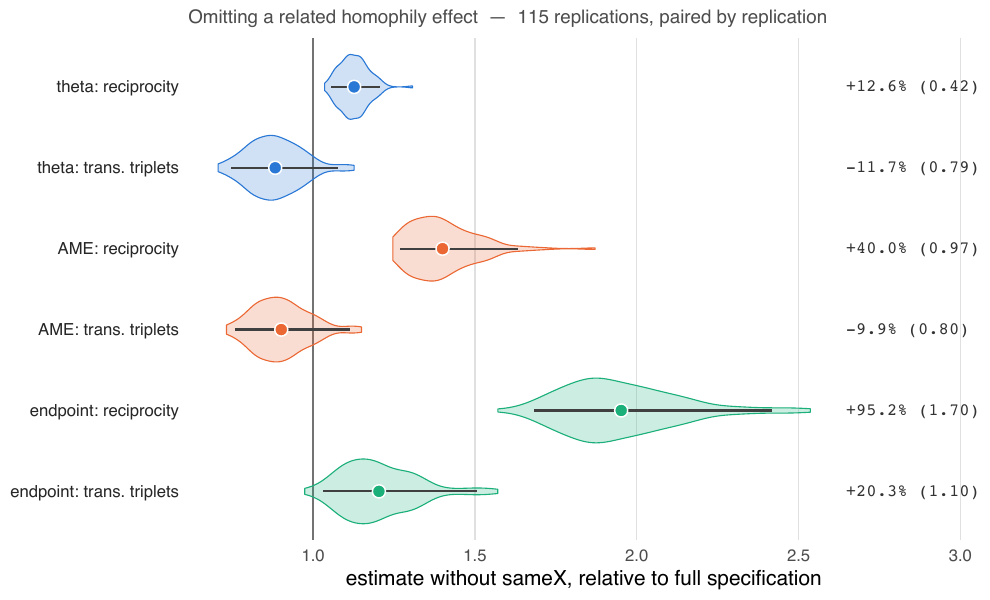}
\caption{Omitting a related homophily effect (\texttt{sameX}), drawn as in
Figure~\ref{fig:scaling_W_experiment_effect_ratios} over \expTwoNreps\ replications.
Here the quantities separate rather than move together: for reciprocity the
coefficient rises by about an eighth, its $\mathrm{AME}^{\mathrm{sim}}$ by
about two fifths, and the period-level endpoint by nearly a factor of two.  For
transitive triplets the coefficient ($\expTwoThetaTransTripPct\%$) and the
$\mathrm{AME}^{\mathrm{sim}}$ ($\expTwoAmeTransTripPct\%$) are indistinguishable
from no change, while the endpoint is displaced by
$\expTwoEndTransTripPct\%$ with only $\expTwoEndTransTripBelow\%$ of
replications on the other side of $1$.  \texttt{sameX} is entangled with both
structural effects---symmetrically with reciprocity, through triads with
transitive triplets---so the two differ in degree rather than in kind.}
\label{fig:scaling_sameX_experiment_effect_ratios}
\end{figure}

The omitted predictor here is an iid three-category monadic covariate $Z_i$,
entering the generating model through a homophily effect (\texttt{sameX}) of
size $\expDgpSameX$.  The covariate is uncorrelated
with the observed network at $t_0$, so the omission is not a matter of
confounding at baseline; what differs from Experiment~1 is that homophily is
entangled with the structural effects as the process runs, symmetrically with
reciprocity and through triads with transitive triplets.  We omit it when
estimating and track coefficients and AMEs.
Figure~\ref{fig:scaling_sameX_experiment_effect_ratios} depicts the
distribution of ratios relative to the full model across \expTwoNreps\ replications,
each ratio formed within a replication so that the two fits are paired.

From a foundational perspective, this omission fundamentally perturbs the data-anchoring 
mechanism established in Section~\ref{S_SAOM}. Because the MoM 
estimation architecture forces the expected value of the time-integrated infinitesimal 
generator $\mathcal{G}$ to align precisely with the observed macroscopic target statistics 
$\mathbb{E}[S]$—anchoring the network trajectory via the identity in 
Equation~\eqref{eq:dynkin}: the system is forced into an immediate state of re-equilibration. 
Because the true data-generating process relied on the missing covariate to 
navigate across network states,  the remaining structural parameters $\beta$ and 
the continuous-time rate parameter $\lambda$ must step in. They systematically distort
their values to absorb the unobserved variance and satisfy the macro-structural
moment conditions. The rate does so to essentially the same extent as in
Experiment~1 ($\expOneRatePct\%$ and $\expTwoRatePct\%$;
Table~\ref{tab:experiment_ratios} reports the levels), so what separates the 
two experiments is not how much opportunity the process gains but what happens 
to the per-step probabilities it is spent on.

Omitting homophily effects even when the covariates are unrelated to the 
network structure at the start of the network evolution process,
introduces complex structural masking rather than simple homogeneous scale inflation. 
Actor homophily is logically and structurally correlated with reciprocity over time 
(reciprocal ties are more often formed and stabilized between similar actors). 
When the homophily effect is omitted, the remaining structural 
parameters compensate for the specific topological gap left behind to 
generate the same target statistics as the full model.
This produces uneven, non-homogeneous coefficient shifts rather than the
homogeneous rescaling of Experiment~1.  Appendix~\ref{app:experiment_tables}
reports the full results for both experiments.

\paragraph{Summary: Scaling and omitted variables in SAOM}. In both analyzed
cases, SAOM coefficients contract when residual variance increases, as in
log-odds models, and probability-scale QOIs do not guarantee invariance: they
represent the magnitude of the effect under the actual model and environment.
What a difference between two effect sizes means therefore depends on what is
being compared, and the two scenarios answer differently.  Across data sets with
different predictability, (A)MEs should not be invariant; they reflect genuine
environmental noise, and removing it would remove part of what is being
measured.  Across nested specifications of one data set the difference is
instead an artefact of the specification, and there (A)MEs standardize impact
under the misspecified model without being invariant to it: omitting an
orthogonal predictor also changes the opportunity timings, because the moment
conditions must still be met and what the choice component no longer explains
has to be absorbed elsewhere, so the scaling operates on the rate as well as on
the choice component.  The period-level
effect $\overline{\mathrm{ME}}^{\pi}$ is the most stable of the summaries we
compare under such an omission, because the rate inflation that the estimator
uses to absorb it enters as additional opportunities rather than as a
multiplier.  Experiment~2 shows that the period level also gives a more complete
picture when the omitted effect is correlated with those retained, where
coefficients and ministep marginal effects mix choice scaling and rate shifts
with genuine confounding, so their change cannot be attributed to confounding
alone.

The same reasoning carries inward.  Across actors and contexts within one data
set, differences in predictability are no less real than differences between
data sets, and marginal effects neither remove them nor should they: this is the
scaling-induced case of Section~\ref{S_Explanation}'s taxonomy, and
Appendix~\ref{app:scaling_details} gives the choice-set geometry behind it.
What differs in kind is the curvature of the mapping itself.  The same latent
contrast moves the probability by different amounts depending on where the
baseline sits, which is enough to make a probability-scale second difference
disagree in sign with its coefficient even where the objective function contains
no interaction at all.  That is the mechanical case of the same taxonomy, it
involves no difference in scale, and the example that follows turns on it. 

\section{Empirical Example}
\label{S_Example}

Two broad regularities recur in adolescent school friendship networks:
ties tend to be reciprocated and to cluster in local groups
\citep{RiveraEtAl2010}, and both mechanisms bring pairs of actors into
shared social situations. This suggests they interact: a friendship 
embedded in a group may need reciprocation less urgently to persist, 
while a mutual pair can generate its own opportunities to meet and is therefore
less reliant on common friends.

\citet{Block2015} reasoned that transitive embedding
primarily \emph{stabilizes} one-sided ties and tested it as an interaction on
the latent preference scale (modelled
first in a longitudinal ERGM by \citet{Igarashi2013}), a logic that bears most
directly on tie \emph{maintenance}. \citet{Block2018} extends the argument: 
in a social-situations model each mechanism channels actors into meeting opportunities,
and because there are decreasing returns to accumulating such opportunities, 
co-embedding is less effective than the sum of its parts. That prediction is framed
explicitly in terms of friendship \emph{probabilities} and bears most
directly on tie \emph{creation}. Whether these interactions are negative
has macro-level consequences: Block's (2018) simulations show the sign of the
closure--reciprocity interaction determines whether the network coheres or fragments, and
\citet{AsikainenEtAl2020} reinforce this from a different angle,
demonstrating that even mild homophily can produce large-scale segregation
when triadic closure is present. Yet existing tests of these predictions rely
on a single pooled interaction coefficient, treating creation and maintenance
as equally affected. \citet{Windzio2020} shows that this assumption can be 
misleading: reciprocity proves a strong driver
of tie creation preference but has no significant effect on maintenance on the latent
scale, a contrast amplified in intra-ethnic dyads and entirely invisible in pooled estimates.

Theoretically, these arguments (especially \citeauthor{Block2018}'s
decreasing-returns model) are often framed in terms of friendship
\emph{probabilities}, yet the previously available tools only permit testing them
through \emph{latent}-scale coefficients (or exponentiated as conditional 
odds ratios). Most prior work, including both Block papers, reports a single pooled
interaction coefficient, implicitly assuming a common directional tendency
regardless of whether a tie is being created or maintained.
We begin by asking whether that overall tendency survives translation to
the probability scale via second differences (Eq.~\eqref{eq:SecondDiff_def}), and 
then examine whether it holds equally across creation and maintenance.  
The same latent-scale contrast can produce very different, or even sign-reversed,
probability-scale shifts when baseline probabilities differ sharply across margins
(the mechanical case of Section~\ref{S_Explanation}'s taxonomy), and the two Block arguments carry somewhat
different implications if maintenance or creation should be most affected.

We use the Glasgow friendship data from the Teenage Friends and Lifestyle Study
\citep{WestSweeting96}, also analyzed by \citet{Block2018}, a longitudinal
study of Scottish secondary school students.\footnote{The data and additional information
are available at \url{https://www.stats.ox.ac.uk/~snijders/siena/Glasgow_data.htm}} 
Friendship nominations were collected at three waves (roughly one year apart 1995-1997)
among students in a single school cohort, who were around 13 years old at the first wave.
Friendship is measured by asking students to nominate up to six friends.\footnote{
The original questionnaire asked students to indicate different types of friendship,
but we use a binary definition where a tie exists if any degree of friendship was described.}
Only within-cohort nominations are used to ensure that the observed network captures
both sides of the friendship relation and we retain the 129 students observed at all
three waves to leave aside issues of composition change in this exposition.
The final network is represented as a directed graph where a tie from student
$i$ to student $j$ indicates that $i$ nominated $j$ as a friend.

\begin{table}[!htbp]
\centering
\caption{Glasgow friendship model: SAOM coefficients and average marginal effects}
\label{tab:glasgow_me}
\footnotesize
\begin{tabular}{lrrrr}
\toprule
  \textbf{Effect} & \textbf{Coef (SE)} & \textbf{AME$^{\text{obs}}$ (SE)} & \textbf{AME$^{\text{sim}}$ (SE)} & \textbf{ME$^{\pi}$ (SE)} \\
\midrule
  Rate period 1 & 12.52 (1.55) & \textemdash & \textemdash & \textemdash \\
  Rate period 2 & 10.08 (0.89) & \textemdash & \textemdash & \textemdash \\
  density & -3.15 (0.12) & \textemdash & \textemdash & \textemdash \\
\midrule
  reciprocity & 2.59 (0.12) & 0.057 (0.003) & 0.060 (0.004) & 0.242 (0.023) \\
  transitive triplets & 1.24 (0.07) & 0.017 (0.001) & 0.018 (0.001) & 0.052 (0.004) \\
  trans. trip. $\times$ recip & -0.56 (0.14) & 0.012 (0.006) & 0.007 (0.005) & 0.134 (0.028) \\
  same gender & 0.76 (0.09) & 0.011 (0.001) & 0.013 (0.002) & 0.015 (0.002) \\
  ego: hanging around & -0.32 (0.11) & -0.001 (0.001) & -0.002 (0.001) & -0.004 (0.002) \\
  alt: hanging around & 0.01 (0.09) & 0.001 (0.001) & 0.001 (0.001) & 0.002 (0.001) \\
  ego $\times$ alt: hanging & 0.17 (0.13) & 0.003 (0.002) & 0.003 (0.002) & 0.003 (0.003) \\
\midrule
\multicolumn{5}{p{0.92\textwidth}}{\footnotesize $\widehat{\text{AME}}^{\text{obs}}$ = unconditional average marginal effect at observed decision situations; $\widehat{\text{AME}}^{\text{sim}}$ = same at model-implied (simulated) decision situations. $\widehat{\text{ME}}^{\pi}$ = period-level endpoint contrast $\overline{\mathrm{ME}}^{\pi}$ of Section~\ref{S_localToIntensity}: the change in the probability that the tie is present at the \emph{end of the period}, rather than after one opportunity. First difference for all effects except trans.\,trip.\,$\times$\,recip. and ego\,$\times$\,alt: hanging (second difference). Effects averaged over all dyad observations and both periods; context-specific effects in Figure~\ref{fig:forest}. SE in parentheses.}\\
\bottomrule
\end{tabular}
\end{table}

The model includes reciprocity (dyadic mutuality), transitive triplets (local
closure\footnote{The transitive-triplets statistic is the elementary two-path
variant (\texttt{transTrip1} in RSiena), which counts for each potential tie
$i\to j$ how many two-paths $i\to h\to j$ exist. This is the natural
building-block closure statistic and forms a proper elementary interaction with
reciprocity.}), their interaction, and
same-gender homophily as a compositional control, overall representing the 
three main established mechanism of friendship dynamics \citep{RiveraEtAl2010}.
All quantities reported below are computed in the RSiena software ecosystem
\citep{SienaManual}.
Estimation constrains the outdegree to at most six, matching the survey's
nomination limit, so an actor already naming six friends faces no tie-creation
alternative at that opportunity and chooses only among maintenance and
no-change.

As a covariate example we include ``hanging around'' (whether a student spends free 
time around the streets at least weekly, entered as a binary\footnote{students were asked to
answer a list of "things you might do only in your free time", including "I hang around the
streets".}).  Out-of-school socialising reflects available time for informal
interaction, and can serve as a proxy for social class \citep{VeenstraSteglich2012}.  
It also has a direct substantive implication: students who spend more time with 
friends outside school may have less need or capacity for in-school friendships, 
particularly with those who do not share this orientation.  Hanging around may
also carry a homophily component: students who hang around outside school may tend to
befriend each other, while those who do not may share a negative view of that
behaviour and cluster among themselves.  Moreover, because hanging around
implies social processes outside the observed school network, it may
introduce additional unobserved heterogeneity in decision-making.  To separate
these channels, we enter sender, receiver, and sender$\times$receiver effects, which decompose
hanging-around into activity, popularity, and matching components.  The
covariate is entered uncentered, so \texttt{egoX} contrasts $(1,0)$
vs.~$(0,0)$, \texttt{altX} contrasts $(0,1)$ vs.~$(0,0)$, and
\texttt{egoXaltX} is the residual $(1,1)$ contrast beyond the additive
combination.  Probability-scale contrasts are interpreted relative to the
$(\text{ego},\text{alter})=(0,0)$ baseline.

Before turning to the results, it is useful to note the baseline asymmetry
that shapes all subsequent comparisons: at the estimated parameter vector,
average friendship creation probability is approximately
$\glasgowPredCreation\,\%$, while average maintenance probability is
approximately $\glasgowPredMaintenance\,\%$.
That means on average creation of ties is very unlikely,
while maintenance of ties is probable, but not near certainty. Over ten percent of ties
are predicted to dissolve. However, that does not necessarily translate into much change,
since the average outdegree is around 3 per student.

\begin{figure}[!ht]
\centering
\includegraphics[width=0.92\textwidth]{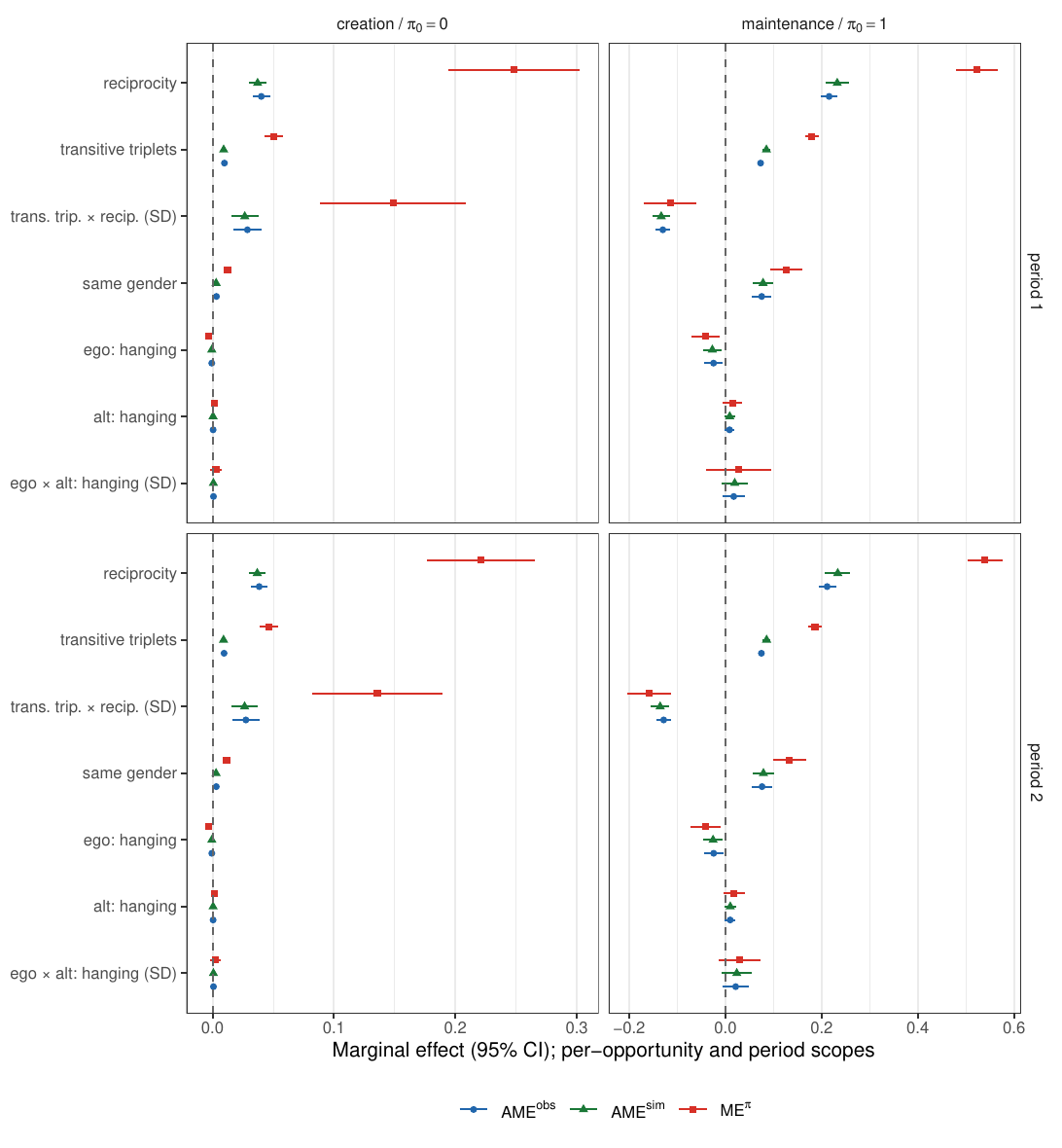}
\caption{Marginal effects~(pp) for all model effects, faceted by creation
versus maintenance and by observation period.  To be able to visually distinguish
creation effects, we do not plot them on the same scale as maintenance effects.  
Circles = observed-wave average ($\mathrm{AME}^{\text{obs}}$); triangles = dynamic average
($\mathrm{AME}^{\text{sim}}$); squares = period-level endpoint contrast
($\overline{\mathrm{ME}}^{\pi}$), which is several times larger because it
cumulates over a whole period rather than one opportunity.  The margin variable
differs with the scope: for the per-opportunity quantities it is the density
change statistic, for the period-level quantity it is $\pi_{0}$, the dyad's
state where the recursion starts, since a period-level quantity has no single
ministep and so no density of its own.  Error bars: 95\,\% simulation-based
intervals.  Main effects: first differences; interactions: second
differences.  Period splitting is shown as a diagnostic check; differences
are small.}
\label{fig:forest}
\end{figure}

Table~\ref{tab:glasgow_me} reports SAOM coefficients alongside
average marginal effects (AMEs), first differences in tie probabilities after
a (potential) ministep in the network evolution process, and second differences,
computed both over all students at observation moments and over model-implied decisions
in continuous time. Figure~\ref{fig:forest} decomposes the same quantities by decision margin
and observation period, allowing a direct check of whether period materially
changes the qualitative pattern. Full estimation output (coefficients, standard errors,
convergence $t$-ratios) is reported in Appendix~\ref{app:estimation_tables}.
Robustness of the structural AMEs under the separate (maxLike) specification is
shown in Appendix~\ref{app:full_vs_separate} (Table~\ref{tab:glasgow_full_vs_separate}
and Figure~\ref{fig:forest_separate}).
We discuss the structural effects
(reciprocity, closure, their interaction) first, then turn to the covariate
effects.

\paragraph{Reciprocity and closure effects on tie probabilities.} Reciprocity is
the dominant mechanism determining friendship probabilities: an overall first difference
of $\glasgowRecipFDoverall$~pp, both across all potential first ministeps at observation moments
and over all simulated ministeps between observations. As Figure~\ref{fig:forest} shows, the
per-margin values differ widely  ($\glasgowRecipFDcreation$~pp for creation and $\glasgowRecipFDmaintenance$~pp
for maintenance), but because about $88\%$ of the alternatives actors actually
face are potential tie creations, the overall AME is close to the creation value. Closure (transitive triplets) contributes about
$\glasgowTransTripFDoverall$~pp overall ($\glasgowTransTripFDcreation$~pp creation,
$\glasgowTransTripFDmaintenance$~pp maintenance). The interaction between
closure and reciprocity is the central quantity of interest: the overall second
difference is $\glasgowInteractionSDoverall$~pp, the probability of a tie after
a ministep is slightly higher for closure embedded reciprocated friendships compared to friendships
that are only reciprocated or only friends of a student's friends. However, this masks a sign reversal 
($\glasgowInteractionSDcreation$~pp for creation, $\glasgowInteractionSDmaintenance$~pp for maintenance),
which we examine next.

\begin{figure}[!ht]
\centering
\includegraphics[width=0.92\textwidth]{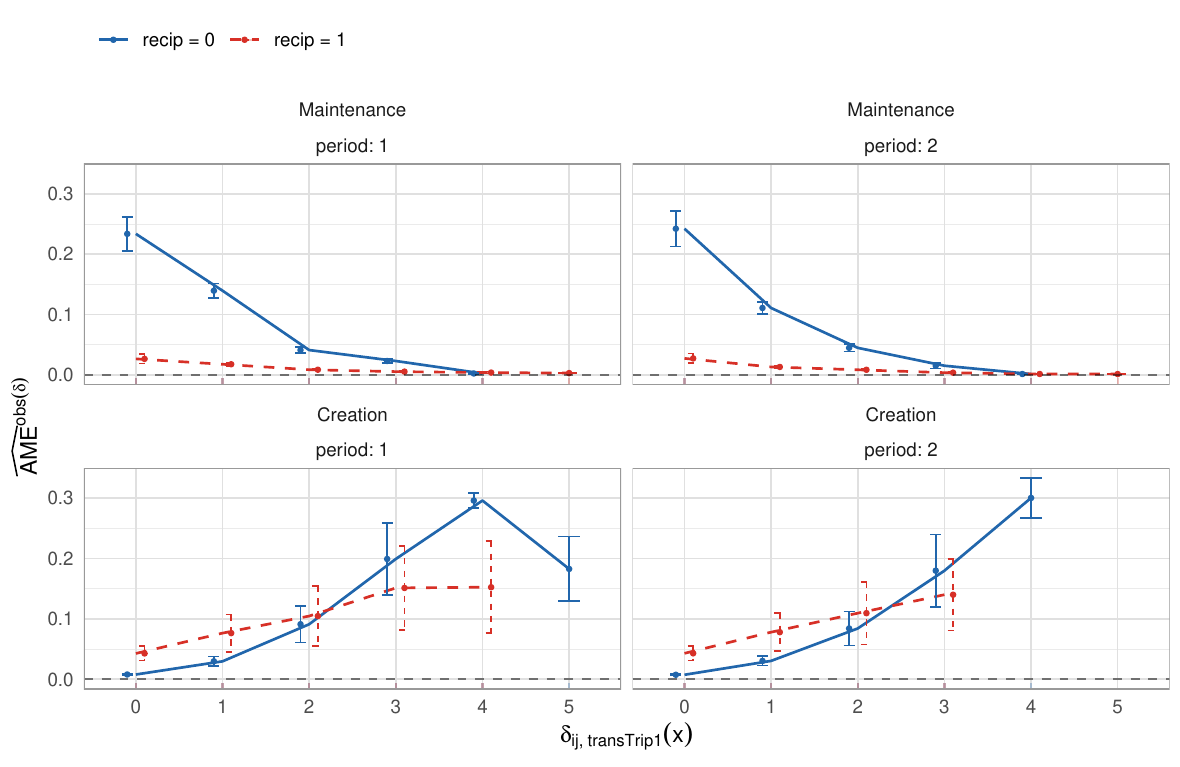}
\caption{Closure first differences~(pp) conditional on reciprocity status,
faceted by creation versus maintenance and period.  The gap between the
reciprocated and non-reciprocated curves is related to, but not identical
with, the counterfactual second difference~\eqref{eq:SecondDiff_def},
because the second difference perturbs reciprocity within the same decision
situation rather than comparing across subpopulations.  Downward bars show
the observed frequency distribution of two-path counts.}
\label{fig:closure-by-recip}
\end{figure}

Figure~\ref{fig:closure-by-recip} plots conditional closure first differences
(the friendship probability shift from one additional common friend) by reciprocity
status, margin, and observation period; effects are stable across periods.
On the \emph{maintenance} margin, closure has a sizable effect for
non-reciprocated ties ($\glasgowTransTripMaintRecipZeroTTzero$~pp at zero
two-paths, declining to below $\glasgowTransTripMaintRecipZeroTTtwo$~pp at
two two-paths) but is negligible for reciprocated ties throughout.  On the
\emph{creation} margin, effects are small and increasing in two-path count;
reciprocated dyads show slightly larger closure effects (from
$\glasgowTransTripCreatRecipOneTTzero$~pp at zero two-paths to
$\glasgowTransTripCreatRecipOneTTfour$~pp at four two-paths), but the
frequency bars confirm that most dyads in any period have zero two-paths,
in particular if $i$ does not already nominate $j$ as a friend.

The pattern visible in these conditional first differences is closely related
to, but not identical with, the counterfactual second differences
(Eq.~\eqref{eq:SecondDiff_def}): the figure conditions on reciprocity status, 
while the second difference perturbs reciprocity within the same decision situation. 
In practice, the two convey a similar qualitative message. On the \emph{maintenance} margin,
closure is larger for non-reciprocated friendships, yielding a second difference of
$\glasgowInteractionSDmaintenance$~pp, the most direct confirmation of the
expectation, that closure does not really matter that much for stabilizing friendships,
that are already mutual.  On the \emph{creation} margin, the sign reverses to
$\glasgowInteractionSDcreation$~pp. This is driven by the difference in baseline probability:
friendship creation is very rare ($\glasgowPredCreation\,\%$ per opportunity),
so any factor jointly raising creation probability moves decisions into a steeper part of the logit curve,
amplifying combined effects even when the latent-scale interaction itself is negative. 
Theoretically, this makes sense as well: new friendships require a lot of effort, so 
multiple factors might have to come together to make it more likely.
Substantively, the Block prediction holds for friendship maintenance probabilities; 
friendship creation shows the opposite, a distinction invisible at the log-odds
coefficient level.

The same decomposition applies to the covariate
effects, and it changes the reading in the same way.  Same-gender homophily
raises the probability of a tie by about one percentage point on average, which
looks negligible beside reciprocity; but that average is dominated by creation
opportunities, where homophily barely moves the probability, while on the
maintenance margin the same effect is worth
$\glasgowHomophilyMaintenance$~pp.  Gender similarity does little to make
friendships and a good deal to keep them, and the hanging-around effects are
small on the tie-presence scale with intervals covering zero in both the sender
and the receiver role, so the covariates reproduce the margin asymmetry of the
structural effects rather than qualifying it.

\paragraph{From one opportunity to a whole period.} The period column of
Table~\ref{tab:glasgow_me} and the squares in Figure~\ref{fig:forest} carry the
same contrasts to the end of an observation period, with the perturbation held
at every opportunity the dyad receives rather than at one.  The baselines move
with the scope: over a period an absent tie ends up present with probability
$\glasgowPredPeriodCreation\,\%$ and a present one survives with probability
$\glasgowPredPeriodMaintenance\,\%$, against $\glasgowPredCreation\,\%$ and
$\glasgowPredMaintenance\,\%$ at a single opportunity.  Both margins grow
under that reading, creation by between $\glasgowAmpCreationLo$ and
$\glasgowAmpCreationHi$ times and maintenance by between
$\glasgowAmpMaintenanceLo$ and $\glasgowAmpMaintenanceHi$.  The multiplier is
otherwise close to uniform within a margin, so the ordering of effects is
preserved and the larger ones simply gain the most.  Sustained reciprocity is
the clearest case, $\glasgowRecipPeriodMaintenance$~pp over the period against
$\glasgowRecipMinistepMaintenance$~pp at a single opportunity.  At the other end
the closure-by-reciprocity interaction does not grow at all on the maintenance
margin, $\glasgowInteractionPeriodMaintenance$~pp against
$\glasgowInteractionMinistepMaintenance$~pp.

Across periods, Figure~\ref{fig:forest} shows only modest movement in these
summaries: the largest period-to-period shift is about 0.45~pp (dynamic
ego$\times$alt interaction on the maintenance margin), with most shifts
substantially smaller. All effects are practically the same at the observed-moments
and over the model-implied decision distribution, reassuring that the observed-moments
summaries are not driven by a few atypical decision situations.

\paragraph{Summary.} The example illustrates three broader points.
First, probability-scale QOIs can differ markedly from coefficient-level patterns, 
with substantively consequential implications. The negative latent preference
interaction is consistent with \citeauthor{Block2015}'s (\citeyear{Block2015}) 
prediction on \emph{maintaining} friendships, where the sign holds also on the probability
scale.  On the \emph{creation} margin, however, the probability-scale second difference
reverses sign, directly contradicting the expectation that decreasing
returns to meeting opportunities also reflect in the joint effect on probability of friendship
formation.  Because creation is the dominant decision context for the vast
majority of ego-alter dyads, this reversal is not a minor qualification: it means that
the overall probability-scale interaction points in the opposite direction
from the substantive hypothesis.  In addition, the creation-maintenance
decomposition reveals asymmetries for most mechanisms in the model
(Figure~\ref{fig:forest}); aggregated averages can mask substantively
meaningful heterogeneity. The framework accommodates both
alter-specific and ego-level summaries, reflecting the different levels at
which network mechanisms operate.
The same approach applies directly to the wider set of mechanism
interactions documented by \citet{Block2018} and to the
interplay of structural and covariate mechanisms that shapes substantive
outcomes in recent applied work
\citep[e.g.,][]{ChenEtAl2026, LosevaEtAl2026}.

\clearpage

\section{Discussion}
\label{S_Discussion}

This paper develops probability-scale effect sizes for the decision model
underlying actor-oriented continuous-time network models like the SAOM.
The quantities of interest are (i)~probabilities of a focal move under clearly
specified decision contexts and (ii)~contrasts of those probabilities under 
counterfactual changes to a focal covariate or network feature, and
(iii)~the same contrast carried to the end of an observation period, where the
opportunity rate re-enters and a per-opportunity difference accumulates.
When the substantive interest is in how network mechanisms change relationships, 
these are the directly informative reporting target. They are also valuable
for researchers who prefer to stay on the latent preference scale: 
inspecting probability-scale implications serves as a diagnostic that reveals
whether coefficient-level patterns survive the nonlinear mapping to decision
predictions across heterogeneous decision situations, as the sign reversal in
Section~\ref{S_Example} illustrates.

One practical implication is comparability. Section~\ref{S_Scaling} shows why
raw log-odds SAOM coefficients can be difficult to compare across model specifications
and data sets: scaling affects both the mapping from the objective function
to probabilities and the distribution of intermediate decision
contexts that receive weight. It is tempting to pool coefficients for
models with very different invisible scaling and network structures that
do not actually correspond to the
same underlying estimand. \citet{DuxburyWertsching2023} demonstrate this
empirically for pooled ERGM, showing that heterogeneous scaling across networks
can produce pooled coefficients with the wrong sign or significance, and that
marginal effects within pooled frameworks substantially reduce that bias, a
finding that might carry over to pooled SAOM settings. This also
suggests why averaged marginal effects may be more stable than coefficients
and highly conditional marginal effects, because they aggregate over heterogeneous
decision contexts.
Whether a particular average marginal effect improves cross-study comparability depends on
how decision contexts are weighted and on how model and
data differences shift the relevant parts of the state space. For concrete
meta-analyses, these issues should be examined with targeted sensitivity checks;
developing practical diagnostic tools and reporting standards for such checks is
an important direction for future research.

The same point applies to interaction terms. Because first differences are
nonlinear and heterogeneous across decision contexts, a negative SAOM
interaction does not by itself determine the sign or shape of the
corresponding probability-scale second difference.  
\citet{Block2018} predicts that co-embedding makes friendship
formation less likely than the additive combination of individual mechanisms, a
claim stated in terms of probabilities. On the maintenance margin the
probability-scale second difference confirms this; for creation probabilities it
reverses, an outcome that is invisible in the log-odds coefficient and that
qualifies how the prediction should be read empirically. More broadly,
the same latent interaction can have quite different impacts across
creation and maintenance probabilities and across the state space.

\section{Extensions, limitations, and outlook}
\label{S_Extensions}

The marginal effects developed here address the choice component of the SAOM, at
a single decision opportunity and carried forward on the probability scale over
a period. Richer summaries
connect to several neighboring literatures. Finite-horizon
effects on expected between-wave outcomes incorporate indirect feedback along
model-implied paths \citep[e.g.][]{PutterEtAl2007}; system-level
effects parallel the direct and indirect effects studied in spatial
interaction and network equilibrium models
\citep[e.g.][]{LeSagePace2009,LeeEtAl2014a,Lin2014}; and summaries on the count
scale, such as the expected number of ties created between waves, remain open:
there the opportunity rate enters without the bound that keeps a presence
probability in check, so any count-level effect mixes differences in opportunity
frequency with differences in decisions given an opportunity. This rate--choice
distinction is equally
central in dynamic network actor models 
\citep[DyNAM;][]{StadtfeldBlock2017}, relational event models 
\citep[e.g.,][]{Butts2008relevent}, and continuous-time graph
process models \citep[]{Butts2023}.
When the target is
tie existence at the end of a period, carrying the tie probability along
simulated trajectories and contrasting its endpoint is the natural bridge, and
Section~\ref{S_Scaling} finds it the summary least displaced by an orthogonal
omission.
The same QOI perspective extends to
co-evolution settings (behavioral dynamics, multiple networks, two-mode and
multiplex relations), where defining local counterfactuals across interacting 
processes requires additional care.

Several limitations deserve acknowledgment. 
First, the core quantities are conditional on a single decision opportunity
in a given network state. The period-level extension
(Section~\ref{S_localToIntensity}) incorporates the opportunity rate and follows
the tie probability along simulated paths, but the rate itself is held at its
fitted value under the counterfactual.  The quantities are developed for
rate functions that are constant within an observation period, as in the
empirical example.  Rate effects conditional on actor covariates or on network
position are not covered: they change which decision situations carry weight in
the observed-wave average, and where the rate varies with the focal statistic
itself they would require explicit treatment of the opportunity process.  The
continuous-time formulation also supports quantities on the time scale rather
than the probability scale, such as expected waiting times between changes,
which we do not develop here.  The period-level quantity further lets only the focal dyad respond,
with the rest of the network held to the trajectory it took without the
change, a combination no single run of the model produces on its own. It also answers a dyad-level question and has no
actor-level counterpart in the present construction.
Second, the counterfactual isolates one mechanism at a time, whereas
in practice network mechanisms co-vary, so the reported effects should be
read as partial, not total, contributions.  Where a higher-order statistic
cannot change without its lower-order parts the two are tied by the network
rather than by the specification, a case \citet{Duxbury2026} treats for the
cross-sectional ERGM and one we leave open for the ministep setting.
Third, uncertainty estimates inherit the estimation uncertainty of the fitted 
SAOM and assume correct model specification; dynamic averages additionally 
require re-simulating full trajectories, which can be computationally demanding 
for large networks.
Finally, the empirical illustration uses a single dataset; the magnitude of 
phenomena such as sign reversals across creation and maintenance will vary 
across networks, and the framework should be applied more broadly before 
general conclusions are drawn. 

We also note that these marginal effects, while providing 
interpretable decompositions of observed differences, are not on their
own sufficient for causal inference: as with marginal effects in other
non-experimental regression settings, they quantify model-implied
associations rather than identified causal contrasts, and a causal
reading would require additional design or identification arguments
beyond the scope of this paper.

The practice of directly comparing SAOM coefficients across model
specifications and data sets has been widespread but is
often not appropriate. The probability-scale marginal effects developed here
provide an interpretable alternative: effect sizes
that stay close to the decision process the model implies, that decompose
naturally by behavioral margin, and that state explicitly what is held fixed,
what is changed, and over which decision situations the effect is averaged.
As the empirical example illustrates, moving to the probability scale can
reveal structural relationships that remain invisible on the coefficient scale.

\clearpage
\section*{Acknowledgements}

\noindent
This paper grew out of a collaboration between Universit\"{a}t Hamburg and the
University of Groningen, supported by their joint funding programme through the
project \emph{Comparing Effects in Stochastic Network Models} (COMMODE), which
was awarded for work on the quantities developed here.  Funded by the Deutsche
Forschungsgemeinschaft (DFG, German Research Foundation) -- project number
447967785, Research Unit~5234 \emph{Multipler Wettbewerb im Hochschulsystem},
subproject~D2.2.  Further funded by the DFG under Germany's Excellence Strategy
-- EXC~2037 ``CLICCS -- Climate, Climatic Change, and Society'' -- project number
390683824.


\clearpage
\begingroup
\doublespacing
\section*{ENDNOTES}
\theendnotes
\endgroup

\appendix

\section{Post-estimation and uncertainty (details)}
\label{A_uncertainty}

Uncertainty for the QOIs defined in the main text can be quantified by simulation from the
estimated sampling distribution of the SAOM parameter estimates (a \emph{parametric}, or
\emph{model-based}, bootstrap; \citealp{EfronTibshirani1994}). This approach is standard for
post-estimation quantities in nonlinear models \citep{Herron1999, KingEtAl2000} and is closely
related to the Krinsky--Robb procedure used to obtain uncertainty for nonlinear functions of
estimated parameters \citep{KrinskyRobb1986,KrinskyRobb1991}.

\begin{itemize}
    \item \textbf{Input.} Point estimates $\widehat{\theta}$ and their estimated covariance
    matrix $\widehat{\Sigma}$ (as returned by RSiena).

    \item \textbf{Point estimate of the QOI (avoid simulation noise if possible).}
    Report $\widehat{\tau}=h(\widehat{\theta};\mathcal{E})$. Following the general advice
    in \citet{Rainey2017}, simulation should primarily be used to characterize uncertainty,
    not to generate the point estimate, whenever $h(\widehat{\theta};\mathcal{E})$ can be
    computed exactly or with negligible numerical error. This avoids adding unnecessary
    Monte Carlo variation to reported effect sizes. We only use model-implied simulations
    when $h(\cdot)$ itself requires simulation (e.g., averaging over simulated
    ministeps). Concretely, observed-wave summaries evaluate effects at observed $x_t$ and
    then average, while model-implied summaries additionally average over simulated
    ministep decision situations between waves.

    \item \textbf{Parameter draws (for uncertainty).} For $b=1,\ldots,B$, draw
    $\theta^{(b)} \sim \mathcal{N}(\widehat{\theta},\widehat{\Sigma})$
    (model-based / parametric bootstrap; \citealp{EfronTibshirani1994}).

    \item \textbf{Compute QOIs per draw.} For each draw, compute
    $\tau^{(b)} = h(\theta^{(b)};\mathcal{E})$.

    \item \textbf{Summarize uncertainty.} Use quantiles of $\{\tau^{(b)}\}_{b=1}^B$
    (e.g., 2.5\% and 97.5\%) and/or their standard deviation as confidence intervals /
    standard errors \citep{Herron1999, KingEtAl2000}.

    \item \textbf{If $h(\cdot)$ itself requires simulation.} When $h(\theta;\mathcal{E})$
    involves simulation, use enough internal draws that Monte Carlo error is small relative
    to the across-draw variation in $\{\tau^{(b)}\}$, and keep the point estimate anchored
    at $\widehat{\tau}=h(\widehat{\theta};\mathcal{E})$ rather than at a noisy simulation
    average \citep{Rainey2017}.
\end{itemize}

\paragraph{Expected values vs.\ predictive draw noise.}
The QOIs defined above—both $\mathrm{AME}^{\text{obs}}$ and
$\mathrm{AME}^{\text{sim}}$—target \emph{expected values}: averages of
the local probability contrasts over decision situations, at a fixed
parameter vector $\theta$. They are \emph{not} predictions for a specific
single network trajectory. This distinction mirrors the fundamental
separation between expected-value (EV) and predicted-value (PV) quantities
in post-estimation \citep{KingEtAl2000}: an EV confidence interval describes
uncertainty about the average behavior of a system across many trajectories,
whereas a PV interval additionally incorporates stochasticity of any one
realization.

In model-implied summaries $\mathrm{AME}^{\text{sim}}$, the SAOM does not
admit a closed-form expression for the expectation over the latent ministep
sequence. Simulation is therefore inherent even for the EV itself: we approximate
$h(\theta;\mathcal{E}) = \EE_X[\tau\mid\theta]$ by averaging over $n_3$ simulated
chains. The resulting approximation
$\widehat{h}_{n_3}(\theta;\mathcal{E}) \approx h(\theta;\mathcal{E})$ carries a
residual Monte Carlo variance $\sigma^2_h/n_3$ (where $\sigma^2_h$ is the
across-chain variance of the QOI at fixed $\theta$). This is PV leakage into an
EV estimator, not fundamental predictive uncertainty; it shrinks as $n_3$
increases and vanishes as $n_3\to\infty$. The recommendation to use ''enough
internal draws'' (\citealp{Rainey2017}; see bullet above) ensures that this
leakage is negligible relative to the parameter uncertainty
$\mathrm{Var}_\theta[\tau^{(b)}]$ across bootstrap draws, so that the resulting
interval width is dominated by sampling uncertainty in $\widehat{\theta}$—the
genuine source of inferential uncertainty—rather than by simulation noise.

For observed-wave summaries ($\mathrm{AME}^{\text{obs}}$), the change
statistics $\delta_{ij,k}(x_t)$ are computed once from the observed
data and reused across all parameter draws $\theta^{(b)}$: only the
softmax weighting changes with $\theta^{(b)}$, not the change
statistics themselves. Here $h(\theta^{(b)};\mathcal{E})$ involves no
simulation, so each $\tau^{(b)}$ is exact and the across-draw variation
$\mathrm{Var}_\theta[\tau^{(b)}]$ reflects pure parameter uncertainty.

For model-implied summaries ($\mathrm{AME}^{\text{sim}}$), each parameter
draw $\theta^{(b)}$ requires re-simulating the ministep chain (because the
chain itself depends on $\theta$), so the change contributions vary across
draws. This makes $\tau^{(b)}$ itself a noisy simulation average: the
across-draw variation now mixes parameter uncertainty with the within-draw
chain variance $\sigma^2_h/n_3$ discussed above. The leakage condition
($\sigma^2_h/n_3 \ll \mathrm{Var}_\theta[\tau^{(b)}]$) is, however,
sometimes satisfied at quite modest $n_3$: because the QOI is an average over
many dyads and decision situations within each chain, within-draw variance
tends to be small relative to the across-parameter-draw variance from
$\widehat{\Sigma}$. Whether a given $n_3$ suffices depends on the QOI
granularity (finer-grained, less-averaged QOIs have larger $\sigma^2_h$),
the network size (smaller networks offer less averaging within a chain),
and the strength of structural effects (which can concentrate probability
mass and amplify chain-to-chain variation). In practice, $n_3$ for
post-estimation uncertainty quantification need not match the $n_3$ used
during model estimation; investing the saved compute in more bootstrap draws
$B$ typically does more to sharpen the reported intervals.

\section{Delta-method standard errors}
\label{app:deltaMethod}

\dmsection{Setup: marginal effects as one differentiable map}
\label{sec:setup}

\dmsubsection{Notation recap}
We briefly restate the notation from the main paper. For later use it is
convenient to define the signed toggle statistic
$\tau_{ij,k}(x) := (1-2x_{ij})\delta_{ij,k}(x)$, so that the \emph{objective
function} for actor $i$ evaluating toggle $(i\!\to\!j)$ at network state $x$
can be written as
\begin{equation}
f_{ij}(x;\beta)
\;=\; \sum_k \beta_k\,\tau_{ij,k}(x).
\label{eq:obj}
\end{equation}
Here $\delta_{ij,k}(x)$ is the underlying change statistic of effect $k$: the
amount by which statistic $k$ increases when tie $x_{ij}$ is created
(and decreases when it is dissolved). The signed form $\tau_{ij,k}(x)$ absorbs
the create/dissolve sign, so both the choice probabilities and the marginal
effects depend on the same utility contribution. The parameter vector is
$\beta \in \mathbb{R}^K$. Choice probabilities follow the conditional
logit\ifstandalone\else~\eqref{eq:clogit}\fi,
\ifstandalone
\begin{equation}
p_{ij}(x;\beta)
\;=\;
\frac{\exp\!\bigl(f_{ij}(x;\beta)\bigr)}
{\displaystyle\sum_{h\in A_i(x)} \exp\!\bigl(f_{ih}(x;\beta)\bigr)},
\label{eq:clogit}
\end{equation}
\fi
where $A_i(x)$ is the choice set of actor $i$: the $n-1$ possible toggles
plus the no-change option $ii$ (for which $f_{ii}(x;\beta)=0$ by convention).
Dyad-level marginal effects are first differences on the toggle probability
scale, e.g.\ the creation $\ME^+_{ij,k}(\Delta\delta_{ij,k}\mid x)$
\ifstandalone\else\eqref{eq:MEdef_creation} \fi and the tie-presence
$\ME^{\mathrm{tie}}_{ij,k}$. The average marginal effects (AMEs) aggregate
these over decision contexts: $\mathrm{AME}^{\text{obs}}_k$ over observed-wave
states $x_t$, $\mathrm{AME}^{\text{sim}}_k$ over simulated ministep states,
and the period-level $\overline{\mathrm{ME}}^{\pi}_k$ obtained by propagating
the tie probability along the simulated path\ifstandalone\else~\eqref{eq:piRecursion}\fi.

\dmsubsection{All QOIs as a single smooth map}

All of the above quantities share a common computational graph. Index by $r$
the rows of a flattened contribution table; each row is a dyad $(i,j)$ at
network state $x^{(r)}$ (and, in the dynamic case, at ministep $m$ of
simulated chain $c$). Write the $K$-vector of signed toggle statistics for row
$r$ as
$\tau_r = (\tau_{ij,1}(x^{(r)}),\ldots,\tau_{ij,K}(x^{(r)}))^\top$
and collect them into the \emph{contribution matrix}
$\mathbf{T} \in \mathbb{R}^{N\times K}$ with $\mathbf{T}_{rk}=\tau_{ij,k}(x^{(r)})$.
Noting that $f_{ij}(x;\beta)=\mathbf{T}_r^\top\beta$, the map from parameters to any
AME is:
\begin{equation}
\beta
\;\xrightarrow{\;f_r=\mathbf{T}_r^\top\beta\;}\;
f
\;\xrightarrow{\;p_r=\mathrm{clogit}_{A_i}(f)\;}\;
p
\;\xrightarrow{\;m_r\;}\;
m
\;\xrightarrow{\;M=\frac{1}{|C|}\sum_{r\in C} m_r\;}\;
M ,
\label{eq:graph}
\end{equation}
where $m_r$ is the row-level outcome ($\ME^+$, $\ME^{\mathrm{tie}}$, a second
difference, etc.) and $C$ is the set of decision contexts over which we
average.

If we assume that $\betah$ is asymptotically normal,
$\betah \approx \mathcal{N}(\beta_\ast,\Sigma)$ with $\Sigma$ estimated by
$\Sigmah$ (for method-of-moments, via the Schweinberger--Snijders Monte-Carlo
derivative estimate of the Jacobian of the moment function;
\citealp{SchweinbergerSnijders07a, Schweinberger2007}), the delta method gives
\begin{equation}
\widehat{\Var}(M) \;=\; \nabla_\beta M(\betah)^\top\,\Sigmah\;\nabla_\beta M(\betah),
\qquad
\mathrm{SE}_\delta(M) \;=\; \sqrt{\widehat{\Var}(M)}.
\label{eq:delta}
\end{equation}
The only object needed beyond the point estimate is the gradient
$\nabla_\beta M$.

\paragraph{Observed-wave versus model-implied (dynamic) quantities.}
For \emph{observed-wave} summaries ($\mathrm{AME}^{\text{obs}}_k$), the
contribution matrix $\mathbf{T}$ is fixed (computed once from the observed network
states $x_t$), so $\beta \mapsto M$ is a deterministic smooth function and
\eqref{eq:delta} is exact up to the normal approximation of $\betah$. For
\emph{model-implied} and \emph{period-level} quantities
($\mathrm{AME}^{\text{sim}}_k$, $\overline{\mathrm{ME}}^{\pi}_k$),
$\mathbf{T}$ is stochastic: it depends on Markov chains simulated at $\beta$, so
the full derivative carries an additional path-distribution term.
Section~\ref{sec:reinforce} handles that term analytically;
Section~\ref{sec:cond} first gives the part that holds the simulated chains
fixed.

\dmsection{Conditional (frozen-chain) Jacobian}
\label{sec:cond}

Condition on the realised contribution matrix $\mathbf{T}$ (for observed-wave
quantities this is unconditional; for dynamic quantities it freezes the $n_3$
chains already simulated at $\betah$ for the point estimate). The map
\eqref{eq:graph} is then a composition of a linear map, a softmax, and an
average, and its Jacobian is closed form.

\paragraph{Softmax Jacobian.}
With $f_r = \mathbf{T}_r^\top\beta$ and \eqref{eq:clogit}
(suppressing the $x^{(r)}$ argument for brevity),
\begin{equation}
\frac{\partial p_{ij}}{\partial\beta_k}
\;=\;
p_{ij}\bigl(\tau_{ij,k}(x)-\bar\tau_{A_i,k}\bigr),
\qquad
\bar\tau_{A_i,k}=\sum_{h\in A_i}p_{ih}\,\tau_{ih,k}(x),
\label{eq:softmaxjac}
\end{equation}
i.e.\ the standard softmax Jacobian composed with the linear objective.
The term $\bar\tau_{A_i,k}$ is the probability-weighted mean toggle
statistic across the choice set: it encodes the \emph{within-choice-set
dependence} between alternatives that distinguishes the conditional logit
from a simple binary logit.

\paragraph{First-difference gradient.}
A first difference perturbs the change statistic of effect $k^\ast$ by
$\Delta\delta_k>0$ while holding the network state $x$ fixed, yielding the
counterfactual preference shift
$\Delta f_{ij} = (1-2x_{ij})\beta_{k^\ast}\Delta\delta_k$ and the
corresponding counterfactual probability
\[
p_{ij}^{\mathrm{cf}}
= \frac{p_{ij}\,e^{\Delta f_{ij}}}{1 - p_{ij} + p_{ij}\,e^{\Delta f_{ij}}}
\]
\citep[the incremental multinomial-logit update,][p.~114]{BenAkivaLerman1985}.
The gradient of $\ME^+_{ij,k^\ast} = p_{ij}^{\mathrm{cf}} - p_{ij}$ (creation
first difference) with respect to $\beta_m$ has two channels: $\beta$ acts on
the baseline $p_{ij}$ through the full softmax \eqref{eq:softmaxjac}, and on
the counterfactual shift $\Delta f_{ij}$ directly through $\beta_{k^\ast}$:
\begin{equation}
\frac{\partial\,\ME^+_{ij,k^\ast}}{\partial\beta_m}
=
\underbrace{\Bigl(\tfrac{\partial p_{ij}^{\mathrm{cf}}}{\partial p_{ij}}-1\Bigr)\,\frac{\partial p_{ij}}{\partial\beta_m}}_{\text{baseline channel}}
\;+\;
\underbrace{\tfrac{\partial p_{ij}^{\mathrm{cf}}}{\partial \Delta f_{ij}}\,(1-2x_{ij})\,\Delta\delta_k\,\mathbf{1}[m=k^\ast]}_{\text{counterfactual channel}},
\label{eq:fdgrad}
\end{equation}
where
\[
\frac{\partial p_{ij}^{\mathrm{cf}}}{\partial p_{ij}}
=
\frac{e^{\Delta f_{ij}}}{\bigl(1-p_{ij}+p_{ij}e^{\Delta f_{ij}}\bigr)^2},
\qquad
\frac{\partial p_{ij}^{\mathrm{cf}}}{\partial \Delta f_{ij}}
=
\frac{p_{ij}(1-p_{ij})e^{\Delta f_{ij}}}{\bigl(1-p_{ij}+p_{ij}e^{\Delta f_{ij}}\bigr)^2}.
\]
The dissolution ME $\ME^-_{ij,k^\ast}$ and the tie-presence
$\ME^{\mathrm{tie}}_{ij,k^\ast}$ follow by sign conventions
(Section~2 of the paper). Averaging \eqref{eq:fdgrad} over $C$ gives
$\nabla_\beta M$ for the $\mathrm{AME}^{\text{obs}}_{k^\ast}$ or
$\mathrm{AME}^{\text{sim}}_{k^\ast}$; second differences
$\ME^{(2)+}_{ij,k\cdot l}$ subtract two such gradients.

\paragraph{What this captures.}
The frozen-chain gradient plugged into \eqref{eq:delta} propagates
\emph{parameter uncertainty} $\Sigmah$ through the choice model while holding
the realised decision contexts fixed. For observed-wave AMEs this is the whole
story. For model-implied and period-level effects it is the leading term but omits
the path-distribution sensitivity $g_{\mathrm{re}}$; the correction
(Section~\ref{sec:reinforce}) is estimated from the \emph{same} frozen chains
and requires no further chain simulation.

\dmsubsection{Period-level endpoint: a linear sensitivity recursion}
\label{sec:periodjac}

The period-level quantity is not a function of one choice probability but the
endpoint of the propagation
recursion\ifstandalone\else~\eqref{eq:piRecursion}\fi. Collecting the terms in
the carried state rewrites that recursion as an affine map,
\begin{equation}
\pi_{ij,k}
\;=\;
c_{ij,k}\;+\;\pi_{ij,k-1}\,\rho_{ij,k},
\qquad
\rho_{ij,k}\;\equiv\;1-c_{ij,k}-d_{ij,k},
\label{eq:piAffine}
\end{equation}
in which each opportunity adds a fresh creation term and retains the incoming
probability with factor $\rho_{ij,k}$. This $\rho$ is the attenuation the
recursion carries: it is near one when both branch probabilities are small, so
that early opportunities are passed forward almost undamped, and shrinks as
either branch becomes likely.

Because \eqref{eq:piAffine} is affine in $\pi_{ij,k-1}$, its derivative is a
\emph{linear} recursion. Writing
$s_{ij,k,m}\equiv\partial\pi_{ij,k}/\partial\beta_m$ and noting that
$\pi_{ij,0}=x_{ij}(t_1)$ is data, so that $s_{ij,0,m}=0$,
\begin{equation}
s_{ij,k,m}
\;=\;
\bigl(1-\pi_{ij,k-1}\bigr)\frac{\partial c_{ij,k}}{\partial\beta_m}
\;-\;
\pi_{ij,k-1}\,\frac{\partial d_{ij,k}}{\partial\beta_m}
\;+\;
\rho_{ij,k}\,s_{ij,k-1,m}.
\label{eq:piSens}
\end{equation}
Unrolling \eqref{eq:piSens} to the endpoint gives the closed form
\begin{equation}
\frac{\partial \pi_{ij,K}}{\partial\beta_m}
\;=\;
\sum_{k=1}^{K}
\Bigl[
\bigl(1-\pi_{ij,k-1}\bigr)\frac{\partial c_{ij,k}}{\partial\beta_m}
-
\pi_{ij,k-1}\,\frac{\partial d_{ij,k}}{\partial\beta_m}
\Bigr]
\prod_{l=k+1}^{K}\rho_{ij,l},
\label{eq:piJac}
\end{equation}
a weighted sum of per-opportunity softmax derivatives, each discounted by the
attenuation accumulated \emph{after} it. The same $\rho$ that damps the
probability therefore damps its sensitivity, and \eqref{eq:piJac} is the one
place in this development where the product $\prod_l\rho_{ij,l}$ is actually
computed rather than merely invoked.

Two features make this cheap. The branch derivatives
$\partial c_{ij,k}/\partial\beta_m$ and $\partial d_{ij,k}/\partial\beta_m$ are
the softmax Jacobian \eqref{eq:softmaxjac} evaluated at opportunity $k$ on the
absent and present branch respectively --- the same objects the ministep
gradient \eqref{eq:fdgrad} already forms, so a table reporting both scopes can
share them. And the homogeneous factor $\rho_{ij,k}$
does not depend on~$m$, so a single scalar per opportunity propagates all $K$
sensitivities at once: one forward pass carrying $K$ sensitivities alongside
$\pi$ replaces the $2K$ passes a finite-difference Jacobian requires, and is
exact rather than $O(\varepsilon^2)$.

The marginal effect contrasts two endpoints computed under their own branch
systems, so its Jacobian is the difference of two applications of
\eqref{eq:piJac}, exactly as a first difference subtracts two applications of
\eqref{eq:softmaxjac}. The derivation assumes only that the utility is linear
in~$\beta$; direction-split and interacted effects break that assumption here
for the same reason they break it for \eqref{eq:fdgrad}, and fall back to
finite differences in the same way.

\dmsection{Dependence-aware correction via the score-function identity}
\label{sec:reinforce}

For model-implied and period-level quantities the averaging measure (the
distribution of ministep states visited by the process) shifts with~$\beta$.
Following \citet{SchweinbergerSnijders07a}, denote by $Z$ the complete data,
consisting of the holding times and the sequence of arc changes between panel
waves, with probability density $p_{\lambda,\beta}(Z)=dP_{\lambda,\beta}/d\mu_Z$
with respect to a reference measure $\mu_Z$. The contribution table $\mathbf{T}(Z)$ is
a deterministic function of $Z$, so any expected-value quantity can be written as a
functional $M(\beta,Z)$ of the complete-data path. The EV target is then
\[
M(\beta) \;=\; \E_{\beta}[M(\beta,Z)]
= \int M(\beta,Z)\,p_{\lambda,\beta}(Z)\,d\mu_Z(Z),
\]
where the expectation is over the latent complete-data path at fixed parameter
$(\lambda,\beta)$.

Differentiating under the integral sign and using the product rule gives
\[
\frac{\partial}{\partial\beta_k} M(\beta)
= \int \frac{\partial M(\beta,Z)}{\partial\beta_k}\,p_{\lambda,\beta}(Z)\,d\mu_Z(Z)
+ \int M(\beta,Z)\,\frac{\partial p_{\lambda,\beta}(Z)}{\partial\beta_k}\,d\mu_Z(Z).
\]
For any differentiable family $p_{\lambda,\beta}$ with
$p_{\lambda,\beta}(Z)>0$ on its support we can write
\[
\frac{\partial p_{\lambda,\beta}(Z)}{\partial\beta_k}
= p_{\lambda,\beta}(Z)\,\frac{\partial\log p_{\lambda,\beta}(Z)}{\partial\beta_k}
= p_{\lambda,\beta}(Z)\,S_k(Z),
\]
where
\[
S_k(Z)
= \frac{\partial\log p_{\lambda,\beta}(Z)}{\partial\beta_k}
\]
is the \emph{score} of the complete-data distribution. Substituting and
re-expressing both integrals as expectations yields the standard
score-function identity
\begin{equation}
\frac{\partial}{\partial\beta_k}\,\E_\beta[M]
\;=\;
\underbrace{\E_\beta\!\Bigl[\tfrac{\partial M}{\partial\beta_k}(\beta,Z)\Bigr]}_{g_{\mathrm{cond},k}\ \text{(frozen-path term)}}
\;+\;
\underbrace{\E_\beta\!\bigl[\,M(\beta,Z)\,S_k(Z)\,\bigr]}_{g_{\mathrm{re},k}\ \text{(path-distribution term)}}.
\label{eq:reinforce}
\end{equation}
The first term differentiates the integrand $M(\beta,Z)$ while holding the
realised path $Z$ fixed (the frozen-path view of Section~\ref{sec:cond}); the
second term captures how the distribution of paths changes with $\beta$
through the score $S_k(Z)$. Equation~\eqref{eq:reinforce} is a general
mathematical identity: it holds for any differentiable family of distributions
and any $\beta$-differentiable integrand $M$
\citep[REINFORCE / score-function estimator;][]{Williams1992}.

In practice, the score correction is estimated with the same Dolby-style
control-variate adjustment used in the implementation. For each parameter
component $k$, the Monte Carlo score term is formed from chainwise quantities
$Q_c$ and $S_{c,k}$, where $Q_c$ is the chain-level contribution to the QoI and
$S_{c,k}$ is the corresponding score contribution. The raw score estimate is
\[
\hat g_{\mathrm{re},k}^{\,\mathrm{raw}}
=
\frac{1}{n_3}\sum_{c=1}^{n_3} Q_c\,S_{c,k},
\]
and the Dolby correction subtracts a regression baseline chosen to minimize
variance:
\begin{equation}
\hat g_{\mathrm{re},k}
=
\frac{1}{n_3}\sum_{c=1}^{n_3}
\bigl(Q_c - a_k\bigr)\,S_{c,k},
\qquad
a_k=\frac{\Cov(Q_cS_{c,k},\,S_{c,k})}{\Var(S_{c,k})},
\label{eq:dolby}
\end{equation}
with the usual zero-variance guard setting $a_k=0$ when needed. This is the
Dolby-corrected REINFORCE term used for \texttt{deltaFull}; it reduces Monte
Carlo variance without changing the target expectation because the score has
mean zero.

\citet{SchweinbergerSnijders07a} apply the same identity to estimate the
Jacobian of the SAOM moment function during MoM estimation. In their setting
$M(Z)=s_k$ is a target statistic (a network statistic evaluated on the
simulated complete-data path, with no direct $\beta$-dependence), so the
conditional term vanishes and the identity reduces to
\[
\frac{\partial}{\partial\beta_k}\E_\beta[z_\ell]
\;=\;
\E_\beta[z_\ell\,S_k(Z)].
\]
This is what RSiena's internal Monte-Carlo derivative estimator computes.

In the SAOM the complete-data chain likelihood factors over ministeps as a
product of conditional choice probabilities,
$p(\text{chain }c\mid\beta) = \prod_{m\in c}p_{i_m j_m}(x^{(m)};\beta)$.
Taking logs and differentiating gives
\[
\begin{aligned}
\log p(\text{chain }c\mid\beta)
&= \sum_{m\in c}\log p_{i_m j_m}(x^{(m)};\beta),\\
S_{c,k}
&= \frac{\partial}{\partial\beta_k}\log p(\text{chain }c\mid\beta)
= \sum_{m\in c}\frac{\partial}{\partial\beta_k}\log p_{i_m j_m}(x^{(m)};\beta).
\end{aligned}
\]
With the conditional logit objective~\eqref{eq:obj} this simplifies to the
familiar ``observed minus expected'' change-statistic form
\begin{equation}
S_{c,k}
= \sum_{m\in c}\Bigl[\,\delta_{i_m j_m,k}(x^{(m)})
- \E_{p_{i_m}}\delta_{i_m\cdot,k}(x^{(m)})\,\Bigr],
\label{eq:ssc}
\end{equation}
the per-chain sum of score contributions at each ministep
\citep[the \texttt{ssc} array;][Appendix~A.9]{SchweinbergerSnijders07a,
Schweinberger2007}. All temporal and structural dependence of the process is
encoded in the states $x^{(m)}$ and the choice probabilities $p_{i_m j_m}$;
the log-likelihood nonetheless decomposes as a sum, so its derivative
decomposes as a sum.

In our setting $M(\beta,Z)$ is an AME that depends on $\beta$ both through the
chain distribution (captured by $g_{\mathrm{re}}$) and through the within-chain
choice probabilities (captured by $g_{\mathrm{cond}}$,
Section~\ref{sec:cond}). Both terms of~\eqref{eq:reinforce} are therefore
present; equation~\eqref{eq:ssc} is borrowed purely for the explicit form of
the score $S_k$ in the second term.

The label \emph{frozen-chain} applies specifically to $g_{\mathrm{cond}}$:
chain paths $Z$ (and hence their contribution tables $\Delta(Z)$) are held
fixed when evaluating the within-path Jacobian~\eqref{eq:fdgrad}. The score
correction $g_{\mathrm{re}}$ does not need new chains: the per-chain pairs
$(Q_c, S_{c,k})$ used in~\eqref{eq:dolby} come from the same $n_3$ chains,
provided scores were recorded at simulation time
(\texttt{includeScores\,=\,TRUE}). Consequently \texttt{deltaFull} computes
the complete gradient $g_{\mathrm{full}}=g_{\mathrm{cond}}+g_{\mathrm{re}}$
entirely from the chains already at hand; no additional simulation is needed
relative to \texttt{delta}.

\paragraph{Variance decomposition.}
By the law of total variance the variance of a model-implied AME splits into
three pieces,
\begin{equation}
\Var[M(\betah)]
=\underbrace{\E_{\Delta}\!\bigl[\Var_\beta[M\mid\Delta]\bigr]}_{\text{(A) parameter unc., frozen chains}}
+\underbrace{\Var_{\Delta}\!\bigl[\E_\beta[M\mid\Delta]\bigr]}_{\text{(B) chain Monte-Carlo noise}}
+\underbrace{2\,\Cov(\cdot,\cdot)}_{\text{(C) path-distribution sensitivity}} .
\label{eq:vardecomp}
\end{equation}

In the main text\ifstandalone\else{} and Appendix~\ref{A_uncertainty}\fi{} we emphasise that
$\mathrm{AME}^{\mathrm{obs}}$ and $\mathrm{AME}^{\mathrm{sim}}$ are
\emph{expected-value} (EV) quantities: $M(\beta)$ stands for
$\E_{\Delta\mid\beta}[M(\beta,\Delta)]$, the average behaviour of the process
at fixed~$\beta$, not the outcome of a particular single trajectory. For this
EV target, term (B) in~\eqref{eq:vardecomp} is pure Monte Carlo approximation
error from using finitely many chains $n_3$ to approximate the expectation and
should be made negligible by choosing $n_3$
large\ifstandalone\else, as discussed in Appendix~\ref{A_uncertainty}\fi.
Terms (A) and (C) then describe how sampling
uncertainty in $\hat\beta$ propagates to the EV itself: (A) via the frozen-chain
Jacobian and (C) via the path-distribution sensitivity captured by the score
term.

\dmsection{Summary of analytic Jacobians}
\label{sec:jacobians}

Table~\ref{tab:jacobians} collects every QoI, its analytic Jacobian, and the
corresponding SE mode.  For aggregate quantities the Jacobian is the
per-dyad expression from Section~\ref{sec:cond} averaged or summed over the
relevant decision contexts; the same vector is plugged into \eqref{eq:delta}.
Mode A\,=\,\texttt{"delta"} (frozen-chain, captures variance channel (A));
mode F\,=\,\texttt{"deltaFull"} (adds the score correction, captures (A)+(C));
FD\,=\,finite-difference fallback currently used in the implementation.

\begin{table}[ht]
\centering\small
\setlength{\tabcolsep}{4pt}
\caption{Analytic Jacobians and SE mode for each QoI.
$\bar\tau_{A_i,m}$ is defined in \eqref{eq:softmaxjac};
  ${}^{\dagger}$Jacobian analytically derived; FD fallback in current code;
  ${}^{\ddagger}$mode~F unavailable for this row, which returns the
  frozen-chain SE (Section~\ref{sec:impl}).}
\label{tab:jacobians}
\begin{tabular}{@{}p{3.1cm}p{5.4cm}p{2.8cm}l@{}}
\toprule
QoI & Jacobian $\partial M/\partial\beta_m$ & SE & Mode \\
\midrule
Toggle prob.\ $p_{ij}$ & 
  $p_{ij}\bigl[\tau_{ij,m}(x)-\bar\tau_{A_i,m}\bigr]$
  \eqref{eq:softmaxjac} &
  $\sqrt{g^\top\Sigmah g}$ &
  A \\[5pt]
Creation\; $\ME^+_{ij,k}$ &
  \eqref{eq:fdgrad} with $d_{ij}=+1$ &
  $\sqrt{g^\top\Sigmah g}$ &
  A \\[5pt]
Dissolution\; $\ME^-_{ij,k}$ &
  Negated \eqref{eq:fdgrad} with $d_{ij}=-1$ &
  $\sqrt{g^\top\Sigmah g}$ &
  A \\[5pt]
Tie-presence\; $\ME^{\mathrm{tie}}_{ij,k}$ &
  $\partial\ME^+/\partial\beta_m$ if $x_{ij}=0$;\;
  $-\partial\ME^-/\partial\beta_m$ if $x_{ij}=1$ &
  $\sqrt{g^\top\Sigmah g}$ &
  A \\[5pt]
Second diff.\ $\ME^{(2)+}_{ij,k\cdot l}$ &
  Difference of two \eqref{eq:fdgrad} gradients &
  $\sqrt{g^\top\Sigmah g}$ &
  A \\[5pt]
\midrule
$\mathrm{AME}^{\mathrm{obs}}_k$ &
  Avg.\ of $\partial\ME^{\mathrm{tie}}_{ij,k}/\partial\beta_m$
  over observed $x_t$ &
  $\sqrt{g^\top\Sigmah g}$, exact &
  A \\[5pt]
$\mathrm{AME}^{\mathrm{sim}}_k$ &
  Avg.\ per-ministep \eqref{eq:fdgrad};
  $+\hat g_{\mathrm{re},m}$ for mode~F &
  $\sqrt{g^\top\Sigmah g}$ &
  A, F \\[5pt]
$\overline{\mathrm{AME}}^{\mathrm{acc}}_k$ &
  Actor-level sum of per-ministep \eqref{eq:fdgrad};
  $+\hat g_{\mathrm{re},m}$ for mode~F &
  $\sqrt{g^\top\Sigmah g}$ &
  A$^\dagger$, F$^\dagger$ \\[5pt]
Period-level\; $\overline{\mathrm{ME}}^{\pi}_k$ &
  Difference of two sensitivity recursions \eqref{eq:piJac}, one per branch
  system &
  $\sqrt{g^\top\Sigmah g}$, frozen-chain &
  A$^\dagger$, FD$^\ddagger$ \\[5pt]
Rate-weighted\; $\ME^q_{ij,k}$ &
  $\lambda_i\,\partial\ME/\partial\beta_m
   +\ME_{ij,k}\,\partial\lambda_i/\partial\beta_m$ &
  $\sqrt{g^\top\Sigmah g}$ &
  FD \\
\bottomrule
\end{tabular}
\end{table}

\dmsection{Implementation in RSiena}
\label{sec:impl}

The procedure is implemented in the postestimation layer and exposed through a
single argument:
\begin{center}
\verb|marginalEffects(fit, data, ...,|\\
\verb|    uncertaintyMode = c("bootstrap","delta","deltaFull"))|.
\end{center}
\begin{itemize}\setlength\itemsep{2pt}
\item \texttt{"bootstrap"} --- the Krinsky--Robb / simulation method used in the
  main paper (draw $\beta^{(b)} \sim \mathcal{N}(\betah,\Sigmah)$, re-evaluate;
  re-simulate chains for dynamic quantities). Captures (A)+(C) for dynamic,
  (A) for observed-wave.
\item \texttt{"delta"} --- frozen-chain analytic SE: build $g_{\mathrm{cond}}$
  from \eqref{eq:softmaxjac}--\eqref{eq:fdgrad} on the contribution matrix and
  apply \eqref{eq:delta}. Captures (A) (and inherits (B)). For observed-wave
  quantities this is the full, essentially exact SE.
\item \texttt{"deltaFull"} --- adds the Dolby-corrected score term 
  corresponding to \eqref{eq:reinforce}, requiring the per-chain scores \eqref{eq:ssc} (obtained
  by simulating the chains with \texttt{includeScores = TRUE}). Captures
  (A)+(C) analytically.
\end{itemize}
Internally the per-row SE is formed as
$\mathrm{SE}_r=\sqrt{\sum_k (J\Sigmah)_{rk}\,J_{rk}}$ from the Jacobian $J$ of
stacked output rows --- i.e.\ $\sqrt{\operatorname{diag}(J\Sigmah J^\top)}$ at
$O(nK^2)$ rather than $O(n^2 K)$ cost. The analytic Jacobian \eqref{eq:fdgrad}
runs in one sweep over the contribution matrix; a finite-difference Jacobian
($2K$ sweeps) is retained as a fallback and is currently used automatically for
the \emph{period-level} effect ($\overline{\mathrm{ME}}^{\pi}_k$) and for
\emph{rate-weighted} quantities.  For the period-level case the finite-difference
route supplies the same quantity at a higher cost rather than a different one:
the sensitivity recursion \eqref{eq:piSens} gives the Jacobian in closed form,
but it is not yet implemented, so the fallback is a matter of implementation
rather than of principle.

The score correction of Section~\ref{sec:reinforce} is a separate matter, and
it does \emph{not} currently reach the period-level rows.  The bucketed
estimator of $\hat g_{\mathrm{re}}$ needs the per-chain totals $(Y_c,N_c)$ of
the quantity being corrected, which it assembles from a row-wise prediction
keyed to chains; the period-level predictor returns a per-dyad endpoint frame
instead, and the bucketing cannot be built from it.  Requesting
\texttt{"deltaFull"} therefore returns the \emph{conditional} standard error
for those rows, with a warning naming them, while the ministep rows in the same
table do receive the correction.  The gap is in how the quantity is handed to
the estimator, not in the identity \eqref{eq:reinforce}, which holds for any
$\beta$-differentiable functional of the path and so covers $\pi_{ij,K}$; what
has not been checked is whether the bucketing granularity and the Dolby
baseline of \eqref{eq:dolby} carry over unchanged to an endpoint quantity that
contributes one value per dyad per chain rather than a count of events.  Until
that is settled, period-level standard errors reported under either delta mode
should be read as frozen-chain quantities, and
Section~\ref{sec:reinforce}'s caution applies with full force: the omitted term
is a covariance, so the conditional standard error is not a bound.  On the
Glasgow model the correction is not negligible where it does apply --- the
ministep standard errors rise by a factor of $1.6$ to $2.3$ when it is
switched on, with the point estimates unchanged.

\section{Scaling geometry in the SAOM choice set}
\label{app:scaling_details}

Section~\ref{S_Scaling} reports that probability-scale quantities do not inherit
the scale-robustness that the binary-logit literature relies on. This appendix
gives the reason.

For a counterfactual that shifts the systematic preference of one alternative
$(i\to j)$ by $\Delta f_{ij}$ while holding all other alternatives fixed, the
induced probability change under an explicit scale $\sigma$ is, to first order,
\begin{equation}
    \Delta p_{ij}(\sigma \mid x, \Delta f_{ij}) \approx
    p_{ij}(x;\sigma)\big(1-p_{ij}(x;\sigma)\big)\frac{\Delta f_{ij}}{\sigma}.
\label{eq:Delta_p_local_short}
\end{equation}
This is the one-alternative case, which is the mode used in the illustrations of
Section~\ref{S_Scaling}. When several alternatives shift together, the
$p_{ij}(1-p_{ij})$ factor is replaced by the deviation of the focal shift from
the probability-weighted mean shift over the choice set, as in the softmax
Jacobian~\eqref{eq:softmaxjac} of Appendix~\ref{app:deltaMethod}.

Equation~\eqref{eq:Delta_p_local_short} has two channels: the direct scaling
channel $1/\sigma$, and the $p_{ij}(1-p_{ij})$ logit-slope channel. Whether
the second offsets the first depends on where added noise drives the
probability, and this is where the binary and the multinomial case part company.
In binary logit the attractor and the point of maximum slope coincide at $0.5$,
so noise always moves a probability toward greater responsiveness, and the
slope channel always works against the attenuation. That is what
licenses the usual robustness claim. In the SAOM the two points differ:
probabilities are driven toward the uniform value $1/K$ over the $K$
alternatives, while the mapping is still steepest at $0.5$. A probability below
$1/K$ is moved up, which does open the channel, but only as far as $1/K$ and
never to $0.5$; and a probability lying between $1/K$ and $0.5$ is moved away
from the steepest point, so that the two channels can compound rather than
offset. Neither case has to be the common one for the conclusion to follow: the
compensation available is capped by the distance from the attractor to the
optimum, and for a choice set of the size a network model generates that
distance is large. At $K=129$ the attractor sits at $p(1-p)\approx 0.008$,
against $0.25$ at $p=0.5$.

\begin{figure}[htbp]
\centering
\includegraphics[width=\textwidth]{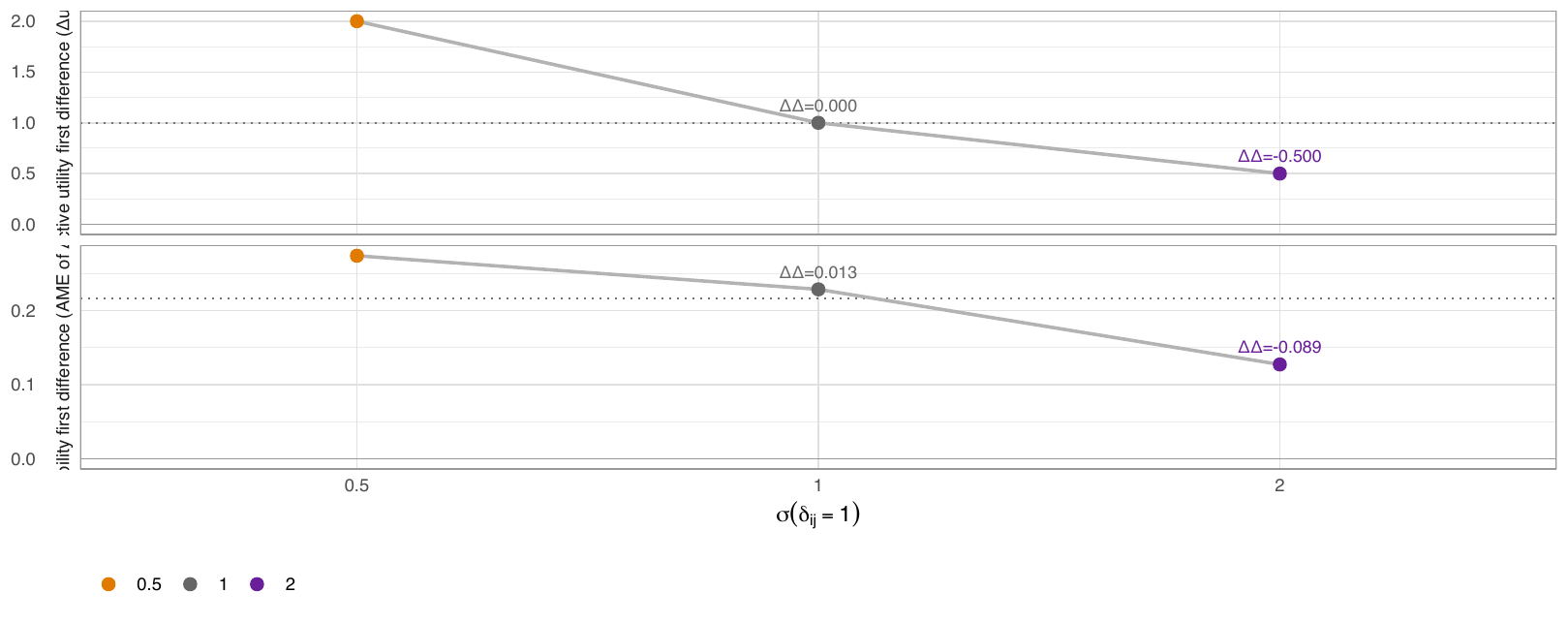}
\caption{Alter-dependent scaling ($\sigma$ depends on $\delta_{ij,\mathrm{altX}}$)
induces moderation of an ego contrast ($\delta_{i,\mathrm{egoX}}\!:\,0\to 1$):
the effective-objective function first difference $\Delta f/\sigma$ and the probability
first difference differ across alter contexts even without an explicit interaction term.}
\label{fig:alterscaling_moderation}
\end{figure}

The same mechanism also operates \emph{within} a data set. If $\sigma$ depends on
an alter attribute, an ego-level contrast is rescaled differently depending on
which alter is considered, so it appears to interact with that attribute on the
probability scale even though the objective function contains no interaction
term. This is the scaling-induced interaction of Section~\ref{S_Explanation}.
Figure~\ref{fig:alterscaling_moderation} shows it for an ego contrast
$(\delta_{i,\mathrm{egoX}}\!:\,0\to 1)$ under alter-dependent scaling.

\clearpage

\section{Omitted-variable experiments: full results}
\label{app:experiment_tables}

\begin{table}[!htbp]
\centering
\caption{Data-generating process for the two omitted-variable experiments}
\label{tab:experiment_design}
\footnotesize
\begin{tabular}{@{}lll@{}}
\toprule
 & Experiment~1 (orthogonal) & Experiment~2 (related) \\
\midrule
\emph{Shared structural core} & \multicolumn{2}{c}{} \\
\quad actors $n$ & \multicolumn{2}{l}{50} \\
\quad observations & \multicolumn{2}{l}{2 (one period)} \\
\quad starting network & \multicolumn{2}{l}{density 0.046, mean degree 2.26} \\
\quad outdegree & \multicolumn{2}{l}{-2.5} \\
\quad reciprocity & \multicolumn{2}{l}{2.0} \\
\quad transitive triplets & \multicolumn{2}{l}{0.5} \\
\quad rate $\lambda$ & \multicolumn{2}{l}{7.0 opportunities per actor} \\
\addlinespace
\emph{Omitted covariate effect} & & \\
\quad covariate & iid $N(0,1)$, dyadic & iid, 3 categories, monadic \\
\quad effect & \texttt{X} (direct dyadic) & \texttt{sameX} (homophily) \\
\quad coefficient & 1.0 & 2.0 \\
\quad covariate sd & $1$ & $\approx0.47$ \\
\quad semi-standardized & $\approx 1$ & $\approx 1$ \\
\addlinespace
\emph{What the process generates} & & \\
\quad endpoint density & 0.066 & 0.145 \\
\quad endpoint mean degree & 3.2 & 7.1 \\
\addlinespace
\emph{Estimation} & & \\
\quad method & \multicolumn{2}{l}{conditional method of moments, 4 subphases, 1000 phase-3 simulations} \\
\quad replications used & 107 & 115 \\
\midrule
\multicolumn{3}{p{0.88\textwidth}}{\footnotesize The coefficients differ because the covariates do---a unit change is one standard deviation of $W$ but a whole same-category indicator---so both carry approximately the same semi-standardized effect.  Homophily adds tie-creation pressure an orthogonal covariate does not, hence the different endpoint densities.}\\
\bottomrule
\end{tabular}
\end{table}

\begin{table}[!htbp]
\centering
\caption{Omitted-variable experiments: levels under each specification and the change between them}
\label{tab:experiment_ratios}
\footnotesize
\begin{tabular}{@{}lrrrr@{}}
\toprule
 & full (sd) & reduced (sd) & \% change (MCSE) & gap/sd (MCSE) \\
\midrule
\multicolumn{5}{@{}l}{\textbf{Experiment 1: orthogonal dyadic covariate ($n_{\mathrm{rep}} = 107$)}} \\
\emph{Coefficients} & \multicolumn{4}{c}{} \\
\quad density & -2.498 (0.149) & -1.919 (0.081) & -23.1 (0.29) & 5.62 (0.40) \\
\quad reciprocity & 1.975 (0.223) & 1.527 (0.166) & -22.6 (0.31) & 4.91 (0.35) \\
\quad transitive triplets & 0.480 (0.114) & 0.364 (0.080) & -23.7 (0.32) & 3.05 (0.23) \\
\quad rate $\lambda$ & 6.948 (0.983) & 9.922 (1.447) & +42.9 (0.55) & 5.19 (0.37) \\
\emph{Per-ministep} $\mathrm{AME}^{\mathrm{sim}}$ & \multicolumn{4}{c}{} \\
\quad reciprocity & 0.060 (0.012) & 0.042 (0.007) & -30.3 (0.37) & 3.55 (0.26) \\
\quad transitive triplets & 0.009 (0.002) & 0.007 (0.002) & -21.7 (0.33) & 2.67 (0.21) \\
\emph{Period-level} $\overline{\mathrm{ME}}^{\pi}$ & \multicolumn{4}{c}{} \\
\quad reciprocity & 0.228 (0.040) & 0.231 (0.043) & +1.3 (0.20) & 0.67 (0.11) \\
\quad transitive triplets & 0.032 (0.009) & 0.031 (0.008) & -3.5 (0.22) & 1.34 (0.13) \\
\addlinespace
\multicolumn{5}{@{}l}{\textbf{Experiment 2: related homophily effect ($n_{\mathrm{rep}} = 115$)}} \\
\emph{Coefficients} & \multicolumn{4}{c}{} \\
\quad density & -2.531 (0.206) & -1.777 (0.105) & -29.6 (0.28) & 5.83 (0.40) \\
\quad reciprocity & 2.034 (0.217) & 2.284 (0.175) & +12.6 (0.42) & 3.63 (0.26) \\
\quad transitive triplets & 0.512 (0.095) & 0.445 (0.044) & -11.7 (0.79) & 1.25 (0.12) \\
\quad rate $\lambda$ & 6.949 (0.627) & 9.976 (1.259) & +43.3 (0.90) & 3.78 (0.27) \\
\emph{Per-ministep} $\mathrm{AME}^{\mathrm{sim}}$ & \multicolumn{4}{c}{} \\
\quad reciprocity & 0.065 (0.011) & 0.090 (0.013) & +40.0 (0.97) & 4.82 (0.33) \\
\quad transitive triplets & 0.010 (0.002) & 0.009 (0.001) & -9.9 (0.80) & 1.11 (0.12) \\
\emph{Period-level} $\overline{\mathrm{ME}}^{\pi}$ & \multicolumn{4}{c}{} \\
\quad reciprocity & 0.254 (0.040) & 0.491 (0.053) & +95.2 (1.70) & 8.24 (0.55) \\
\quad transitive triplets & 0.043 (0.006) & 0.051 (0.004) & +20.3 (1.10) & 2.18 (0.17) \\
\midrule
\multicolumn{5}{p{0.92\textwidth}}{\footnotesize Levels are means over replications (sd); changes are formed within replication, then averaged, with their Monte Carlo standard error $\mathrm{sd}/\sqrt{n_{\mathrm{rep}}}$ in parentheses \citep{MorrisEtAl2019}.  The final column scales that gap by how much it varies from one dataset to the next, which is what a standard error of the difference estimates: it is the expected signal-to-noise of the comparison a researcher with one dataset and two nested specifications would make, so a value near $1$ or below means that comparison carries little power.  It too has a Monte Carlo standard error.  Change is in magnitude, so the negative density entry means a coefficient closer to zero.  Coefficients and the rate have generating values in Table~\ref{tab:experiment_design}; $\mathrm{AME}^{\mathrm{sim}}$ and $\overline{\mathrm{ME}}^{\pi}$ are evaluated at estimated parameters and do not.}\\
\bottomrule
\end{tabular}
\end{table}

\clearpage

\section{Full model vs.\ separate (maxLike) specification}
\label{app:full_vs_separate}

\begin{figure}[!htbp]
\centering
\includegraphics[width=0.92\textwidth]{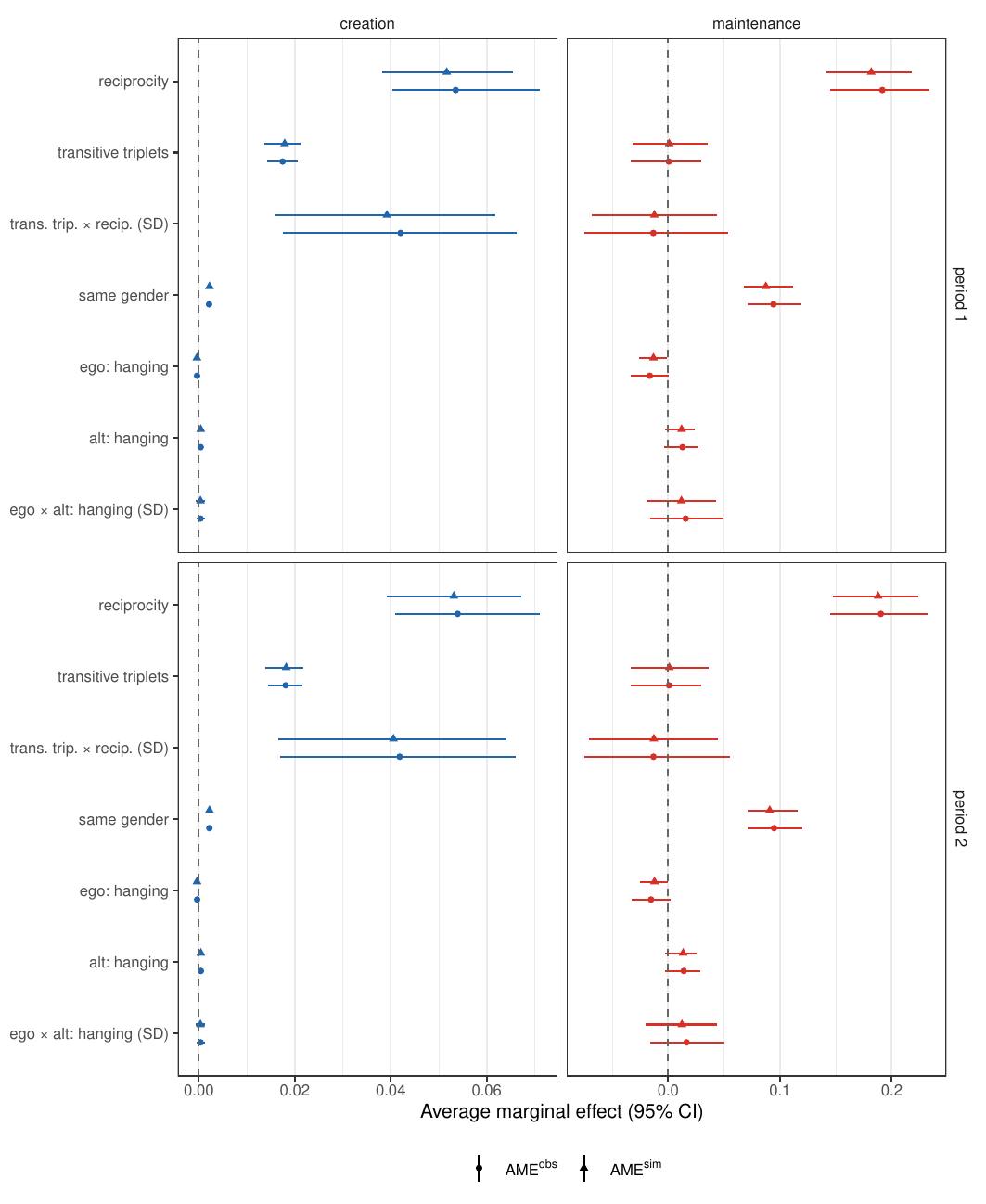}
\caption{Separate (maxLike) specification: AME$^{\text{obs}}$ and
AME$^{\text{sim}}$ by context (creation vs.\ maintenance) and period.
Layout mirrors Figure~\ref{fig:forest} for the combined model.}
\label{fig:forest_separate}
\end{figure}

Table~\ref{tab:glasgow_full_vs_separate} compares marginal effects for the
three structural effects under the combined (MoM) and separate (maxLike)
specifications; full coefficient estimates appear in
Table~\ref{tab:glasgow_estimation_table} (combined) and
Table~\ref{tab:glasgow_estimation_table_separate} (separate).
Figure~\ref{fig:forest_separate} shows the creation vs.\ maintenance
decomposition for the separate model, analogous to Figure~\ref{fig:forest}.

\begingroup
\footnotesize
\begin{table}[!htbp]
\centering
\caption{Glasgow friendship model: AMEs for structural effects --- full (MoM) vs.\ separate (maxLike) specification}
\label{tab:glasgow_full_vs_separate}
\footnotesize
\begin{tabular}{lrr|rr}
\toprule
  \multicolumn{1}{c}{} & \multicolumn{2}{c}{\textbf{Full model}} & \multicolumn{2}{c}{\textbf{Separate (maxLike)}} \\
\cmidrule(lr){2-3} \cmidrule(lr){4-5}
  \textbf{Effect} & \textbf{ME$^{\text{obs}}$ (SE)} & \textbf{ME$^{\text{sim}}$ (SE)} & \textbf{ME$^{\text{obs}}$ (SE)} & \textbf{ME$^{\text{sim}}$ (SE)} \\
\midrule
  reciprocity & 0.0568 (0.0035) & 0.0598 (0.0044) & 0.0571 (0.0077) & 0.0554 (0.0075) \\
  transitive triplets & 0.0174 (0.0008) & 0.0184 (0.0009) & 0.0174 (0.0015) & 0.0176 (0.0017) \\
  trans.~trip.\ $\times$ recip & 0.0120 (0.0056) & 0.0070 (0.0049) & 0.0407 (0.0123) & 0.0388 (0.0125) \\
\midrule
\multicolumn{5}{p{0.90\textwidth}}{\footnotesize $\widehat{\text{ME}}^{\text{obs}}$ = unconditional marginal effect at observed decision situations; $\widehat{\text{ME}}^{\text{sim}}$ = dynamic marginal effect at model-implied decisions. SE in parentheses. \textemdash\ = effect not included in this specification.}\\
\bottomrule
\end{tabular}
\end{table}

\endgroup

\clearpage

\section{Estimation results}
\label{app:estimation_tables}

The following tables report full estimation output (coefficients, standard
errors, convergence $t$-ratios) for the combined model, the
covariate-only specification, and the separate (maxLike) specification.
The structural-only model is omitted to save space.

\begin{table}[!htbp]
\centering
\caption{Glasgow friendship model: estimation output for the combined (MoM) specification.}
\label{tab:glasgow_estimation_table}
\footnotesize
\begin{tabular}{l | r@{.}l r@{.}l  | }
\hline
\rule{0pt}{2ex}\relax
Effect &\multicolumn{2}{c}{par.}&\multicolumn{2}{c }{ (s.e.) }  \\[0.5ex]
\hline
\rule{0pt}{2ex}\relax
constant friendship rate (period 1)    &   12 & 52 & (1 & 55)\\
constant friendship rate (period 2)    &   10 & 08 & (0 & 89)\\
\hline
\rule{0pt}{2ex}\relax
outdegree (density)                    &  --3 & 15$^{\ast\ast\ast}$ & (0 & 12)\\
reciprocity                            &    2 & 59$^{\ast\ast\ast}$ & (0 & 12)\\
transitive triplets (1)                &    1 & 24$^{\ast\ast\ast}$ & (0 & 07)\\
same fem                               &    0 & 76$^{\ast\ast\ast}$ & (0 & 09)\\
hng alter                              &    0 & 01                  & (0 & 09)\\
hng ego                                &  --0 & 32$^{\ast\ast}$     & (0 & 11)\\
hng ego x hng alter                    &    0 & 17                  & (0 & 13)\\
transitive triplets (1) x reciprocity  &  --0 & 56$^{\ast\ast\ast}$ & (0 & 14)\\
\hline
\multicolumn{5}{l}{\footnotesize{$^\dagger$ $p$ $<$ 0.1;
				$^\ast$ $p$ $<$ 0.05; $^{\ast\ast}$ $p$ $<$ 0.01;
				$^{\ast\ast\ast}$ $p$ $<$ 0.001;}}\\
\multicolumn{5}{l}
   {\footnotesize{convergence $t$ ratios all $<$ 0.1.}}\\
\multicolumn{5}{l}{\footnotesize{Overall maximum convergence ratio 0.12.}}
\end{tabular}

\end{table}

\begin{table}[!htbp]
\centering
\caption{Estimation output for the covariates-only specification.}
\label{tab:glasgow_estimation_table_exo}
\footnotesize
\begin{tabular}{l | r@{.}l r@{.}l  | }
\hline
\rule{0pt}{2ex}\relax
Effect &\multicolumn{2}{c}{par.}&\multicolumn{2}{c }{ (s.e.) }  \\[0.5ex]
\hline
\rule{0pt}{2ex}\relax
constant friendship rate (period 1)  &   5 & 93 & (0 & 34)\\
constant friendship rate (period 2)  &   5 & 24 & (0 & 33)\\
\hline
\rule{0pt}{2ex}\relax
outdegree (density)                  & --2 & 55$^{\ast\ast\ast}$ & (0 & 10)\\
same fem                             &   1 & 26$^{\ast\ast\ast}$ & (0 & 08)\\
hng alter                            & --0 & 16$^\dagger$        & (0 & 09)\\
hng ego                              & --0 & 35$^{\ast\ast\ast}$ & (0 & 10)\\
hng ego x hng alter                  &   0 & 42$^{\ast\ast\ast}$ & (0 & 13)\\
\hline
\multicolumn{5}{l}{\footnotesize{$^\dagger$ $p$ $<$ 0.1;
				$^\ast$ $p$ $<$ 0.05; $^{\ast\ast}$ $p$ $<$ 0.01;
				$^{\ast\ast\ast}$ $p$ $<$ 0.001;}}\\
\multicolumn{5}{l}
   {\footnotesize{convergence $t$ ratios all $<$ 0.08.}}\\
\multicolumn{5}{l}{\footnotesize{Overall maximum convergence ratio 0.1.}}
\end{tabular}

\end{table}

\begin{table}[!htbp]
\centering
\caption{Estimation output for the separate (maxLike) specification.}
\label{tab:glasgow_estimation_table_separate}
\footnotesize
\begin{tabular}{l | r@{.}l r@{.}l  | }
\hline
\rule{0pt}{2ex}\relax
Effect &\multicolumn{2}{c}{par.}&\multicolumn{2}{c }{ (s.e.) }  \\[0.5ex]
\hline
\rule{0pt}{2ex}\relax
constant friendship rate (period 1)               &   8 & 12 & (0 & 67)\\
constant friendship rate (period 2)               &   7 & 33 & (0 & 64)\\
\hline
\rule{0pt}{2ex}\relax
outdegree (density)                               & --3 & 31$^{\ast\ast\ast}$ & (0 & 09)\\
reciprocity maintenance                           &   1 & 62$^{\ast\ast\ast}$ & (0 & 26)\\
reciprocity creation                              &   3 & 69$^{\ast\ast\ast}$ & (0 & 20)\\
transitive triplets (1) maintenance               &   0 & 05                  & (0 & 19)\\
transitive triplets (1) creation                  &   2 & 20$^{\ast\ast\ast}$ & (0 & 15)\\
same fem                                          &   0 & 69$^{\ast\ast\ast}$ & (0 & 08)\\
hng alter                                         &   0 & 11$^\dagger$        & (0 & 06)\\
hng ego                                           & --0 & 12$^\dagger$        & (0 & 06)\\
hng ego x hng alter                               &   0 & 10                  & (0 & 12)\\
transitive triplets (1) x reciprocity maintenance & --0 & 11                  & (0 & 25)\\
transitive triplets (1) x reciprocity creation    & --1 & 37$^{\ast\ast\ast}$ & (0 & 22)\\
\hline
\multicolumn{5}{l}{\footnotesize{$^\dagger$ $p$ $<$ 0.1;
				$^\ast$ $p$ $<$ 0.05; $^{\ast\ast}$ $p$ $<$ 0.01;
				$^{\ast\ast\ast}$ $p$ $<$ 0.001;}}\\
\multicolumn{5}{l}
   {\footnotesize{convergence $t$ ratios all $<$ 0.14.}}\\
\multicolumn{5}{l}{\footnotesize{Overall maximum convergence ratio 0.23.}}
\end{tabular}

\end{table}

\clearpage

\bibliography{bib/CSbib, bib/RSiena, bib/Methods}

\end{document}